\documentclass[times, final, oneside]{elsarticle}
\usepackage{jcomp}
\usepackage{framed,multirow}
\usepackage{multicol}
\usepackage{amssymb}
\usepackage{amsmath}
\usepackage{amsthm}
\usepackage{amsfonts}
\usepackage{latexsym}
\usepackage{fontawesome}
\usepackage[%
    colorlinks=true,
    pdfborder={0 0 0},
    linkcolor=blue
]{hyperref}
\usepackage[dvipsnames]{xcolor}

\usepackage{algorithm}
\usepackage{algpseudocode}

\DeclareFixedFont{\ttb}{T1}{txtt}{bx}{n}{8} % for bold
\DeclareFixedFont{\ttm}{T1}{txtt}{m}{n}{8}  % for normal

\usepackage[dvipsnames]{xcolor}
\usepackage{listings}

\lstnewenvironment{python}[1][]
{
\pythonstyle
\lstset{#1}
}
{}

\theoremstyle{plain}

\theoremstyle{definition}

\theoremstyle{remark}

\usepackage{subcaption}
\usepackage{wrapfig}
\usepackage{mathtools}
\newcommand\vertarrowbox[3][16ex]{%
  \begin{array}[t]{@{}c@{}} #2 \\
  \left\downarrow\vcenter{\hrule height #1}\right.\kern-\nulldelimiterspace\\
  \makebox[0pt]{\scriptsize#3}
  \end{array}%
}

\makeatletter
\newcommand{\xdashrightarrow}[2][]{\ext@arrow 0359\rightarrowfill@@{#1}{#2}}
\newcommand{\xdashleftarrow}[2][]{\ext@arrow 3095\leftarrowfill@@{#1}{#2}}
\newcommand{\xdashleftrightarrow}[2][]{\ext@arrow 3359\leftrightarrowfill@@{#1}{#2}}
\def\rightarrowfill@@{\arrowfill@@\relax\relbar\rightarrow}
\def\leftarrowfill@@{\arrowfill@@\leftarrow\relbar\relax}
\def\leftrightarrowfill@@{\arrowfill@@\leftarrow\relbar\rightarrow}
\def\arrowfill@@#1#2#3#4{%
  $\m@th\thickmuskip0mu\medmuskip\thickmuskip\thinmuskip\thickmuskip
   \relax#4#1
   \xleaders\hbox{$#4#2$}\hfill
   #3$%
}
\@mparswitchfalse%
\makeatother

\reversemarginpar %if you want them on the left of your twosided document. 

\usepackage{booktabs}
\usepackage{colortbl}

\usepackage{pst-node}
\usepackage{tikz-cd}
\usepackage{tikz}

\usepackage{pifont}% http://ctan.org/pkg/pifont

\usepackage{url}
\usepackage{xcolor}
\definecolor{newcolor}{rgb}{.8,.349,.1}

\newif\ifincludecomments
\includecommentstrue
\includecommentsfalse

\ifincludecomments
  \newlength{\extramargin}
  \usepackage[color=red!20, textwidth=\extramargin, textsize=footnotesize]{todonotes}
  \newcommand{\jdm}[1]{\todo[color=red!20]{JDM: #1}}
  \newcommand{\map}[1]{\todo[color=red!20]{MAP: #1}}
\else
  \newcommand{\todo}[1]{}
  \newcommand{\jdm}[1]{}
  \newcommand{\map}[1]{}
\fi

\usepackage{soul}

\newcommand{\eqn}[1]{Eq.~#1}

\newcommand{\tbl}[1]{Table~#1}

\newcommand{\fig}[1]{Fig.~#1}

\newcommand{\sectn}[1]{Sec.~#1}

\newcommand{\img}{\ensuremath{\text{i}}}
\renewcommand{\exp}[1]{\ensuremath{\text{e}^{#1}}}
\newcommand{\dx}{\ensuremath{\text{d}}}
\newcommand{\sphere}{\ensuremath{{\mathbb{S}^2}}}
\newcommand{\sothree}{\ensuremath{{\text{SO}(3)}}}

\journal{Journal of Computational Physics}
\begin{document}

\verso{Price \& McEwen}

\begin{frontmatter}

  \title{Differentiable and accelerated spherical harmonic and Wigner transforms}%
  \author[1]{Matthew~A.~{Price}\corref{cor1}}
  \author[1,2]{Jason~D.~{McEwen}\corref{cor1}}
  \cortext[cor1]{Correspondence e-mail: \{m.price.17, jason.mcewen\}@ucl.ac.uk}

  \address[1]{Mullard Space Science Laboratory, University College London, Dorking, RH5 6NT, UK.}
  \address[2]{Alan Turing Institute, London, NW1 2DB, UK.}
  % \address[3]{Advanced Research Computing, University College London, Gower Street, London WC1E 6BT, UK.}

  \received{---}
  \finalform{---}
  \accepted{---}
  \availableonline{---}
  \communicated{---}

  \begin{abstract}
    %%%
    Many areas of science and engineering encounter data defined on spherical manifolds. Modelling and analysis of spherical data often necessitates spherical harmonic transforms, at high degrees, and increasingly requires efficient computation of gradients for machine learning or other differentiable programming tasks.
    %%%
    We develop novel algorithmic structures for accelerated and differentiable computation of generalised Fourier transforms on the sphere $\sphere$ and rotation group $\sothree$, \textit{i.e.}\ spherical harmonic and Wigner transforms, respectively.  We present a recursive algorithm for the calculation of Wigner $d$-functions that is both stable to high harmonic degrees and extremely parallelisable. By tightly coupling this with separable spherical transforms, we obtain algorithms that exhibit an extremely parallelisable structure that is well-suited for the high throughput computing of modern hardware accelerators (\textit{e.g.}\ GPUs).  We also develop a hybrid automatic and manual differentiation approach so that gradients can be computed efficiently.  Our algorithms are implemented within the \texttt{JAX} differentiable programming framework in the \texttt{S2FFT} {\href{https://github.com/astro-informatics/s2fft}{\faGithub}} software code.  Numerous samplings of the sphere are supported, including equiangular and \texttt{HEALPix} sampling.   Computational errors are at the order of machine precision for spherical samplings that admit a sampling theorem.  When benchmarked against alternative C codes we observe up to a 400-fold acceleration.  Furthermore, when distributing over multiple GPUs we achieve very close to optimal linear scaling with increasing number of GPUs due to the highly parallelised and balanced nature of our algorithms.  Provided access to sufficiently many GPUs our transforms thus exhibit an unprecedented effective linear time complexity.
    %%%
  \end{abstract}


\end{frontmatter}

\section{Introduction}

Research efforts in many fields of science and engineering, both in industrial and academic settings, are increasingly considering the analysis of data that live on spherical manifolds, or variants thereof. The diversity in applications is remarkable, ranging from quantum chemistry \cite{ritchie:1999, choi:1999}, molecular modelling \cite{boomsma:2017:spherical,kondor:2018:clebsch}, biomedical imaging \cite{tuch:2004, daducci:ssdmri, mcewen:s2let_ridgelets, goodwin-allcock:dmri_spherical_cnn}, geophysical inverse problems \cite{audet:2010, simons:2011, simons:2011a, marignier:s2proxmcmc}, general relativity and gravitational waves \cite{thorne:1980, beyer:2014:numerical, boyle:2016:transformations}, to the wider cosmos \cite{wallis:mass_mapping_sks, price:massmapping_uq_sphere, loureiro:2022:almanac, atkins:2023:atacama} and beyond. Reflecting the growth of scientific interest, consortiums within these fields have grown in both size and scope, with international missions including, \textit{e.g.}, ESA's Planck \cite{planck2016-l01} and Euclid \cite{pires:2020:euclid} satellite missions, the Dark Energy Survey \cite[DES;][]{jeffrey:2021:dark}, the Rubin Observatory Legacy Survey of Space and Time \cite[LSST;][]{ivezic:2019} and soon the Laser Interferometer Space Antenna \cite[LISA;][]{amaro:2017:laser}. As such, a substantial global community of researchers have an interest in the development and use of foundational technology for the analysis and modelling of spherical data.

For many of these applications a harmonic analysis of data can be insightful, in some cases critical.  Many physical models are best described in harmonic space (often due to the independent physical evolution of different harmonic modes) necessitating spherical harmonic transforms, for example in cosmology \citep{dodelson:2003} and seismology \citep{dahlen:1998}.  Furthermore, many analyses either rely on spherical harmonic transforms for a direct analysis or as an integral component of other analysis techniques, for example wavelets on the sphere \cite[\textit{e.g.}][]{mcewen:2006:fcswt,mcewen:s2let_localisation,mcewen:s2let_spin} or geometric machine learning on the sphere \cite[\textit{e.g.}][]{cohen:2018:spherical,kondor:2018:clebsch, esteves:2018, cobb:efficient_generalized_s2cnn,mcewen:scattering}.

Widespread uses of the harmonic analysis of spherical data, combined with increasingly large data volumes, heightens the need for scalable and efficient algorithms, with readily accessible and easy to use software, to compute generalised fast Fourier transforms on both the sphere $\sphere$ and rotation group $\sothree$.  Modern hardware accelerators, such as GPUs and TPUs, provide a great opportunity to leverage high throughput computing to address the increasing computational challenge.
With the growth of differentiable programming opening up many new types of analysis, many applications also require spherical transforms that are differentiable.  Machine learning models on the sphere require differentiable transforms so that the models may be trained by gradient-based optimisation algorithms.  Emerging physics-enhanced machine learning approaches \cite{karniadakis:2021} also require differentiable physical models, which in many cases themselves require differentiable spherical transforms, for hybrid data-driven and model-based approaches \cite[\textit{e.g.}][]{liaudat:2021, mars:learned_imaging_spider, mcewen:proxnest_learned_maxent,campagne:2023, piras:2023}.
While a number of software codes are available to compute spherical harmonic transforms, these do not typically run on hardware accelerators (\textit{e.g.}\ GPUs) or support differentiable transforms.

\begin{figure}
   \centering
   \begin{subfigure}{.23\textwidth}
      \centering
      \includegraphics[width=0.8\linewidth, trim={13cm 13cm 13cm 13cm},clip]{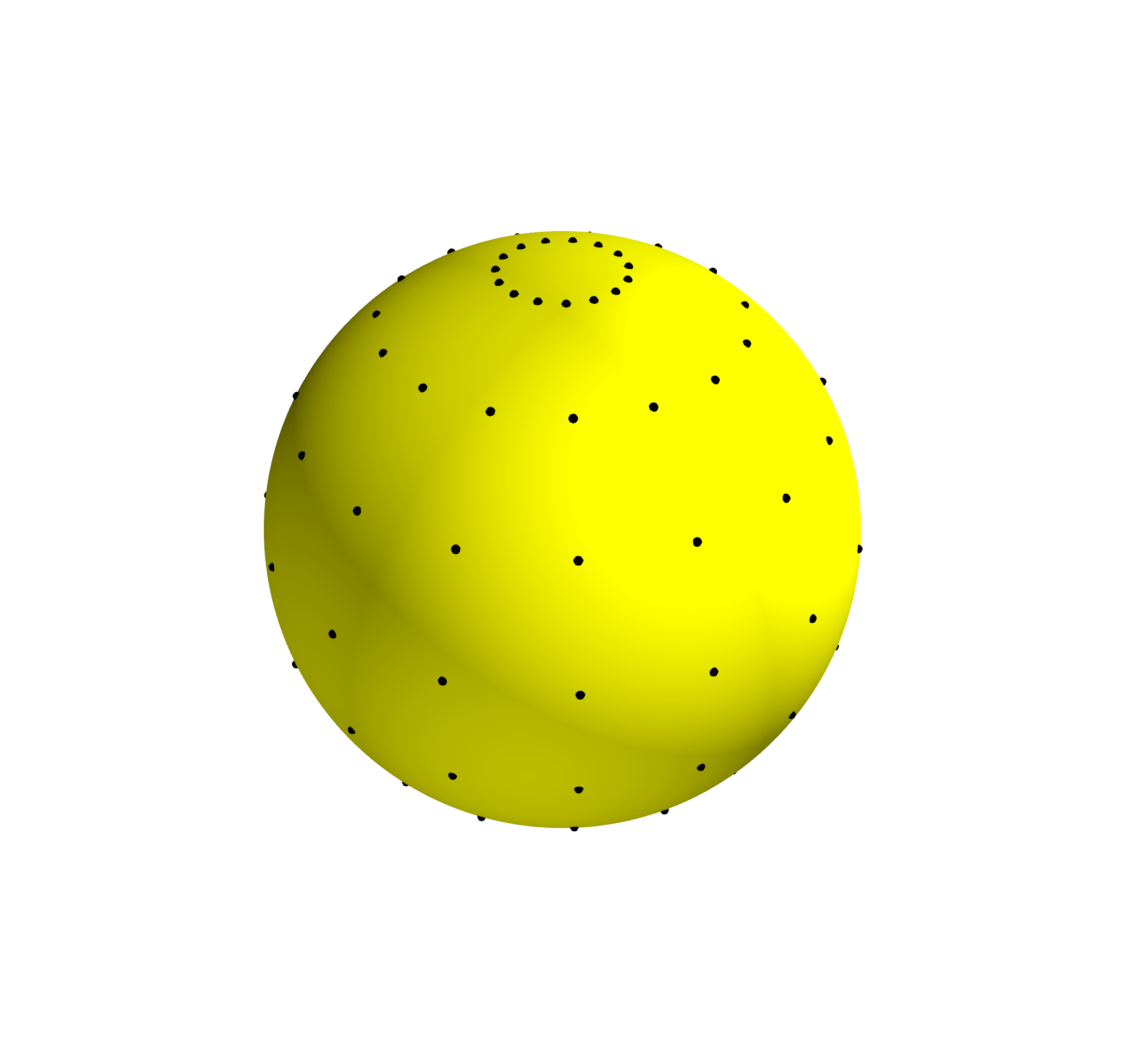}
      \includegraphics[width=0.8\linewidth, trim={13cm 13cm 13cm 13cm},clip]{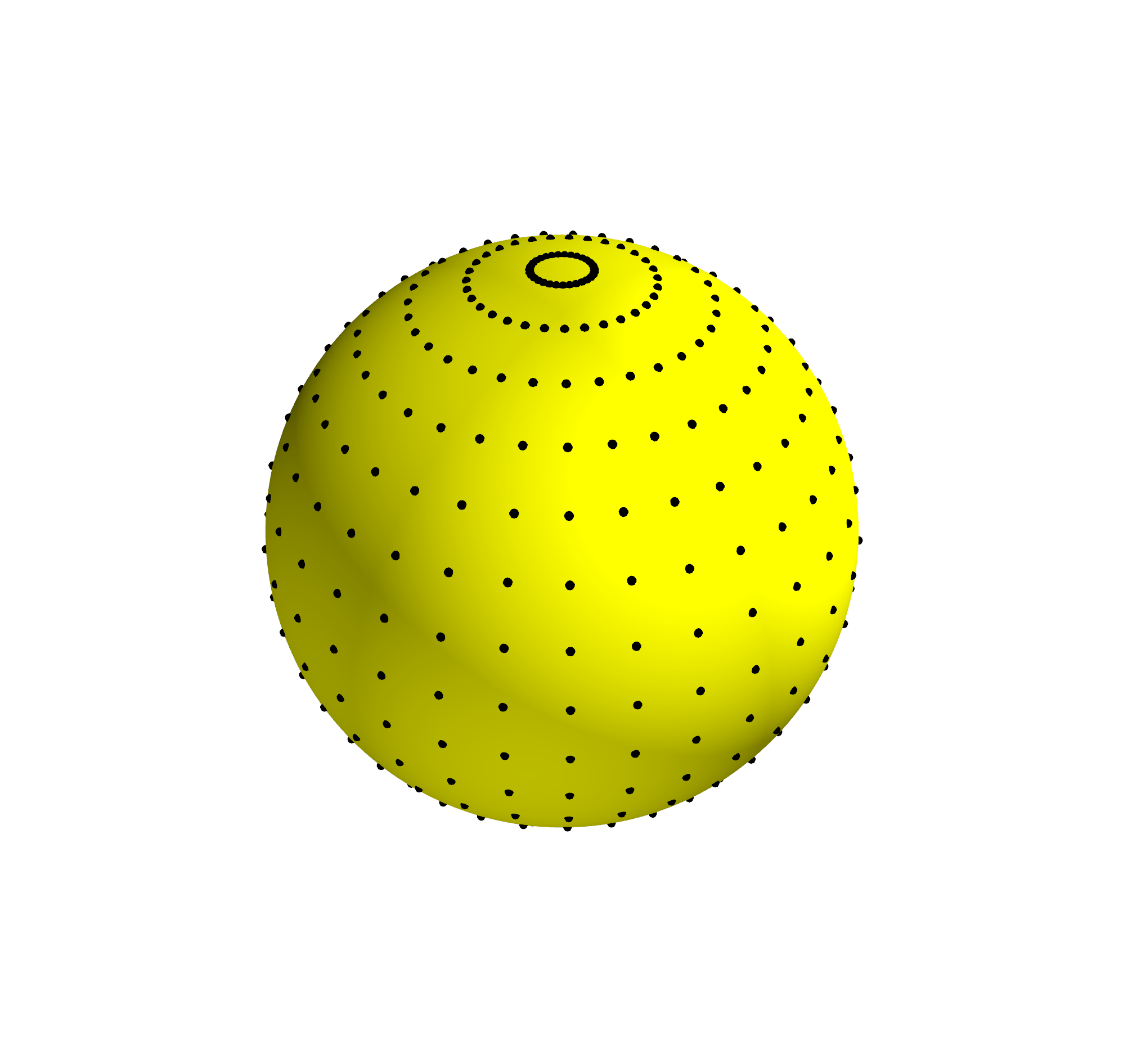}
      \caption{McEwen \& Wiaux (MW) \cite{mcewen:fssht}}
      \label{fig:mw_sampling}
   \end{subfigure}%
   % \begin{subfigure}{.18\textwidth}
   %   \centering
   %   \includegraphics[width=0.8\linewidth, trim={13cm 13cm 13cm 13cm},clip]{figures/sampling_patterns/mwss_sampling_8.png}
   %   \includegraphics[width=0.8\linewidth, trim={13cm 13cm 13cm 13cm},clip]{figures/sampling_patterns/mwss_sampling_16.png}
   %   \caption{McEwen \& Wiaux* \cite{mcewen:fssht}}
   %   \label{fig:mwss_sampling}
   % \end{subfigure}%
   \begin{subfigure}{.23\textwidth}
      \centering
      \includegraphics[width=0.8\linewidth, trim={13cm 13cm 13cm 13cm},clip]{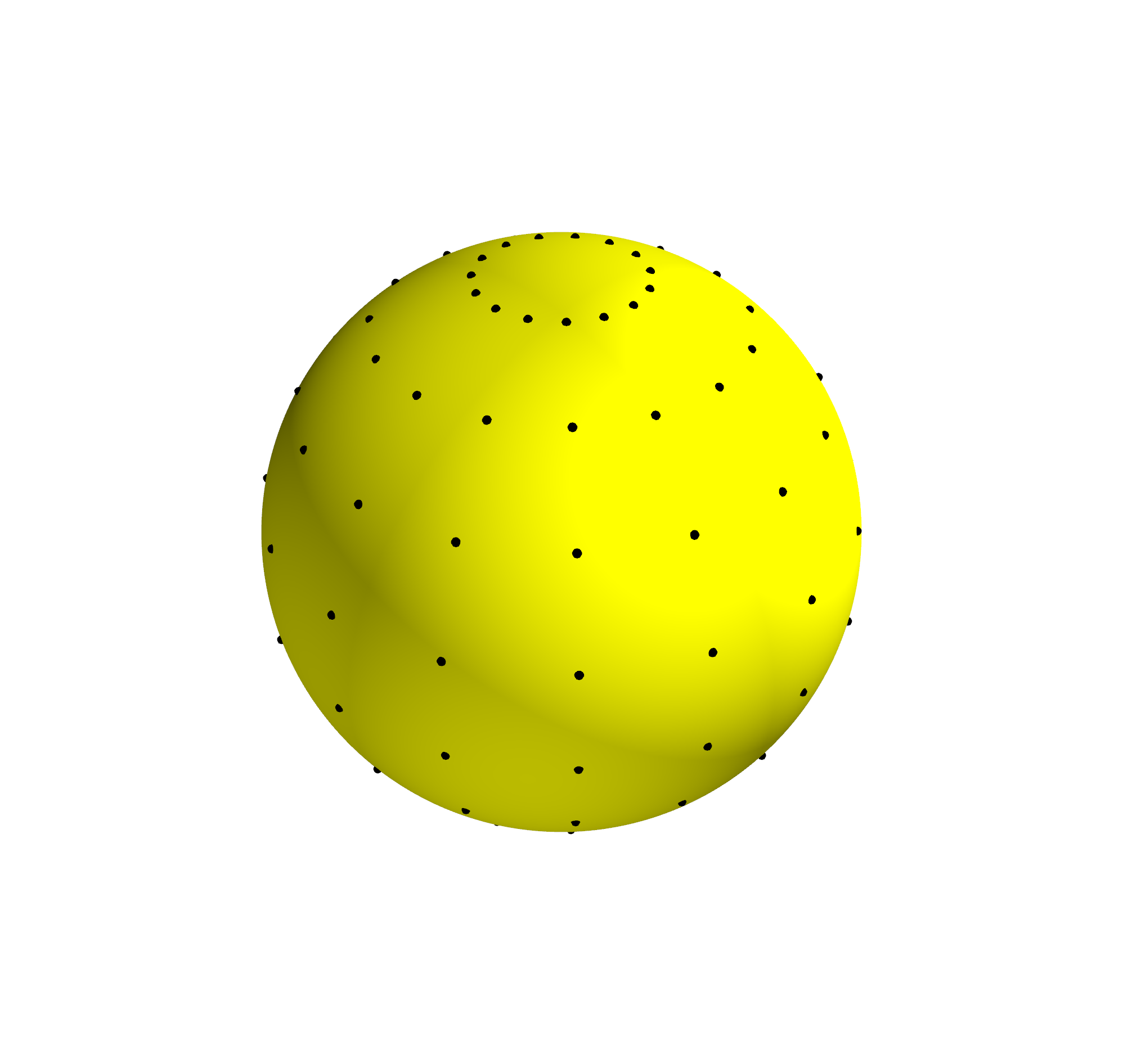}
      \includegraphics[width=0.8\linewidth, trim={13cm 13cm 13cm 13cm},clip]{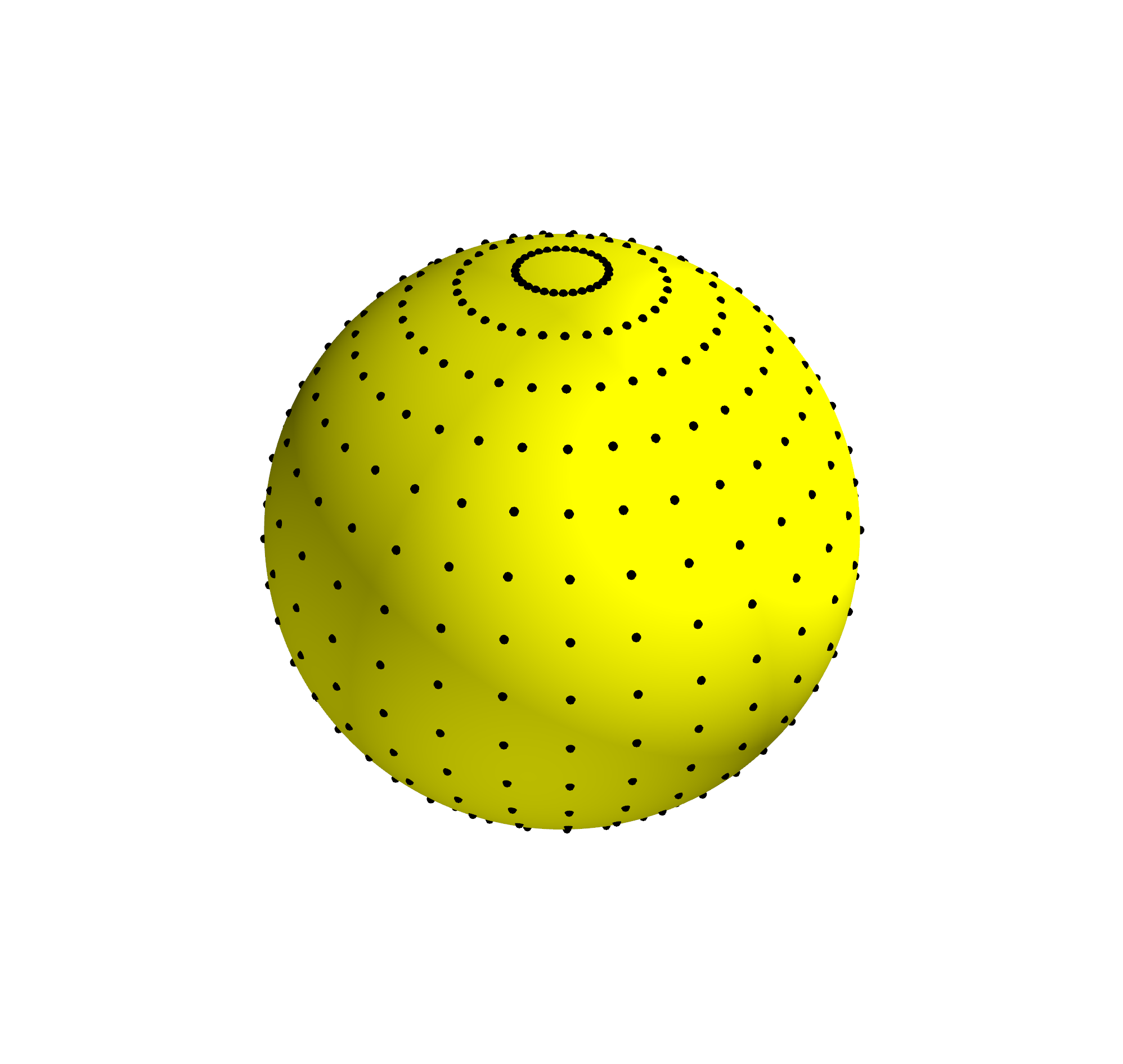}
      \caption{Gauss-Legendre (GL) \cite{mcewen:fssht, schaeffer:2013:efficient}}
      \label{fig:gl_sampling}
   \end{subfigure}
   \begin{subfigure}{.23\textwidth}
      \centering
      \includegraphics[width=0.8\linewidth, trim={13cm 13cm 13cm 13cm},clip]{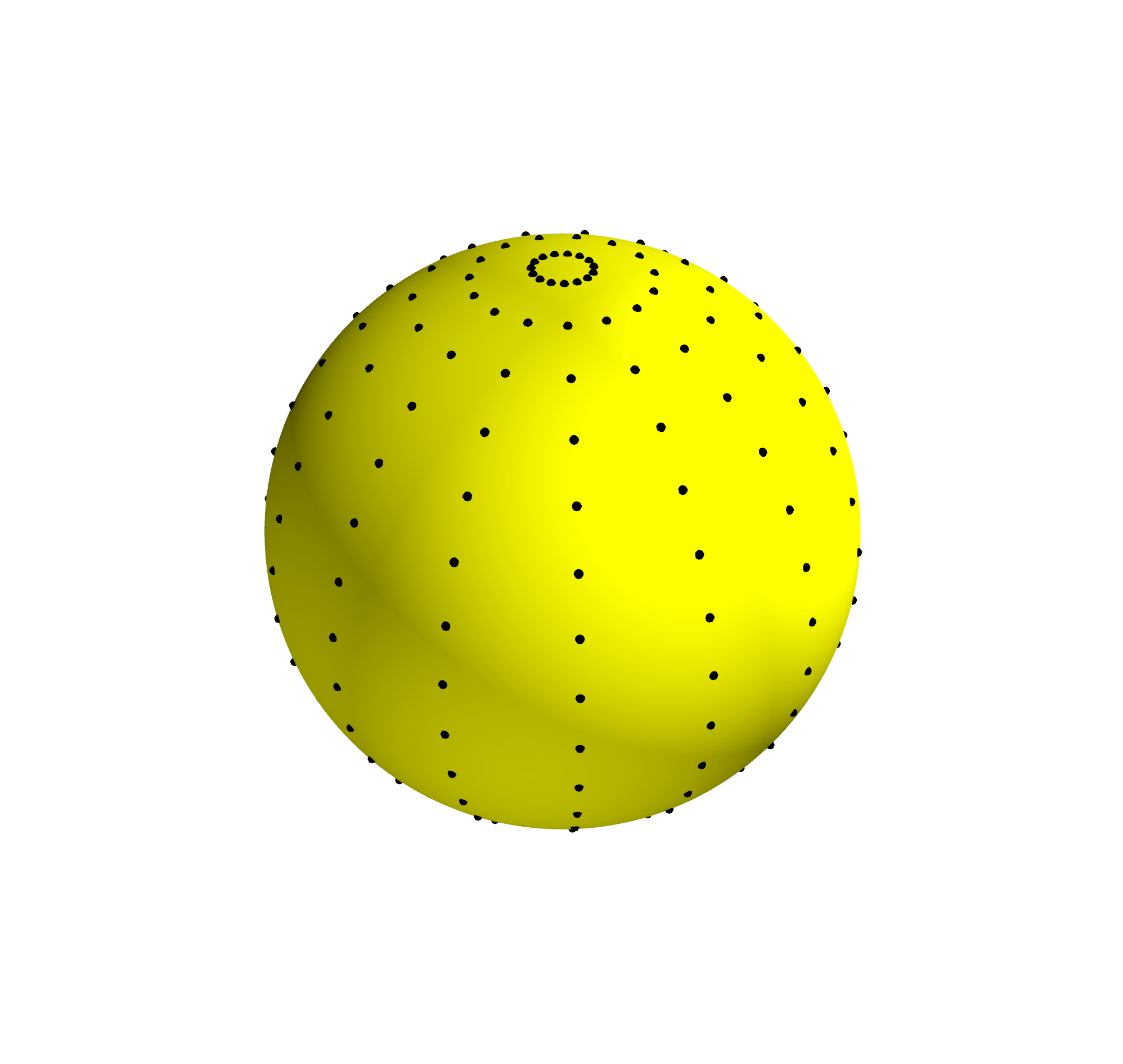}
      \includegraphics[width=0.8\linewidth, trim={13cm 13cm 13cm 13cm},clip]{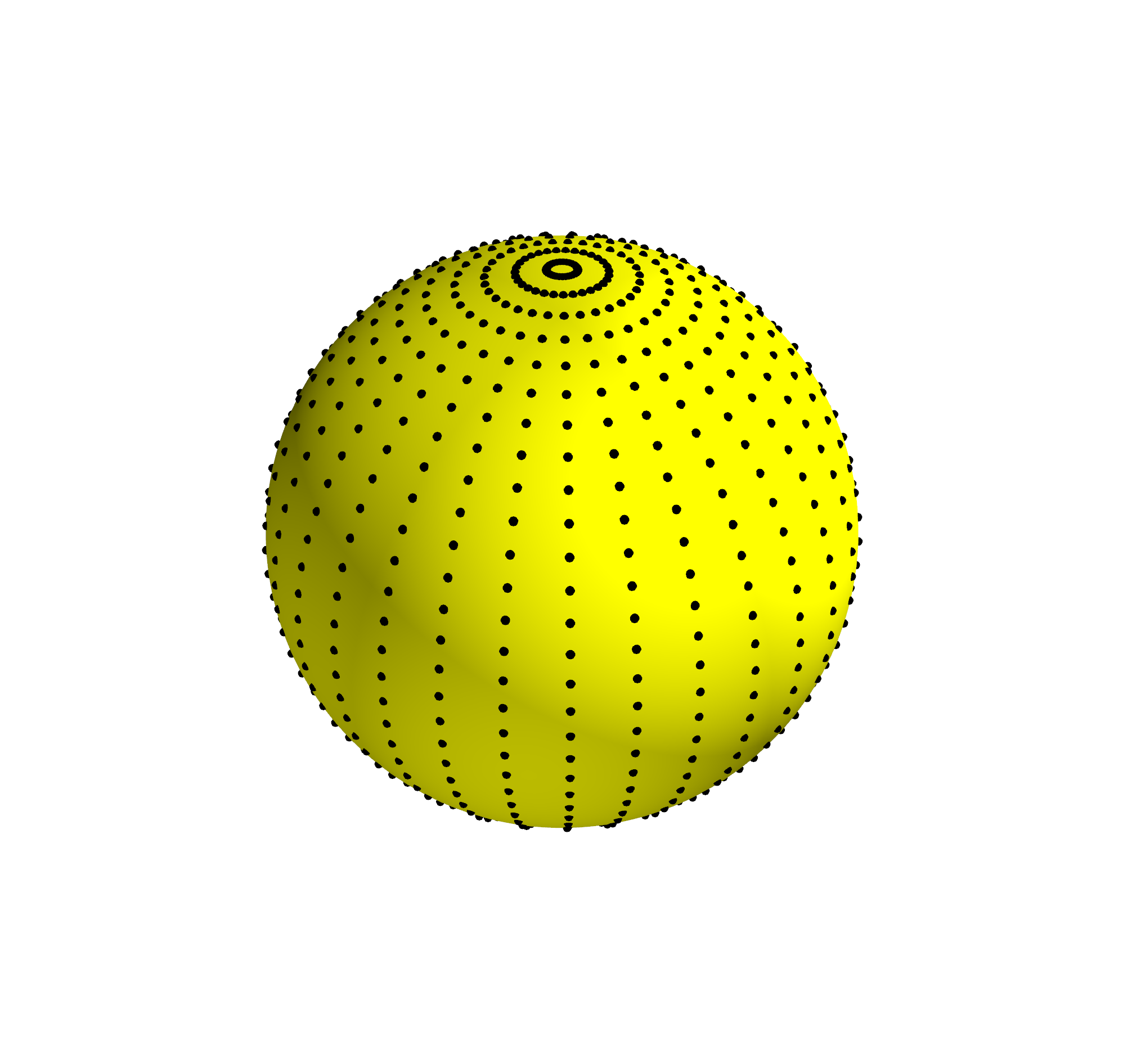}
      \caption{Driscoll \& Healy (DH) \cite{driscoll:1994}}
      \label{fig:dh_sampling}
   \end{subfigure}
   \begin{subfigure}{.23\textwidth}
      \centering
      \includegraphics[width=0.8\linewidth, trim={13cm 13cm 13cm 13cm},clip]{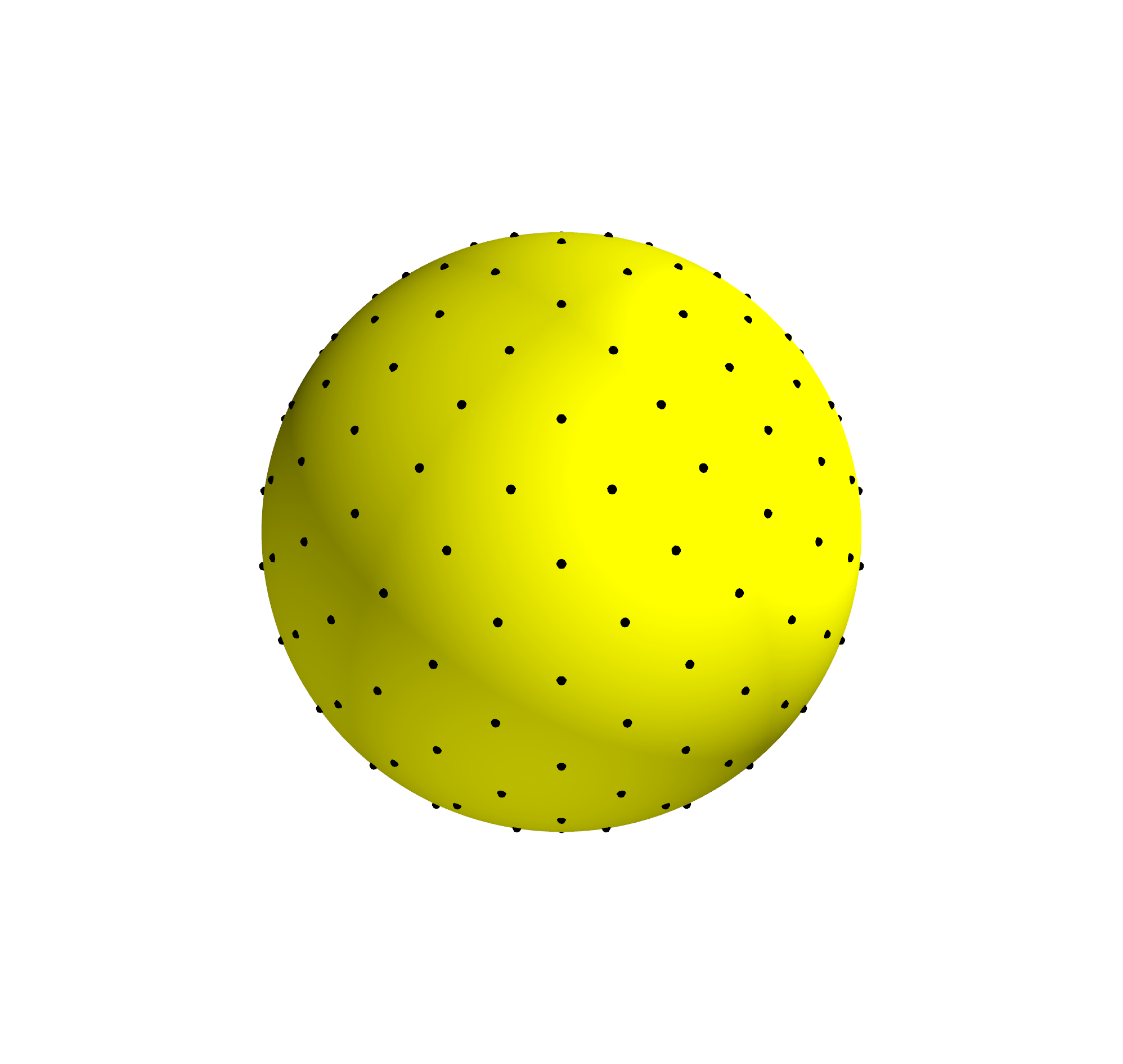}
      \includegraphics[width=0.8\linewidth, trim={13cm 13cm 13cm 13cm},clip]{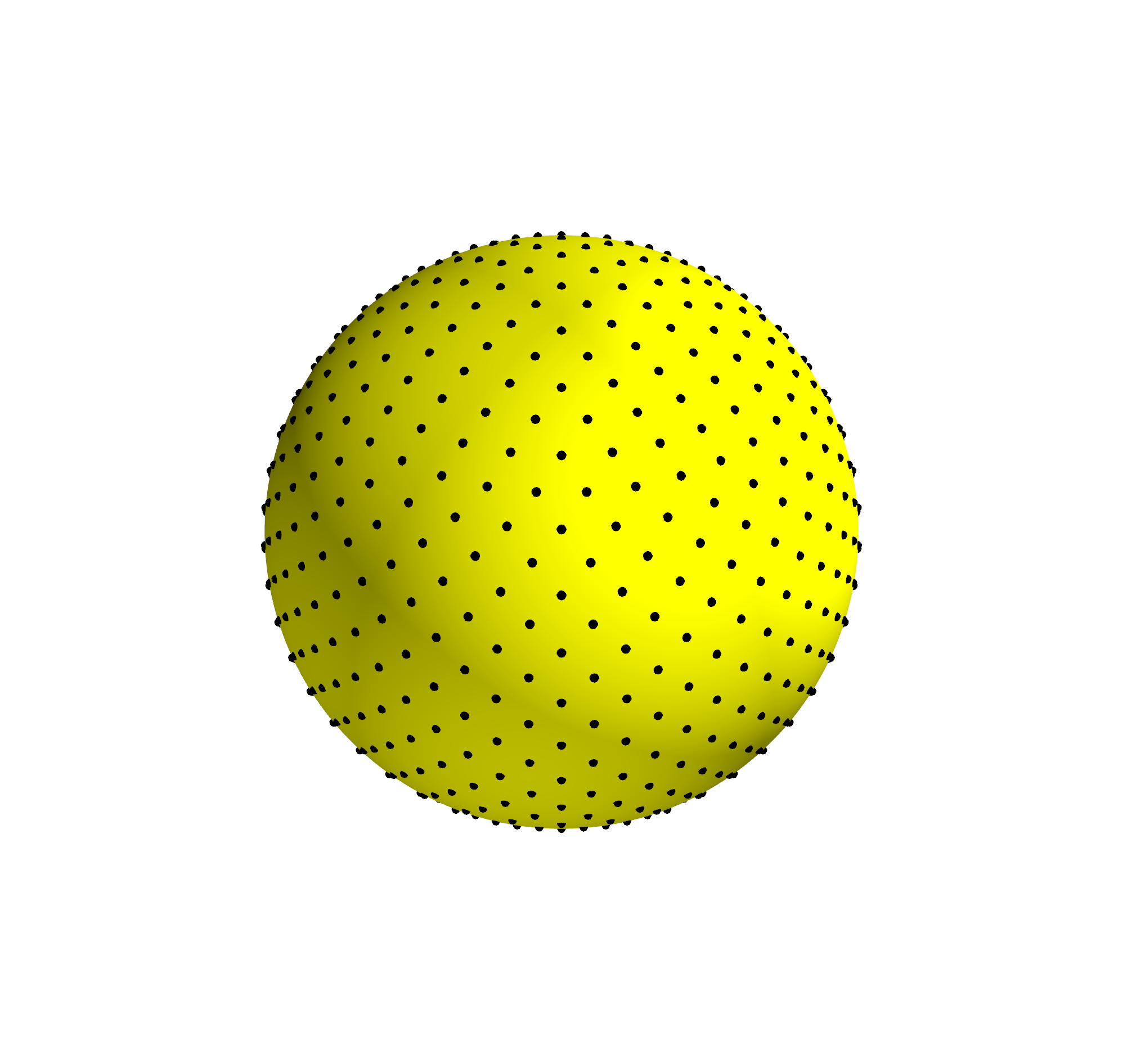}
      \caption{\texttt{HEALPix} \cite{gorski:2005}}
      \label{fig:healpix_sampling}
   \end{subfigure}
   \caption{Visualisation of sample positions on the sphere for various sampling schemes.  The Driscoll \& Healy \cite[DH;][]{driscoll:1994}, McEwen \& Wiaux \cite[MW;][]{mcewen:fssht}, and Gauss-Legendre \cite[GL;][]{mcewen:fssht, schaeffer:2013:efficient} samplings all admit sampling theorems with exact spherical harmonic transforms, although \texttt{HEALPix} does not.  Both MW and GL sampling require $\sim 2L^2$ samples (although the MW sampling is slightly more efficient), whereas DH \cite{driscoll:1994} sampling requires $\sim 4L^2$, and \texttt{HEALPix} considers $\sim kL^2$ samples, where $k$ typically ranges from $3$ to $12$. The \texttt{HEALPix} scheme has the practical advantage of equal-area pixels and is widely adopted in cosmology.  Modern computer vision, machine learning and geophysical applications typically adopt equiangular sampling and thus map most efficiently and straightforwardly onto the MW sampling.}
   \label{fig:figures/sampling_patterns}
\end{figure}

In this work we address precisely this need by developing spherical transforms that are both differentiable and accelerated.  Specifically, we develop algorithms to compute generalised Fourier transforms on the sphere and also on the rotation group, also referred to as spherical harmonic and Wigner transforms, respectively. These algorithms are constructed upon novel stable Wigner $d$-function recursions that recurse along harmonic order $m$ alone, therefore permitting extreme parallelisation across all other indices. To avoid overflowing numerical precision we also introduce an on-the-fly normalisation scheme. Such algorithmic structures are designed to be highly parallelised in order to effectively exploit very high parallel throughput on hardware accelerators like GPUs.  We pay careful attention to the computation of gradients through a hybrid automatic and manual differentiation approach, avoiding the memory overhead of full automatic differentiation while also avoiding the complexities of full manual differentiation.  We implement these algorithms in a new open-source software code called \texttt{S2FFT} {\href{https://github.com/astro-informatics/s2fft}{\faGithub}}, which is developed in the \texttt{JAX} differentiable programming framework.
Our algorithms and code support a number of different approaches to sample the sphere: including various equiangular samplings that admit sampling theorems \cite{driscoll:1994,mcewen:fssht,mcewen:so3} and thus theoretically exact spherical harmonic transforms; and also the popular \texttt{HEALPix} scheme \cite{gorski:2005}, which, although it does not support exact harmonic transforms, has the practical advantage of pixels of equal area.  Various sampling schemes are illustrated in \fig{\ref{fig:figures/sampling_patterns}}.

The remainder of this article is structured as follows. In \sectn{\ref{sec:background}} we review the mathematical background that underpins themes throughout this article. Following this, in \sectn{\ref{sec:recursions}} we present a recursive algorithm suitable for the parallel calculation of the Wigner $d$-functions.  Algorithms to compute the spherical harmonic transform for arbitrary spin and the Wigner transform are presented in \sectn{\ref{sec:algorithms}}.  In \sectn{\ref{sec:gradients}} we present various approaches to the efficient computational of the gradients of our spherical transforms. Penultimately, we provide an overview of our software implementation with benchmarking results in \sectn{\ref{sec:software}}, before drawing concluding remarks in \sectn{\ref{sec:conclusion}}.

\section{Fourier transforms on the sphere and rotation group}
\label{sec:background}

We concisely review the generalised Fourier transform on the two-sphere $\mathbb{S}^2$ and three-dimensional rotation group $\text{SO}(3)$.  We first review the scalar spherical harmonic transform, after which we generalise to spin functions.  We describe the generalised Fourier transform on the rotation group, the so-called Wigner transform, and draw connections to spin spherical harmonic transforms.  Finally, we discuss sampling spherical signals for the computation of spherical transforms, highlighting the distinction between sampling schemes and sampling theorems (the latter of which provide exact quadrature and exact harmonic transforms), and discuss a number of common spherical samplings that our algorithms support.

\subsection{Spherical harmonic transforms for scalar functions}
\label{sec:background:spherical_transform_scalar}

The scalar spherical harmonic functions $Y_{\ell m} : \mathbb{S}^2 \rightarrow \mathbb{C}$ form the canonical orthogonal basis for the space of square integrable scalar functions on the sphere and are defined by
\begin{equation} \label{eq:scalar_sh}
   Y_{\ell m}(\vartheta,\varphi)
   = \sqrt{\frac{2\ell+1}{4\pi}\frac{(\ell-m)!}{(\ell+m)!}}
   \: P^m_{\ell}(\cos\vartheta) \: \exp{\img m\varphi},
\end{equation}
for spherical coordinates with colatitude $\vartheta \in [0, \pi]$ and longitude $\varphi \in [0, 2\pi)$, natural order $\ell \in \mathbb{N}$ and integer degree $m\in\mathbb{Z}$, $\vert m \vert \leq \ell$, and where $P^{m}_{\ell}(\cdot)$ are the associated Legendre functions. Throughout this article we adopt the Condon-Shortley phase convention such that the conjugate symmetry relation \mbox{$Y^{*}_{\ell m}(\vartheta, \varphi) = (-1)^m\,Y_{\ell, -m}(\vartheta, \varphi)$} holds.  The orthogonality and completeness relations for the spherical harmonics read
\begin{equation} \label{eq:orthogonality_spherical}
   \langle Y_{\ell m}, Y_{\ell^{\prime} m^{\prime}} \rangle
   = \int_{\mathbb{S}^2} \dx\mu(\vartheta, \varphi) \: Y_{\ell m}(\vartheta, \varphi) \: Y_{\ell^{\prime} m^{\prime}}^\ast(\vartheta, \varphi)
   = \delta_{\ell\ell^{\prime} }\delta_{mm^{\prime}}
\end{equation}
and
\begin{equation} \label{eq:completeness_spherical}
   \sum_{\ell = 0}^\infty \sum_{m=-\ell}^\ell
   Y_{\ell m}(\vartheta,\varphi) \:
   Y_{\ell m}^\ast(\vartheta^\prime,\varphi^\prime)
   =
   (\sin\vartheta)^{-1} \:
   \delta(\vartheta - \vartheta^\prime) \:
   \delta(\phi-\phi^\prime),
\end{equation}
respectively, where $\langle \cdot , \cdot, \rangle$ denotes the inner product, $\dx\mu(\vartheta, \varphi) = \sin\vartheta \, d\vartheta \, d\varphi$ is the rotationally invariant Haar measure on the sphere, $\delta_{ij}$ is the Kronecker delta symbol and $\delta(\cdot)$ is the Dirac delta function.

By the orthogonality and completeness of the spherical harmonic functions, any square integrable function $f \in \text{L}^2(\mathbb{S}^2)$ may be represented by its spherical harmonic expansion
\begin{equation} \label{eq:sht_cts_inverse}
   f(\vartheta, \varphi)
   = \sum_{\ell =0}^\infty
   \sum_{m=-\ell}^\ell
   \hat{f}_{\ell m} \: Y_{\ell m}(\vartheta, \varphi),
\end{equation}
where the square summable spherical harmonic coefficients $\hat{f} \in l^2$ are given by the projection
\begin{equation} \label{eq:sht_cts_forward}
   \hat{f}_{\ell m}
   = \langle f, Y_{\ell m} \rangle
   = \int_{\mathbb{S}^2} \dx\mu(\vartheta, \varphi) \:
   f(\vartheta, \varphi) \:
   Y^{\ast}_{\ell m}(\vartheta, \varphi) .
\end{equation}
We denote the forward spherical harmonic transform by $\mathcal{F} : \text{L}^2(\mathbb{S}^2) \rightarrow l^2$, which maps functions on the sphere to their spherical harmonic coefficients $\mathcal{F} : f \mapsto \hat{f}$.  Similarly, we denote the inverse spherical harmonic transform by $\mathcal{F}^{-1} :  l^2 \rightarrow \text{L}^2(\mathbb{S}^2)$, which maps spherical harmonic coefficients to the corresponding function on the sphere $\mathcal{F}^{-1} : \hat{f} \mapsto f$.
For real signals, note that the Hermitian symmetry relation $\hat{f}^{*}_{\ell m} = (-1)^{m}\,\hat{f}_{\ell, -m}$ holds.

\subsection{Spherical harmonic transforms for spin functions}
\label{sec:background:spherical_transform_spin}

The (scalar) spherical harmonic functions are sufficient to characterise the harmonic representation of scalar functions on the sphere, however they do not fully capture the symmetries associated with spin functions.  Square integrable spin-$s$ functions on the sphere ${}_sf \in \text{L}^2(\mathbb{S}^2)$ exhibit an additional symmetry such that they transform by ${}_sf^{\prime} = \exp{-s\chi} \! {}_sf$ under right-handed rotations $\chi$ in the local tangent plane.  The spin spherical harmonics ${}_sY_{\ell m} : \mathbb{S}^2 \rightarrow \mathbb{C}$ form the canonical orthogonal basis for $\text{L}^2(\mathbb{S}^2)$ spin-$s$ functions on the sphere.

Spin spherical harmonics may be defined through actions of spin raising operator $\eth$ and lowering operator $\bar{\eth}$ on the scalar spherical harmonics, which are given by \cite{newman:1966, goldberg:1967,kamionkowski:1996}
\begin{equation}
   \eth = -\sin^s\vartheta \: \Bigg( \partial_\vartheta + \frac{i \partial_\varphi}{\sin\vartheta}\Bigg) \: \sin^{-s}\vartheta
   % \end{equation}
   % and
   % \begin{equation}
   \quad\text{and}\quad
   \bar{\eth} = -\sin^{-s}\vartheta \: \Bigg( \partial_\vartheta - \frac{i \partial_\varphi}{\sin\vartheta}\Bigg ) \: \sin^{s}\vartheta .
\end{equation}
The action of the spin raising and lowering operators on the spin spherical harmonics ${}_sY_{\ell m}$ is given, respectively, by
\begin{equation}
   \eth {}_sY_{\ell m} = \sqrt{(\ell-s)(\ell+s+1)} \: {}_{s+1}Y_{\ell m}
   % \end{equation}
   % and
   % \begin{equation}
   \quad\text{and}\quad
   \bar{\eth} {}_sY_{\ell m} = -\sqrt{(\ell+s)(\ell-s+1)} \: {}_{s-1}Y_{\ell m}.
\end{equation}
It therefore follows that any spin spherical harmonic basis function can be generated from $s$ repeated applications of the spin raising or lowering operators on the scalar (spin $s = 0$) spherical harmonics:
\begin{equation}
   {}_sY_{\ell m}
   =
   \begin{cases}
      (-1)^s\sqrt{\frac{ (\ell+s)! }{ (\ell-s)! }} \bar{\eth}^{-s}Y_{\ell m} & \text{for} \ -\ell \leq s \leq 0 \\
      \sqrt{\frac{ (\ell-s)! }{ (\ell+s)! }} \: \eth^{s} Y_{\ell m}          & \text{for} \ 0 \leq s \leq \ell
   \end{cases} .
\end{equation}

The spin spherical harmonics, for a given spin, satisfy identical orthogonality and completeness relations as the scalar spherical harmonics, \textit{cf.} \eqn{\ref{eq:orthogonality_spherical}} and \eqn{\ref{eq:completeness_spherical}}, respectively.  Consequently, the forward and inverse spin spherical harmonic transforms are also similar to the scalar transforms,  \textit{cf.} \eqn{\ref{eq:sht_cts_forward}} and \eqn{\ref{eq:sht_cts_inverse}} respectively, but with the scalar spherical harmonics replaced by the spin spherical harmonics.
We denote forward and inverse spin spherical harmonic transforms by ${}_s \mathcal{F}$ and ${}_s \mathcal{F}^{-1}$, respectively, often dropping the explicit reference to spin $s$ for notational brevity.
The Hermitian symmetry relation of the spin spherical harmonic coefficients is given by ${}_sf_{\ell m}^\ast = (-1)^{s+m} \: {}_{-s}f_{\ell,-m}$ for a function satisfying ${}_sf^\ast={}_{-s}f$ (which for a spin $s=0$ function equates to the usual reality condition).

\subsection{Wigner transforms for functions on the rotation group}
\label{sec:background:wigner_transform}

In many settings we encounter functions defined over the space of three-dimensional rotations. The Wigner $D$-functions $D^{\ell}_{mn} : \text{SO}(3) \rightarrow \mathbb{C}$ form an irreducible unitary representation of $\text{SO}(3)$, the group of three-dimensional rotations, for natural $\ell \in \mathbb{N}$ and integers $m,n \in \mathbb{Z}$ such that $|m|,|n| \leq \ell$ \cite{varshalovich:1989}.
The Wigner $D$-functions are related to the spin spherical harmonics by
\begin{equation} \label{eq:wigner_spherical_harmonics}
   D^{\ell}_{mn}(\alpha,\beta,\gamma)
   = (-1)^n \sqrt{\frac{4\pi}{2\ell+1}} \:
   {}_{-n}Y_{\ell m}^\ast(\beta,\alpha) \:
   \exp{-\img n \gamma} ,
\end{equation}
where $(\alpha,\beta,\gamma)$ are the Euler angles parameterising $\text{SO}(3)$, with $\alpha \in [0,2\pi)$, $\beta \in [0,\pi]$ and $\gamma \in [0,2\pi)$.  We adopt the $zyz$ Euler convention corresponding to the rotation of a physical body in a \textit{fixed} co-ordinate system about the $z$, $y$ and $z$ axes by $\gamma$, $\beta$ and $\alpha$ respectively.
The Wigner $D$-functions may be decomposed in terms of the real Wigner $d$-functions by \cite{varshalovich:1989}
\begin{equation} \label{eq:wigner_d_decomposition}
   D^{\ell}_{mn}(\alpha,\beta,\gamma)
   =
   \exp{-\img m \alpha} \:
   d^{\ell}_{mn}(\beta) \:
   \exp{-\img n \gamma}.
\end{equation}

The orthogonality and completeness relations for the Wigner $D$-functions read, respectively,
\begin{equation}
   \langle D^{\ell}_{mn},  D^{\ell^{\prime}}_{m^{\prime}n^{\prime}}  \rangle
   = \int_{\text{SO}(3)} \dx\mu(\alpha,\beta,\gamma)
   D^{\ell}_{mn}(\alpha,\beta,\gamma) \:
   D^{\ell^{\prime}\ast}_{m^{\prime}n^{\prime}}{}(\alpha,\beta,\gamma)
   = \frac{8\pi^2}{2\ell+1}  \:
   \delta_{\ell\ell^\prime}  \delta_{mm^\prime}  \delta_{nn^\prime},
\end{equation}
and
\begin{equation}
   \sum_{\ell=0}^{\infty}  \sum_{m=-\ell}^{\ell} \sum_{n=-\ell}^{\ell} D^{\ell}_{mn}(\alpha, \beta, \gamma) \: D^{\ell\ast}_{mn}(\alpha^{\prime}, \beta^{\prime}, \gamma^{\prime}) = \delta(\alpha-\alpha^\prime) \: \delta(\cos\beta-\cos\beta^\prime) \: \delta(\gamma-\gamma^\prime),
\end{equation}
where we overload the notation $\langle \cdot, \cdot \rangle$ to denote inner products on both the sphere and rotation group (note that the case considered can be easily inferred from the context) and $\dx\mu(\alpha,\beta,\gamma) = \sin\beta \dx\alpha \dx\beta \dx\gamma$ is the rotationally invariant Haar measure on the rotation group.

The space of square integrable functions $f \in \text{L}^2(\text{SO}(3))$ admits the Wigner expansion
\begin{equation} \label{eq:wt_cts_inverse}
   f(\alpha, \beta, \gamma)
   = \sum_{\ell = 0}^\infty \:
   \frac{2\ell + 1}{8\pi^2} \:
   \sum_{m=-\ell}^{\ell} \:
   \sum_{n=-\ell}^{\ell} \:
   \hat{f}^{\ell}_{mn} \:
   D^{\ell \ast}_{mn}(\alpha, \beta, \gamma),
\end{equation}
where the square summable Wigner coefficients $\hat{f} \in l^2$ are given by the projection
\begin{equation} \label{eq:wt_cts_forward}
   \hat{f}^{\ell}_{mn}
   = \langle f, D^{\ell\ast}_{mn} \rangle
   = \int_{\text{SO(3)}} \: \dx\mu(\alpha,\beta,\gamma) \: f(\alpha,\beta,\gamma) \: D^{\ell}_{mn}(\alpha,\beta,\gamma).
\end{equation}
As before, we introduce an operator notation with forward Wigner transform $\mathcal{W} : \text{L}^2(\text{SO}(3)) \rightarrow l^2$, which maps functions on the rotation group to their harmonic coefficients $\mathcal{W} : f \mapsto \hat{f}$.  Similarly, we denote the inverse Wigner transform by $\mathcal{W}^{-1} : l^2 \rightarrow \text{L}^2(\text{SO}(3))$, which maps Wigner harmonic coefficients to the corresponding function on the rotation group $\mathcal{W}^{-1} : \hat{f} \mapsto f$.  Note that we use $f$ to denote functions on both the sphere and rotation group, with corresponding harmonic coefficients $\hat{f}$.
For real signals, note that the Hermitian symmetry relation $\hat{f}^{\ell\ast}_{mn} = (-1)^{m+n} \hat{f}^{\ell}_{-m,-n}$ holds.

Due to the relationship between the Wigner $D$-functions and the spin spherical harmonics (Eq.~\ref{eq:wigner_spherical_harmonics}), Wigner transforms can be expressed as a series of spin spherical harmonic transforms and Fourier transforms \cite{mcewen:so3}.
For completeness, the forward Wigner transform may be expressed as
\begin{equation} \label{eq:wigner_sht_forward}
   \hat{f}^{\ell}_{mn} =
   (-1)^n
   \sqrt{\frac{4\pi}{2\ell+1}}
   % \int_0^\pi \dx\beta \sin\beta
   % \int_0^{2\pi} \dx \alpha
   \int_{\mathbb{S}^2} \dx\mu(\beta, \alpha) \:
   f_n(\alpha,\beta) \:
   {}_{-n}Y_{\ell m}^\ast(\beta, \alpha)
   ,
\end{equation}
where
\begin{equation}
   f_n(\alpha, \beta) = \int_0^{2\pi} \dx\gamma
   f(\alpha, \beta, \gamma) \:
   \exp{-\img n \gamma}
   .
\end{equation}
Note that \eqn{\ref{eq:wigner_sht_forward}} corresponds to a (scaled) spin spherical harmonic transform with spin $-n$.  The inverse Wigner transform may be expressed as
\begin{equation}
   f(\alpha, \beta, \gamma)
   =
   \sum_{n=-\ell}^\ell
   f_n(\alpha, \beta) \:
   \exp{\img n \gamma} ,
\end{equation}
where
\begin{equation} \label{eq:wigner_sht_inverse}
   f_n(\alpha, \beta)
   =
   (-1)^n
   \sum_{\ell=0}^\infty
   \sum_{m=-\ell}^\ell
   \sqrt{\frac{2\ell+1}{16\pi^3}} \:
   \hat{f}^\ell_{mn} \:
   {}_{-n}Y_{\ell m}(\beta, \alpha) .
\end{equation}
Note that \eqn{\ref{eq:wigner_sht_inverse}} corresponds to an inverse spin spherical harmonic transform (of scaled harmonic coefficients) with spin $-n$.  These expressions may be leveraged to compute Wigner transforms simply through spin spherical harmonic transforms and standard 1D Fourier transforms \cite{mcewen:so3}.

\subsection{Sampling schemes versus sampling theorems on the sphere and rotation group}
\label{sec:background:sampling}

Up until this point we have considered the continuous form of generalised Fourier transforms on the sphere and rotation group.  In practice, however, one recovers sampled values of the signal of interest at a finite number of positions.  Consequently, sampling schemes must be adopted to pixelise the sphere and rotation group.  Generalised Fourier transforms must then be computed from these sampled values.

Often we consider the space of bandlimited signals $\mathcal{B}_L(\Omega)$, where $\Omega$ represents either the sphere or rotation group, \textit{i.e.} $\Omega = \{\mathbb{S}^2, \text{SO}(3) \}$. Bandlimited signals are those whose harmonic coefficients are zero for degrees $\ell$ greater than or equal to the bandlimit $L$.  A bandlimited signal on the sphere has $\hat{f}_{\ell m} = 0$ for all $\ell \geq L$, while a bandlimited signal on the rotation group has $\hat{f}^{\ell}_{m n} = 0$ for all $\ell \geq L$.  Consequently, harmonic coefficients for bandlimited signals on the sphere and rotation group live, respectively, in $\mathbb{C}^{L^2}$ and $\mathbb{C}^{(4L^3-L)/3}$ \cite{mcewen:so3,khalid:so3_gl}.  For bandlimited signals, summations over degree $\ell$ (\textit{cf.} \eqn{\ref{eq:sht_cts_inverse}} and \eqn{\ref{eq:wt_cts_inverse}}) may be truncated to $L-1$. In many settings azimuthal bandlimits also arise, such that harmonic coefficients are zero for $m \geq M$ and/or $n \geq N$ (summations over $m$ and $n$ may than also be truncated). Real-world signals may be approximated accurately by bandlimited signals for a suitable bandlimit.

For the spatial signal representation, we distinguish between sampling \textit{schemes} and \textit{theorems}.
For \textit{sampling schemes}, the sphere or rotation group may be sampled in any arbitrary manner and harmonic coefficients computed by approximate quadrature.  This provides a great deal of flexibility in how sample positions are set.  However, this flexibility comes at a cost: for an arbitrary sampling, harmonic transforms can only be computed approximately and in some cases can be quite inaccurate.  Furthermore, for an arbitrary sampling the computation of the spherical harmonic transform scales as $\mathcal{O}(L^4)$.  By adopting an isolatitude sampling, where $\varphi$ samples are collected in isolatitudinal rings for each colatitude $\vartheta$, one may apply a separation of variables to reduce computational complexity to $\mathcal{O}(L^3)$ \cite[\textit{e.g.}][]{gorski:2005, mcewen:fssht, mcewen:so3, mcewen:2013:waveletsxv}.  For this reason isolatitudinal sampling is typically adopted; hence, we focus on isolatitudinal sampling in the remainder of this article.
A \textit{sampling theorem}, in contrast, provides a way to capture all of the information content of a bandlimited signal in a finite set of sampled values.  Since all information content of the underlying continuous signal is captured, a sampling theorem implicitly provides a quadrature rule to compute harmonic coefficients exactly.  Sampling theorems are thus tied to particular structured samplings of the sphere.

For both arbitrary sampling schemes and sampling theorems, the forward spin spherical and Wigner transforms of \eqn{\ref{eq:sht_cts_forward}} and \eqn{\ref{eq:wt_cts_forward}}, respectively, may be discretised by
\begin{equation} \label{eq:sht_cts_forward_discrete}
   {}_s \hat{f}_{\ell m}
   \simeq
   \sum_{\theta, \phi}
   q(\theta, \phi) \:
   {}_s f(\theta, \phi) \:
   {}_s Y^{\ast}_{\ell m}(\theta, \phi)
\end{equation}
and
\begin{equation}
   \hat{f}^{\ell}_{mn}
   \simeq
   \sum_{\alphaup,\betaup,\gammaup}
   q(\alphaup,\betaup,\gammaup) \:
   f(\alphaup,\betaup,\gammaup) \:
   D^{\ell}_{mn}(\alphaup,\betaup,\gammaup),
\end{equation}
where we have introduced an upright notation to denote sampled angles, \textit{i.e.} $(\vartheta, \varphi) \mapsto (\theta, \phi)$ and $(\alpha, \beta, \gamma) \mapsto (\alphaup, \betaup, \gammaup)$.  We adopt the symbol $\simeq$ to denote that the computation is approximate for arbitrary sampling schemes but is exact for sampling theorems.  Inverse spherical and Wigner transforms of \eqn{\ref{eq:sht_cts_inverse}} and \eqn{\ref{eq:wt_cts_inverse}}, respectively, involve summations only and so can be computed directly (up to a particular degree $\ell$).

We again adopt an operator notation overloading the previous notation to now consider bandlimited signals with forward spin spherical harmonic transform $\mathcal{F} : \mathcal{B}_L(\mathbb{S}^2) \rightarrow \mathbb{C}^{L^2}$ and inverse
$\mathcal{F}^{-1} : \mathbb{C}^{L^2} \rightarrow \mathcal{B}_L(\mathbb{S}^2)$.  Similarly, the forward Wigner transform is denoted $\mathcal{W} : \mathcal{B}_L(\text{SO}(3)) \rightarrow \mathbb{C}^{(4L^3-L)/3}$, with inverse $\mathcal{W}^{-1} : \mathbb{C}^{(4L^3-L)/3} \rightarrow \mathcal{B}_L(\text{SO}(3))$.

\subsection{Specific isolatitudinal sampling schemes and theorems on the sphere and rotation group} \label{sec:background:specific_samplings}

To complete this section, let us briefly discuss the sampling schemes and theorems considered throughout this article and in the associated software implementation.
One of the most common spherical sampling schemes is the Hierarchical Equal Area isoLatitude Pixelisation of the sphere, abbreviated \texttt{HEALPix} \cite{gorski:2005}.\footnote{\url{https://healpix.sourceforge.io/}}  \texttt{HEALPix} has numerous practical advantages, such as a hierarchical representation and equal area pixels.  Consequently, it is widely used in many application domains \cite[\textit{e.g.}][]{planck2016-l01}.  However, since \texttt{HEALPix} does not exhibit a sampling theorem, spherical harmonic transforms are necessarily approximate.  In fact, harmonic transforms can be quite inaccurate with absolute errors of $\sim 10^{-1}$ in some settings \cite{leistedt:s2let_axisym}.  For this reason, iterations are usually performed to increase accuracy.

The canonical sampling theorems on the sphere for equiangular samplings are those of Driscoll \& Healy \cite[DH;][]{driscoll:1994} and McEwen \& Wiaux \cite[MW;][]{mcewen:fssht}.
Spherical sampling theorems differ significantly to the Euclidean setting.  The Nyquist-Shannon sampling theorem for Euclidean spaces results in an identical number of samples in both the spatial and harmonic domains.  In contrast, no sampling theorem on the sphere achieves the optimal dimensionality of bandlimited spherical signals in harmonic space, given by $L^2$.  The MW sampling theorem provides the most efficient equiangular sampling, with $\sim 2L^2$ spatial samples, reducing the Nyquist rate on the sphere by a factor of two compared to the DH sampling scheme, which requires $\sim 4L^2$ spatial samples.\footnote{Note that an equiangular sampling \textit{scheme} that achieves the optimal number of spatial samples has been developed \cite{khalid:optimal_sampling}. A sampling theorem with exact quadrature is not recovered but high accuracy is nevertheless obtained.}  Fast algorithms for both the DH and MW sampling theorems have been developed.  While a number of algorithms have been introduced for the DH sampling theorem \cite{driscoll:1994,healy:2003}, some of which have computational complexity of $\mathcal{O}(L^2 \log L)$, the only variant is that is universally stable is the so-called semi-naive algorithm, with complexity of $\mathcal{O}(L^3)$.  Fast algorithms for the MW sampling theorem also achieve complexity of $\mathcal{O}(L^3)$ \cite{mcewen:fssht}.  While equiangular samplings are the most common structured sampling, for example matching the acquisition format of 360 panoramic photos and videos, spherical sampling theorems based on Gauss-Legendre (GL) quadrature have also been proposed \cite[\textit{e.g.}][]{shukowsky:1986,mcewen:fssht}, with samples defined on isolatitudinal rings with $\theta$ specified by the roots of the Legendre polynomials of order $L$.  GL sampling exhibits a similar, although slightly greater, number of samples as the MW sampling theorem.

Both the DH and MW sampling theorems have been extended to the rotation group $\text{SO}(3)$ \cite{kostelec:2008,mcewen:so3}.  Both exhibit fast algorithms with complexity $\mathcal{O}(L^4)$, while the MW sampling theorem again achieves a reduction in the Nyquist rate on $\text{SO}(3)$ by a factor of two, requiring $\sim 4 L^3$ spatial samples \cite{mcewen:so3} rather than $\sim 8 L^3$ samples \cite{kostelec:2008}.  Note that in many practical settings low azimuthal bandlimits arise, reducing a factor of $L$ in both number of samples and complexity to $N \ll L$.  The Gauss-Legendre sampling theorem has also been extended to  $\text{SO}(3)$ \cite{khalid:so3_gl}, requiring a similar (although slightly larger) number of samples as the MW sampling theorem on $\text{SO}(3)$.

Throughout the remainder of this article we develop general spherical transforms that can be applied to any isolatitudinal sampling that exhibits an explicit quadrature (either exact or approximate).  In particular, our software implementation supports the \texttt{HEALPix} sampling scheme, due to its widespread use and practical advantages, and the DH and MW sampling theorems, due to the exact spherical transforms afforded.

\section{Parallelised and differentiable Wigner computation} \label{sec:recursions}

To compute spin spherical harmonic and Wigner transforms it is necessary to compute the Wigner $D$-functions, or equivalently the real Wigner $d$-functions (\eqn{\ref{eq:wigner_d_decomposition}}).  Since we target accelerated computation on modern hardware accelerators, such as GPUs and TPUs that can execute a very large number of threads in parallel, it is desirable to compute Wigner functions in a highly parallelised manner.  Furthermore, since we also target differentiable transforms, we must compute Wigner functions in a manner that lends itself to efficient automatic differentiation.  Finally, we are interested in use cases that reach very high harmonic degrees $\ell$, hence careful attention must be given to ensure stable computation up to very high $\ell$.  In this section we first define explicit expressions for the Wigner functions, before discussing in greater detail the specific requirements that must be met to achieve a high degree of parallel computation and efficient automatic differentiation.  We then present a recursive algorithm to compute Wigner functions that meets these requirements, discussing specific considerations to ensure our approach is suitable for differentiable programming frameworks such as \texttt{JAX} \cite{jax:2018:github}.

\subsection{Explicit Wigner representations}

We relate the Wigner $D$-functions to the spin spherical harmonics above (\eqn{\ref{eq:wigner_spherical_harmonics}}), and also to the Wigner $d$-functions (\eqn{\ref{eq:wigner_d_decomposition}}). However, we of course require an explicit representation for computational purposes.  We leverage the decomposition of the Wigner $D$-functions in terms of the $d$-functions and thus focus on the computation of $d^{\ell}_{m n}(\beta)$.

A number of explicit representations of the $d$-functions can be considered, such as the differential representation \cite[][\sectn{4.3.2}, p.\ 77, \eqn{7}]{varshalovich:1989}
\begin{equation}
  d^{\ell}_{m n}(\beta) =
  \frac{(-1)^{\ell-n}}{2^\ell}
  \sqrt{\frac{(\ell+m)!}{(\ell-m)! (\ell+n)! (\ell-n)!}}
  (1-\cos\beta)^{-\frac{(m-n)}{2}} (1+\cos\beta)^{-\frac{(m+n)}{2}}
  \frac{\dx^{\ell-m}}{(\dx \cos\beta)^{\ell-m}}
  \bigl[ (1-\cos\beta)^{\ell-n}  (1+\cos\beta)^{\ell+n} \bigr]
  ,
\end{equation}
or the expression in terms of Jacobi polynomials $P_{s}^{\mu,\nu}(\cdot)$ given by \cite[][\sectn{4.3.4}, p.\ 78, \eqn{13--15}]{varshalovich:1989}
\begin{equation}
  \label{eq:wignerd_jacobi}
  d^{\ell}_{m n}(\beta) =
  \zeta_{m n}
  \sqrt{\frac{s! (s+\mu+\nu)!}{(s+\mu)! (s-\nu)!}}
  \Biggl( \sin\frac{\beta}{2} \Biggr)^{\mu}
  \Bigl( \cos\frac{\beta}{2} \Bigr)^{\nu}
  P_{s}^{\mu,\nu}(\cos\beta)
  ,
\end{equation}
where $\mu=\vert m-n \vert$, $\nu=\vert m+n\vert$, $s=\ell-(\mu+\nu)/2$ and
\begin{equation}
  \zeta_{m n} =
  \Biggl \{
  \begin{array}{ll}
    1          & \text{if} \:\: n \geq m \\
    (-1)^{n-m} & \text{if} \:\: n < m
  \end{array}
  .
\end{equation}
However, these expressions do not lend themselves to convenient computation (indeed the Jacobi representation simply shifts the task to the computation of the Jacobi polynomials).  Instead, it is typical to compute Wigner $d$-functions by recursion.

\subsection{Requirements for parallel computation and automatic differentiation}

While the Wigner $d$-functions can be computed by recursion, our aim is to deploy our algorithms efficiently on hardware accelerators that provide very high levels of parallelisation and to ensure they are stable to very high harmonic degrees $\ell$.  Careful attention therefore must be given to the recursion used to ensure these requirements are met.

A number of recursions, while stable, recurse over both harmonic degree $\ell$ and order $m$ \cite[\textit{e.g.}][]{risbo:1996,trapani:2006}.  This limits the degree of parallelisation that can be achieved.  Three-term recurrences typically recurse over a single index only, which is well-suited for high levels of parallelisation, although they can be unstable with errors compounding and resulting in highly inaccurate calculation \cite{gautschi:1967}.  Nevertheless, careful attention can be given to ensure stability can be achieved; for example, often recursing in some directions can be highly unstable, while other directions may be stable \cite[\textit{e.g.}][]{gautschi:1967,prezeau:2010,reinecke:2013:libsharp}.

For spin spherical harmonic transforms we need to compute $d^{\ell}_{m n}(\beta)$ for all $\ell$, $m$, and $\beta$, but only for a single $n$ (the spin number of interest).  We can compute Wigner transforms in terms of a series of spin spherical harmonic transforms (as discussed in \sectn{\ref{sec:background:wigner_transform}}).  We therefore target a recursion over $m$ only that we can compute for arbitrary $\ell$, $n$, and $\beta$.  Such a recursion supports high degrees of parallelisation, where computations for $\ell$, $n$, and $\beta$ are independent and so can be parallelised across the very large number of threads available on modern hardware accelerators, such as GPUs and TPUs.  Furthermore, this avoids the need for any recursion to compute transforms for all spins $\leq n$, providing an efficiency saving over alternative approaches that often compute spin transforms through $n$ repeated application of spin lowering/raising operations.  We pay careful attention in the next section (\sectn{\ref{sec:recursions:recursion}}) to how such a recursion can be computed in a stable manner up to very high degrees $\ell$.

Another key aim of our approach is to provide differentiable transforms.  A final requirement is that our recursive algorithm must therefore be structured in such a way as to provide computationally efficient automatic differentiation.  We pay careful attention below (\sectn{\ref{sec:recursions:jax}}) to how the recursion we consider can be computed efficiently and stably in a differentiable programming framework such as \texttt{JAX}.  Further details regarding automatic differentiation of not only the computation of the Wigner functions but the full spherical transforms are discussed in \sectn{\ref{sec:gradients}}.

\subsection{Recursion} \label{sec:recursions:recursion}

Our algorithm is built upon the three-term Wigner $d$-function recursion \cite[][\sectn{4.8.3}, p.\ 93, \eqn{17}]{varshalovich:1989}
\begin{equation} \label{eq:vard_recursion}
  \frac{n - m\cos\beta}{\sin\beta}\:d^{\ell}_{mn}(\beta)
  = \frac{1}{2}\sqrt{(\ell + m)(\ell - m + 1)} \: d^{\ell}_{m-1,n}(\beta)
  + \frac{1}{2}\sqrt{(\ell - m)(\ell + m + 1)} \: d^{\ell}_{m+1,n}(\beta).
\end{equation}
This recursion iterates across order $m$ independently of $\ell$, $n$, and $\beta$, thereby supporting an algorithmic structure that can be highly parallelised.  While computing this recursion for increasing $m$ is unstable, it can be computed stably for decreasing $m$ (as discussed in detail in \citet{prezeau:2010}, although they consider recursion over $n$).  One may straightforwardly rearrange \eqn{\ref{eq:vard_recursion}} to recover
\begin{equation} \label{eq:wigner-d-recursion}
  d^{\ell}_{m-1,n}(\beta)
  = \lambda_{m}\,a_{m-1} \: d^{\ell}_{mn}(\beta)
  - \frac{a_{m-1}}{a_{m}} \: d^{\ell}_{m+1,n}(\beta) ,
\end{equation}
where, for notational brevity, the recursion coefficients are concisely refactored into
\begin{equation}
  \lambda_{m} = \frac{n-m\cos\beta}{\sin\beta}
  \quad\text{and}\quad
  a_{m} = \frac{2}{\sqrt{(\ell - m)(\ell + m + 1)}}.
\end{equation}

The recursion needs to be initialised at $d^{\ell}_{\ell n}(\beta)$ so that we can recurse down in $m$, starting from $m=\ell$. For this particular relation, recursing along decreasing harmonic order $m$ is stable, whereas the increasing $m$ direction causes numerical errors to compound. We therefore initialise the algorithm using the relation
\begin{equation} \label{eq:initial_values}
  d^{\ell}_{\ell n}(\beta)
  =
  % (-1)^{\ell-n}
  \sqrt{\frac{(2\ell)!}{(\ell+n)!(\ell-n)!}}\:\Bigg(-\sin\frac{\beta}{2} \Bigg )^{\ell - n} \: \Bigg( \cos \frac{\beta}{2} \Bigg)^{\ell+n},
\end{equation}
which follows by the expression for the Wigner $d$-functions in terms of Jacobi polynomials of \eqn{\ref{eq:wignerd_jacobi}}. To sketch how one reaches this result, consider the case in which the spin number $n=\ell$, the Jacobi polynomial term then reduces to unity, and the symmetry relation $d^{\ell}_{mn}(\beta) = (-1)^{n-m}d^{\ell}_{nm}(\beta)$ \cite[][\sectn{4.4}, p.\ 79, \eqn{1}]{varshalovich:1989} can then be applied to reorder the $m$ and $n$ indices, following which the trigonometric expressions and prefactor follow straightforwardly.  We calculate these terms in log-space to mitigate numerical issues inherent in evaluating large factorials.  Also note that $d^{\,\ell}_{\ell+1, n}(\beta)$ is trivially zero as $d^{\,\ell}_{m n}(\beta) = 0$ for all $m > \ell$.

\begin{figure}
  \centering
  \begin{tikzpicture}
    \node[] () at (0,0)
    {
      \includegraphics[width=0.6\linewidth, trim={5cm 10cm 5cm 10cm},clip]{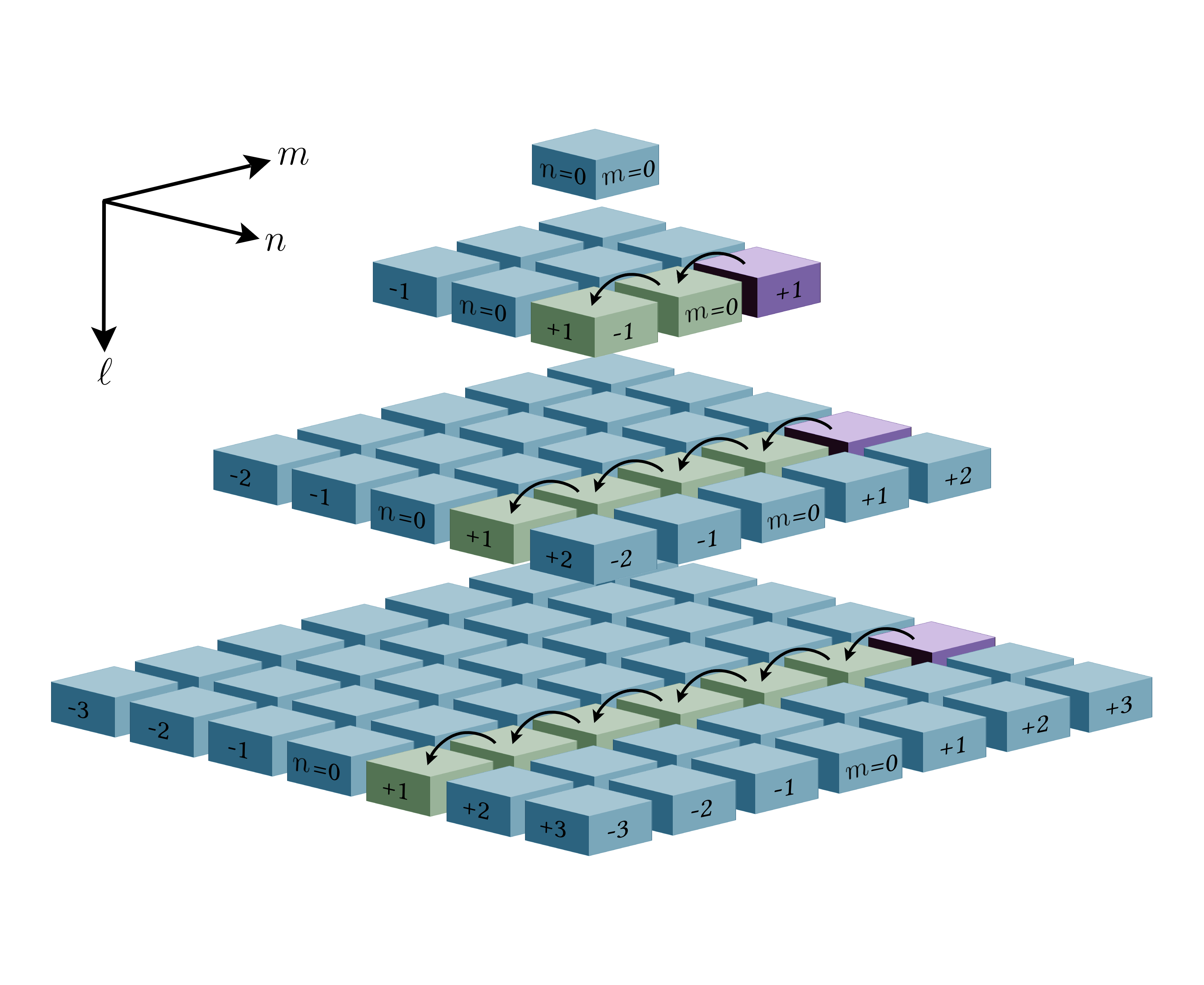}
    };
    \node[] () at (0,0)
    {
      \put(80,100){\fcolorbox{black}{gray!10}{
          \begin{tabular}{@{}l@{}}
            \scriptsize \includegraphics[width=15pt, trim={0 2cm 0 -1cm},clip]{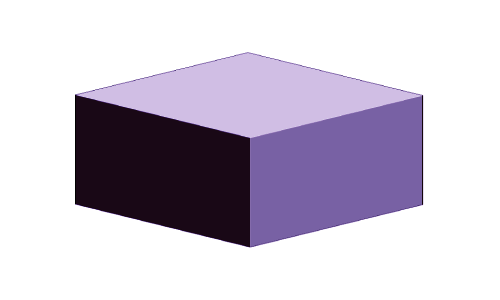} Initialisation (see Eq. \ref{eq:initial_values}) \\
            \scriptsize \includegraphics[width=15pt, trim={0 1cm 0 -2cm},clip]{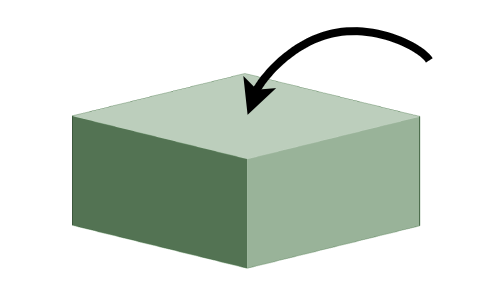} Recursion (see Eq. \ref{eq:wigner-d-recursion})
          \end{tabular}
        }}
    };
  \end{tikzpicture}
  \caption{Overview of the recursive strategy that we adopt to calculate the Wigner $d$-functions $d^{\ell}_{m n}(\beta)$, which underpin both the spin spherical harmonic and Wigner transforms.  The recursion iterates across order $m$ independently of $\ell$, $n$, and $\beta$, in the direction of reducing $m$ (to ensure stability).  Such a recursion supports high degrees of parallelisation, where computations for $\ell$, $n$, and $\beta$ are independent and so can be parallelised across the very large number of threads available on modern hardware accelerators, such as GPUs and TPUs.  Furthermore, this approach avoids the need for any recursion to compute transforms for different spins $n$, providing an efficiency saving over alternative approaches that often compute spin transforms through repeated application of spin lowering/raising operations.  The algorithm is initialised for $d^{\ell}_{\ell n}(\beta)$.  We also consider a non-conditional renormalisation approach to avoid under- and over-flow in a manner that is applicable to differentiable programming frameworks such as \texttt{JAX}.  The resulting recursive algorithm is stable to at least $\ell \sim 10,000$ and likely well beyond.}
  \label{fig:wigner_d_recursion_schematic}
\end{figure}

While some spherical samplings avoid samples at the north and south poles, others do have samples on the poles.  Since we aim to support a number of different spherical samplings (as discussed in \sectn{\ref{sec:background:specific_samplings}}), including those with samples defined on the poles, we must be able to evaluate Wigner $d$-functions on the poles.  The initialisation and recursion presented above are not suitable on the poles  ($\beta=\{0, \pi\}$) since the recursion coefficient $\lambda_m$ is undefined due to the $\sin\beta$ term in the denominator. Fortunately, simple forms for these elements of the Wigner $d$-functions are available, which we implement explicitly. Specifically, at the north pole
\cite[][\sectn{4.16}, p.\ 112, \eqn{1}]{varshalovich:1989}
\begin{equation}
  d^\ell_{mn}(0) = \delta_{mn}
\end{equation}
and at the south pole
\begin{equation}
  d^\ell_{mn}(\pi) = (-1)^{\ell+m} d^\ell_{m,-n}(0) = (-1)^{\ell+m} \delta_{m,-n} ,
\end{equation}
which follows from \citet[][\sectn{4.4}, p.\ 79, \eqn{1}]{varshalovich:1989}.
Our recursive algorithm to compute the Wigner $d$-functions in a highly parallelised and stable manner, and its initialisation, is outlined graphically in \fig{\ref{fig:wigner_d_recursion_schematic}} for clarity.

Additionally, note that the Wigner $d$-functions exhibit an extreme dynamic range. For example, the magnitude of these coefficients $d^{\ell}_{m n}(\beta)$ often ranges in excess of many hundreds of orders of magnitude.  Hence, not only must one recurse along stable paths but one must also develop protocols to avoid under- and over-flowing, which can be equally catastrophic. A common approach to address this issue is to implement a renormalisation scheme, which conditionally compresses the dynamic range during the recursion to an acceptable interval. However, while this may solve the dynamic range problem in theory, algorithms with this conditional structure do not permit efficient automatic differentiation. In fact, due to branching trees, such an algorithm can become extremely expensive to differentiate. Alternative methods exist, such as continued fractions \citep[\textit{e.g.}][]{gautschi:1967}, however, as we discuss below in \sectn{\ref{sec:recursions:jax}}, with some careful software engineering it is possible to develop non-conditional algorithms for renormalisation that are not only feasible, but exhibit preferable complexity scaling.

\subsection{Stable \texttt{JAX} recursion} \label{sec:recursions:jax}

Beginning from the initial values given by \eqn{\ref{eq:initial_values}} sequentially evaluating the recursion presented in \eqn{\ref{eq:wigner-d-recursion}} can rapidly result in iterates $d^{\ell}_{m n}(\beta)$ that under- or over-flow 64-bit representations. One may diligently check the iterates magnitude at recursive step $j$ against some large scaling factor $C$ and, if larger, update a running normalisation constant $C^{\ell \beta} \rightarrow C^{\ell \beta}/C$ before retrospectively renormalising all $d^{\ell}_{\ell-k, n} \rightarrow d^{\ell}_{\ell-k, n}/C$ for $k \leq j$ therefore avoiding problematic numerical overflows. This approach is simple and effective for serial evaluation, however it creates two issues for accelerated differentiable programming.

Firstly, suppose whilst recursing the iterate magnitude becomes larger than $C$ after each $j$ iterations, at which point the above renormalisation scheme is executed. For a given harmonic degree $\ell$ one would expect to encounter $\sim \ell / j$ renormalisation nested loops. The average number of iterates that must be renormalised within each of these loops is straightforwardly $\sim \ell/2$. Compounding these factors it is clear that this approach to renormalisation changes the overall complexity of the recursion from $\mathcal{O}(\ell)$ to $\mathcal{O}(\ell^2)$. Adopting such an approach for a spherical harmonic transform with bandlimit $L$ results in $\mathcal{O}(L^4)$ complexity which is prohibitive for high $L$.

Secondly, although control flow primitives acting on tracer objects, such as checking the magnitude of iterates, are supported by \texttt{JAX}, they are inefficient in this case. Both branches of the conditional flow must be compiled and executed at runtime, whether the condition is met or not. The renormalisation strategy involves two such primitives within every iteration, one to determine whether the renormalisation condition is met and another to determine which elements to which to apply this renormalisation. Not only do such nested conditional flow primitives require exponential memory allocation, they also require CPU scheduling incurring further communication overhead. Similar issues are prevalent when evaluating the gradients of such primitives.

We propose, perhaps counter-intuitively, to instead update the normalisation after every recursive step, in effect performing a running renormalisation by default, rather than dynamically renormalising when under- or over-flow conditions are met. In this manner we avoid passing back over terms many times to renormalise, thus avoiding the associated computation cost and conditional branching that is otherwise required.  Explicitly, we begin by normalising $d^{\ell}_{\ell, n}$ to unity and store this value as our initial normalisation. We then iterate once in descending $m$ calculating the normalised version of $d^{\ell}_{\ell-1, n}$, before undoing the normalisation and reading off $d^{\ell}_{\ell-1, n}$. At this point the absolute value of the normalised $d^{\ell}_{\ell-1, n}$ is computed and aggregated into the normalisation factor. Hence, each recursive step begins normalised to unity with an appropriate running normalisation. Updating the normalisation in this way is computationally inexpensive and is a fair price to pay to retain $\mathcal{O}(\ell)$ complexity. Moreover, as we automatically renormalise after each iteration, no control flow primitives are needed, and thus issues pertaining to automatic differentiation and nested branching trees are avoided entirely.  Throughout we compute and store terms in log-space to achieve a larger dynamic range, further mitigating the risk of under- and over-flow.

Through the approaches presented, we are able to compute Wigner $d$-functions in a manner that can be efficiently deployed on modern hardware accelerators exhibiting high degrees of parallelisation, that provides efficient automatic differentiation, and that are stable to very high degrees $\ell$, to at least $\ell \sim 10,000$ and likely well beyond.  In the following section we leverage these approaches to compute spin spherical harmonic and Wigner transforms.

\section{Fast Fourier transforms on the sphere and rotation group}
\label{sec:algorithms}

We present in this section algorithms for the fast computation of Fourier transforms on the two-sphere $\sphere$ and three-dimensional rotation group $\sothree$: namely, spherical harmonic transforms (for any arbitrary spin) and Wigner transforms, respectively.  We consider a simple general algorithmic structure that is well suited to many different sampling approaches and for deployment on modern hardware accelerators, such as GPUs and TPUs, leveraging the parallelised and differentiable Wigner $d$-function computation presented in \sectn{\ref{sec:recursions}}.  We discuss in \sectn{\ref{sec:gradients}} the gradient computation for the spherical transforms presented in this section.

In much the same way that the standard fast Fourier transform \cite[FFT;][]{cooley:1965} is reliant on uniform sampling, fast generalised Fourier transforms on the sphere and rotation group are reliant on isolatitudinal sampling.  Consequently, we focus our attention solely to isolatitudinal sampling (as discussed in \sectn{\ref{sec:background:sampling}}), which includes the majority of commonly adopted spherical sampling approaches (as discussed in \sectn{\ref{sec:background:specific_samplings}}).  The key advantage of isolatitudinal samplings is that a separation of variables may be performed to reduce computational complexity \cite[\textit{e.g.}][]{gorski:2005, mcewen:fssht, mcewen:so3, mcewen:2013:waveletsxv}.

Since our aim is to support many different spherical sampling approaches (\textit{e.g.}\ DH, MW, GL, \texttt{HEALPix}), we adopt a general algorithmic structure for the computation of spherical transforms that can be applied to any isolatitudinal sampling that exhibits an explicit quadrature for the spherical transform.  Not all sampling approaches necessarily exhibit an explicit quadrature and we discuss some tricks to address this.

We focus first on the general algorithmic form of the spherical harmonic transform and then the specifics of different sampling schemes.  We then discuss a precomputation approach that is highly efficient for moderate bandlimits, before presenting an alternative optimisation applicable for high bandlimits that is achieved by interleaving the Wigner $d$-function recursion with the computation of the harmonic transform.  Finally, we discuss the application of the techniques presented to also compute Wigner transforms.

\subsection{Spin spherical harmonic transforms}
\label{sec:algorithms:spherical_harmonic_transforms}

We consider the computation of the spin spherical harmonic transform of bandlimited signals $f \in \mathcal{B}_L(\mathbb{S}^2)$ denoted by $\mathcal{F} : \mathcal{B}_L(\mathbb{S}^2) \rightarrow \mathbb{C}^{L^2}$ and the inverse denoted by
$\mathcal{F}^{-1} : \mathbb{C}^{L^2} \rightarrow \mathcal{B}_L(\mathbb{S}^2)$.  Since we consider isolatitudinal samplings where the transform can be expressed through an explicit quadrature, a separation of variables can be applied to the discretised representation of the forward transform given by \eqn{\ref{eq:sht_cts_forward_discrete}}, yielding
\begin{equation} \label{eq:disc_sh_expanded}
  {}_s\hat{f}_{\ell m}
  \simeq
  (-1)^s
  \sqrt{\frac{2\ell+1}{4\pi}}
  \sum_{\theta}
  q_\Theta(\theta) \:
  d^{\ell}_{m, -s}(\theta) \:
  {}_s\tilde{f}_m(\theta) ,
\end{equation}
with
\begin{equation} \label{eq:disc_sh_fft}
  {}_s\tilde{f}_m(\theta)
  =
  \sum_{\phi}
  q_\Phi(\phi) \:
  {}_sf(\theta, \phi) \:
  \exp{- \img m\phi}.
\end{equation}
Without loss of generality we have expressed the quadrature weights in the separable form $q(\theta, \phi) = q_\Theta(\theta) \: q_\Phi(\phi)$ (merely for notational convenience for comparison with the inverse transform subsequently).  Recall that we adopt the symbol $\simeq$ to denote that the computation is approximate for arbitrary sampling schemes (\textit{e.g.} \texttt{HEALPix}) but is exact for sampling theorems (\textit{e.g.} DH, MW, GL).

Note that \eqn{\ref{eq:disc_sh_expanded}} is a projection onto the Wigner $d$-functions that can be computed with complexity $\mathcal{O}(L^3)$, while \eqn{\ref{eq:disc_sh_fft}} is simply a one-dimensional Euclidean Fourier transform that can be computed by an FFT for each isolatitudinal ring at $\theta$ with overall complexity $\mathcal{O}(L^2\log L)$.  The separation of variables therefore results in an overall complexity of $\mathcal{O}(L^3 + L^2\log L) = \mathcal{O}(L^3)$, in contrast with the $\mathcal{O}(L^4)$ complexity of direct computation by \eqn{\ref{eq:sht_cts_forward_discrete}}.  Since \eqn{\ref{eq:disc_sh_fft}} is simply computed by FFTs, the main computational challenge is the efficient computation of \eqn{\ref{eq:disc_sh_expanded}}, which is discussed in greater detail in \sectn{\ref{sec:algorithms:precomputation}} and \sectn{\ref{sec:algorithms:interleaving}}.

A similar approach through a separation of variables may also be applied to compute the inverse transform of \eqn{\ref{eq:sht_cts_inverse}}, with a further reordering of summations, yielding
\begin{equation}  \label{eq:disc_ish_fft}
  {}_s f(\theta, \phi)
  =
  \sum_m {}_s \tilde{f}_m(\theta) \: \exp{\img m \phi} ,
\end{equation}
with
\begin{equation}  \label{eq:disc_ish_expanded}
  {}_s \tilde{f}_m(\theta)
  =
  \sum_{\ell}
  (-1)^s \sqrt{\frac{2\ell+1}{4\pi}} \:
  {}_s\hat{f}_{\ell m} \:
  d^{\ell}_{m, -s}(\theta) .
\end{equation}
Again, \eqn{\ref{eq:disc_ish_fft}} may be computed by FFTs with overall complexity $\mathcal{O}(L^2\log L)$ and \eqn{\ref{eq:disc_ish_expanded}} may be computed with complexity $\mathcal{O}(L^3)$, resulting in an overall complexity of $\mathcal{O}(L^3 + L^2\log L) = \mathcal{O}(L^3)$, in contrast with the $\mathcal{O}(L^4)$ complexity of direct computation by \eqn{\ref{eq:sht_cts_inverse}}.  The efficient computation of \eqn{\ref{eq:disc_ish_expanded}} is discussed further in \sectn{\ref{sec:algorithms:precomputation}} and \sectn{\ref{sec:algorithms:interleaving}}.

For now we simply note that since we develop methods to compute the Wigner $d$-functions that recurse over $m$ only (see \sectn{\ref{sec:recursions}}) that we can compute independently for all $\ell$ and $\theta$, and typically fixed $n$, \eqn{\ref{eq:disc_sh_expanded}} and \eqn{\ref{eq:disc_ish_expanded}} can be computed in a highly parallelised manner on hardware accelerators, \textit{i.e.}\ GPUs and TPUs, before reducing over $\theta$ or $\ell$.

Next we discuss how to map various sampling approaches onto the general algorithmic structure for the forward and inverse spherical harmonic transform presented here.  In many cases this is straightforward, although in some cases it is non-trivial.

\subsection{Specifics of different sampling approaches}
\label{sec:algorithms:samplings}

We summarise the details required to support a number of different sampling approaches, including the DH, MW, and GL sampling theorems, which admit theoretically exact spherical transforms, and the \texttt{HEALPix} sampling scheme, which admits only approximate spherical transforms but exhibits the practical advantage of equal-area pixels.  All sampling approaches are mapped onto the general algorithmic structure presented in \sectn{\ref{sec:algorithms:spherical_harmonic_transforms}} for the computation of spherical harmonic transforms through a separation of variables with an explicit quadrature for the spherical transform.

\subsubsection{Driscoll \& Healy (DH) sampling theorem}

The DH \cite[Driscoll \& Healy;][]{driscoll:1994} sampling theorem adopts equiangular sampling position on the sphere defined by
\begin{equation}
  \theta_t = \frac{(2t+1)\pi}{4L}
  \quad\text{for}\quad t \in \{0, 1, \ldots, 2L-1\}
\end{equation}
and
\begin{equation}
  \phi_p = \frac{2\pi p}{2L-1}
  \quad\text{for}\quad p \in \{0, 1, \ldots, 2L-2\},
\end{equation}
resulting in $\sim 4L^2$ samples on the sphere.\footnote{Note that slight variants of the DH sample positions have been considered \cite{driscoll:1994,healy:2003}.  We adopt the $\theta$ sample positions of \cite{healy:2003} but adopt a (slightly) more efficient sampling of $\phi$ that matches the $\phi$ sampling adopting subsequently for other equiangular sampling theorems.} These sample locations afford a sampling theorem with explicit quadrature weights for spherical transforms given by \cite{driscoll:1994,mcewen:2011:waveletsxiv,healy:2003}
\begin{equation}
  q(\theta,\phi)
  = \frac{4\pi}{L(2L-1)} \sin\theta
  \sum^{L-1}_{k=0}
  \frac{\sin \big((2k+1)\,\theta\big)}{2k+1}.
\end{equation}
Note that the quadrature weights are constant with respect to $\phi$.
Spherical harmonic transforms can thus be computed straightforwardly for the DH sampling theorem, following the algorithmic structure outlined above in \sectn{\ref{sec:algorithms:spherical_harmonic_transforms}}.

\subsubsection{McEwen \& Wiaux (MW) sampling theorem}

The MW \cite[McEwen \& Wiaux;][]{mcewen:fssht} sampling theorem provides an alternative equiangular sampling theorem requiring only $\sim 2L^2$ samples on the sphere, providing a reduction in the Nyquist rate on the sphere by a factor of two compared to the DH sampling theorem.

There are two variants of the MW sampling scheme: the original scheme which provides the minimal Nyquist sampling on the sphere \cite{mcewen:fssht}; and a variant which slightly over samples (but yet still with $\sim 2L^2$ samples) to yield sample locations with antipodal symmetry that is useful for many practical applications \cite{daducci:ssdmri,ocampo:disco}, denoted MWSS (MW with symmetric sampling).
The MW sample positions are given by
\begin{equation}
  \theta_t = \frac{(2t+1)\pi }{2L-1} \quad\text{for}\quad t \in \{0, 1, \ldots, L-1\}
\end{equation}
and
\begin{equation}
  \phi_p = \frac{2 \pi p}{2L-1} \quad\text{for}\quad p \in \{0, 1, \ldots, 2L-2\}
  ,
\end{equation}
whereas the MWSS sample positions are given by
\begin{equation}
  \theta_t = \frac{2\pi t}{2L} \quad\text{for}\quad t \in \{0, 1, \ldots, L\}
\end{equation}
and
\begin{equation}
  \phi_p = \frac{2 \pi p}{2L} \text{ for } p \in \{0, 1, \ldots, 2L-1\}
  .
\end{equation}

Quadrature weights have been presented for both the MW \cite{mcewen:fssht} and MWSS \cite{ocampo:disco} cases, which we unify into a single expression.  The MW sampling theorem is based on an extension of the sphere to the torus through careful period extensions of the function of interest to the $\theta$ domain $[0,2\pi)$; recall that on the sphere $\theta \in [0, \pi]$.  Consider the bandlimited representation of the function defined by $\sin\theta$ on $[0,\pi]$ and zero on $(\pi, 2\pi)$, given by
\begin{equation}
  w(\theta_t) = \sum_{m} \hat{w}(m^\prime) \exp{i m^\prime \theta_t} ,
\end{equation}
with Fourier coefficients
\begin{equation}
  \hat{w}(m^\prime)
  =
  \int_0^\pi \sin(\theta) \exp{i m^\prime \theta} \text{d}\theta
  =
  \begin{cases}
    \: \pm i \pi/2        & m^\prime=\pm 1                           \\
    \: 0                  & m^\prime \text{ odd}, m^\prime \neq \pm1 \\
    \: 2/(1-{m^\prime}^2) & m^\prime \text{ even}
  \end{cases}
  .
\end{equation}
The unified quadrature weights are then given by folding back contributions from $(\pi, 2\pi)$ onto $(0,\pi)$ and read
\begin{equation}
  q(\theta_t, \phi) = \frac{2\pi}{T^2} \bigl ( w(\theta_t) + \zeta_t (-1)^s w(\theta_{t}^\text{reflect}) \bigr ) ,
\end{equation}
where
\begin{equation}
  T
  =
  \begin{cases}
    \: 2L-1 & \text{for MW}   \\
    \: 2L   & \text{for MWSS} \\
  \end{cases} ;
  \quad
  \theta_t^{\text{reflect}}
  =
  \begin{cases}
    \: \theta_{2L-2-t} & \text{for MW}   \\
    \: \theta_{2L-t}   & \text{for MWSS} \\
  \end{cases} ;
  \quad
  \zeta_t
  =
  \begin{cases}
    \: (1-\delta_{t,L-1})             & \text{for MW}   \\
    \: (1-\delta_{t0})(1-\delta_{tL}) & \text{for MWSS} \\
  \end{cases}
  .
\end{equation}
The quadrature weights are constant with respect to $\phi$.

Note, however, that these quadrature weights are for the integration of a function bandlimited at $L$ and not for the computation of the spherical harmonic transform of \eqn{\ref{eq:sht_cts_forward}}, the integrand of which is bandlimited at $2L$.  While one could derive an explicit quadrature weight associated with the spherical transforms, we simply upsample signals on the sphere by a factor of two and consider quadrature weights appropriate for the integration of a function bandlimited at $2L$.  Note that the signal on the sphere is simply upsampled and the resulting spherical harmonic transform is still performed at bandlimit $L$.

Signals sampled following either of the MW sampling approaches can first be extended to the torus by a reflection in $\theta$, combined with the introduction on the extended domain of a shift of $\pi$ in $\phi$ and a $(-1)^s$ sign shift \cite{mcewen:fssht}.  While this extension can be computed in harmonic space for either MW or MWSS sampling theorems, it can also be straightforwardly computed in the spatial domain for MWSS due to the symmetric nature of the sample positions.  Once the extended function on the torus is computed it can be upsampling simply by zero padding in the Fourier domain and then truncated back to the spherical domain $\theta \in [0, \pi]$.
Following this approach the algorithmic structure of the spherical harmonic transform algorithms presented in \sectn{\ref{sec:algorithms:spherical_harmonic_transforms}} can be followed directly.

\subsubsection{Gauss-Legendre (GL) sampling theorem}

An alternative sampling theorem on the sphere can be built on Gauss-Legendre quadrature, where sample positions in $\theta$ are defined by the roots of the Legendre polynomials of order $L$ \cite[\textit{e.g.}][]{shukowsky:1986,mcewen:fssht}.  Samples positions in $\phi$ may be defined by
\begin{equation}
  \phi_p = \frac{2 \pi p}{2L-1} \quad\text{for}\quad p \in \{0, 1, \ldots, 2L-2\}.
\end{equation}
While equiangular sampling theorems are usually preferred for practical purposes due to the regular nature of sample positions in $\theta$, the GL sampling theorem on the sphere recovers the same asymptotic number of samples on the sphere as the MW sampling theorem ($\sim 2L^2$), although in practice slightly more samples are required by the GL sampling compared to MW.
Explicit quadrature weights for the computation of spherical transforms are given by \cite{press:1992}
\begin{equation}
  q(\theta, \phi) = \frac{4 \pi \sin^2\theta}{(2L-1) \bigl(P_L{}^\prime(\cos\theta)\bigr)^2} ,
\end{equation}
where above the prime denotes the derivative of the Legendre polynomial. The quadrature weights are constant with respect to $\phi$.
Spherical harmonic transforms can thus by computed straightforwardly for the GL sampling theorem, following the algorithmic structure outlined above in \sectn{\ref{sec:algorithms:spherical_harmonic_transforms}}.

\subsubsection{\texttt{HEALPix} sampling scheme}

\texttt{HEALPix} sampling is designed to provide pixels of equal area, which can be a very useful practical property. We refer the reader to \citet{gorski:2005} for full details regarding \texttt{HEALPix}, including the precise definition of sample positions, and focus here on the implications for computing spherical harmonic transforms.  Since pixels have equal area, approximate quadrature weights are simply given by $q(\theta,\phi) = 4 \pi / N_\text{pix}$ where $N_\text{pix}$ denotes the total number of samples over the sphere. The quadrature weights are constant with respect to both $\theta$ and $\phi$.
Before proceeding, note that for \texttt{HEALPix} it is customary to compute transforms to a maximum degree denoted by $\ell_\text{max}$, which is related to our definition of bandlimit $L$ by $\ell_\text{max} = L-1$.

By construction \texttt{HEALPix} samples are defined on isolatitudinal \textit{rings} with two additional attributes to support a hierarchical tessellation of the sphere: (i) every other ring is shifted by a fixed longitudinal offset; and (ii) the total number of samples per ring is not constant. The ring offsets are managed straightforwardly by a phase correction term which is very cheap and easily accounted for within the Fourier transform by modification of the exponential term.

For the forward spherical harmonic transform a 1D Fourier transform of $f(\theta,\phi)$ over $\phi$ for a ring of constant $\theta$ is performed to compute the intermediate functions $\hat{f}_m(\theta)$ of \eqn{\ref{eq:disc_sh_fft}}.  In theory, each ring contributes information to all $|m| \leq \ell$ for $\ell < L$, and when in the equatorial region (the region near the equator) \texttt{HEALPix} collects sufficient samples to capture this information. In the polar regions the number of longitudinal samples per ring is dynamic and (potentially much) less than $m$, hence information is lost. This is in fact precisely why \texttt{HEALPix} provides only approximate spherical harmonic transforms.

\begin{figure}%[!b]
  \centering
  \includegraphics[width=\linewidth]{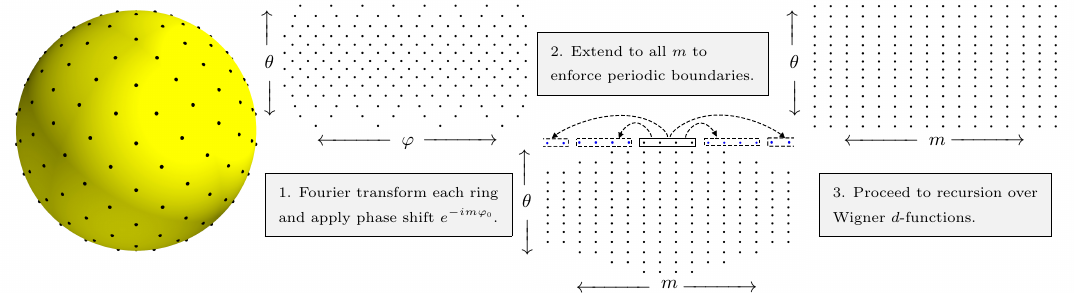}
  \caption{Overview of the additional steps required for \texttt{HEALPix} sampling \cite{gorski:2005} when computing Fourier transforms over $\phi$.  For \texttt{HEALPix} one must: (i) compute the Fourier transform each ring at varying resolution, determined by $n_{\phi}$ per ring; then (ii) generate synthetic orders $m^\prime$ by extending periodically available orders $|\,m\,| \leq |\,m^\prime\,|$. For the inverse spherical harmonic transform the order is reversed and step (ii) is simply replaced by aliasing higher orders $m^\prime$ $|\,m^\prime\,| > |\,m\,|$ into lower orders $m$.  This is in fact precisely why \texttt{HEALPix} provides only approximate spherical harmonic transforms.  For equirectangular sampling schemes this entire step can be computed simply by an FFT for each $\phi$ for all rings in a single line of code.}
  \label{fig:healpix_schematic}
\end{figure}

To mitigate this effect in the polar regions it is customary to first perform a fast Fourier transform over $\phi$ before \textit{repeating} the intermediate function $\hat{f}^{\,\text{polar}}_{m^\prime \leq m}(\theta)$, extending the coefficients periodically to fill out the Fourier $m$ line.  Correspondingly, during the inverse transform one \textit{folds} the full order $\hat{f}_m(\theta)$ onto $\hat{f}^{\,\text{polar}}_{m^\prime \leq m}(\theta)$,
aliasing high frequency information into the low frequency components.  As this nuance is perhaps not obvious, we provide an illustration in \fig{\ref{fig:healpix_schematic}} for clarity.

From a software engineering perspective, this raises some further complications. Ideally, the FFTs of \eqn{\ref{eq:disc_sh_fft}} and \eqn{\ref{eq:disc_ish_fft}} would be implemented as a single function call to \texttt{jax.numpy.fft} across a regular array. However, \texttt{jax.numpy.fft} does not support ragged arrays and therefore each ring within the polar region must computed by individual function calls within a scheduled loop. As the arrays are dynamic this cannot be replaced by, \textit{e.g.}, a \texttt{lax} scan, and therefore suffers from compile time bloat due to unrolled loops in \texttt{XLA} compilation.  Moreover, such serial evaluation of fast Fourier transforms for each isolatitudinal ring does not fully exploit the distributed compute of GPU devices. Nevertheless, the compute time of the overall transform is dominated by the Wigner $d$-functions summations and therefore this inefficiency of \texttt{HEALPix} is relatively minor, and may be solved by future developments of core \texttt{JAX} primitives.
In summary, \texttt{HEALPix} just-in-time compilation is therefore slow, although this only needs to be performed on the first function call (further optimisations to improve this will be considered in future).  Once \texttt{HEALPix} transforms are just-in-time compiled, execution is a little slower than for equiangular samplings but nevertheless still relatively fast.

\subsection{Full precomputation}
\label{sec:algorithms:precomputation}

We now consider how best to evaluate the summations involving the Wigner $d$-functions presented in \eqn{\ref{eq:disc_sh_expanded}} and \eqn{\ref{eq:disc_ish_expanded}}.  For moderate bandlimits ($L \lesssim 2048$) the Wigner $d$-functions may be precomputed, yielding highly efficient harmonic transforms, although at a cost of increased memory usage.
One may precompute and store all required elements $d^{\ell}_{m, -s}(\theta)$, the weighted sum of which can then be implemented extremely straightforwardly. Explicitly, for a forward spherical harmonic transform, given the precomputation and storage of the Wigner $d$-functions $\boldmath{D} = d^{\ell}_{m,-s}(\theta)$ for all $\ell,m,\theta$ at a fixed spin $s$, the transform is simply given by a series of 1D FFTs over $\phi$ for each $\theta$, followed by multiplication by the quadrature weights, followed by multiplication by matrix $\boldmath{D}$, followed by an $\ell$-dependent scaling.  The core implementation of the spherical harmonic transform therefore collapses effectively to a single line of code \texttt{jnp.einsum(`tlm, tm -> lm', $\boldmath{D}$, jnp.fft(f))}.  The inverse transform follows analogously, with minor changes (different ordering, forward 1D FFTs replaced by inverse FFTs, and different scalings).  While this precompute approach is highly computationally efficient, it requires $\mathcal{O}(L^3)$ memory, which is not feasible at high bandlimits $L$.  We next consider an approach that avoids fully precomputing Wigner $d$-functions,  eliminating $\mathcal{O}(L^3)$ memory requirements, and thus is scalable to high degrees $L$.  In this approach we do consider an optional precomputation of some other internal terms, although in a manner that limits memory requirements to $\mathcal{O}(L^2)$.

\subsection{Interleaving recursion and computation for on-the-fly transforms}
\label{sec:algorithms:interleaving}

\begin{figure}%[!b]
  \centering
  \begin{tikzpicture}
    \node[] () at (0,0)
    {
      \includegraphics[width=1.0\linewidth, trim={6cm 5cm 4cm 5cm},clip]{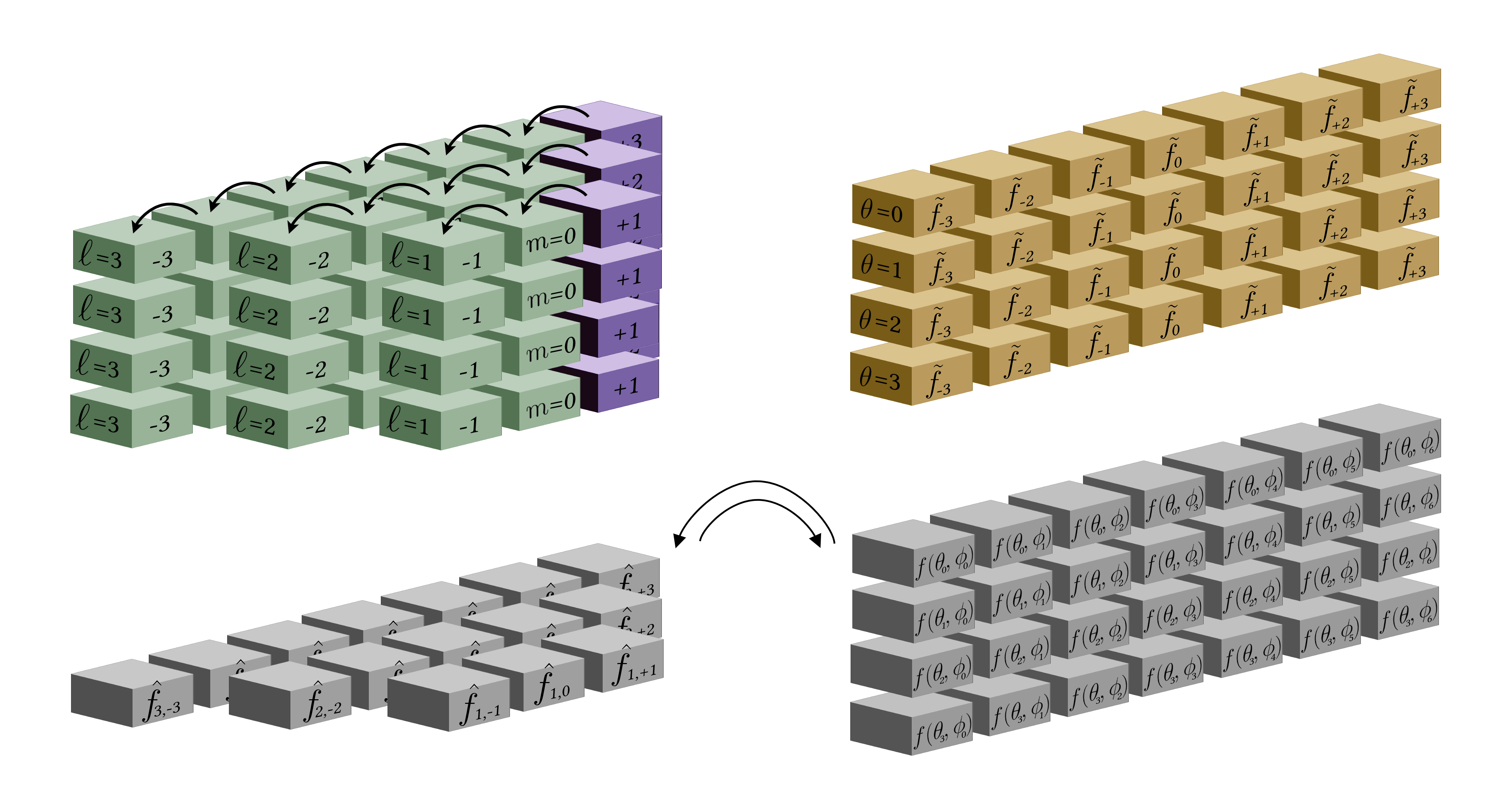}
      \put(-360,230){\large $d^{\ell}_{mn}(\theta)$}
      \put(-360,-2){\large ${}_s\hat{f}_{\ell m}$}
      \put(-120,230){\large ${}_s\tilde{f}_m(\theta)$}
      \put(-120,-2){\large ${}_sf(\theta, \phi)$}
      \put(-155,105){Eq. \ref{eq:disc_ish_fft} \rotatebox{90}{$\bf \longleftarrow$}}
      \put(-85,123){Eq. \ref{eq:disc_sh_fft} \rotatebox{90}{$\bf \longrightarrow$}}
      \put(-254,160){\rotatebox{0}{$\bf \xleftarrow{\makebox[1cm]{}}$}}
      \put(-254,154){\rotatebox{0}{$\bf \xrightarrow{\makebox[1cm]{}}$}}
      \put(-241,165){$\times$}
      \put(-242,147){$\sum\limits_{\ell}$}
      \put(-251,130){Eq. \ref{eq:disc_ish_expanded}}
      \put(-330,73){\rotatebox{90}{$\bf \xrightarrow{\makebox[.5cm]{}}$}}
      \put(-321,80){$\times$}
      \put(-428,55){\rotatebox{90}{$\bf \xleftarrow{\makebox[1cm]{}}$}}
      \put(-421,73){$\sum\limits_{\theta}$ \hspace{2pt}Eq. \ref{eq:disc_sh_expanded}}
      \put(-241,97){$\mathcal{F}$}
      \put(-243,72){$\mathcal{F}^{-1}$}
    };
  \end{tikzpicture}
  \caption{Diagrammatic overview of our forward and inverse spin spherical harmonic transforms. The forward transform $\mathcal{F}$ begins from a bandlimited signal on the sphere ${}_sf(\theta,\phi)$ (bottom right) from which an intermediate function ${}_s \tilde{f}_m(\theta)$ (top right) is calculated by performing a 1D FFT over $\phi$ for each $\theta$. This intermediate function ${}_s\tilde{f}_m(\theta)$ is subsequently projected onto $d^{\ell}_{mn}$ (top left) before being summed over $\theta$ to produce harmonic coefficients ${}_s\hat{f}_{\ell m}$ (bottom left). The inverse spin spherical harmonic $\mathcal{F}^{-1}$ follows straightforwardly in the reverse direction as illustrated. We provide support for two operational modalities: precompute and on-the-fly. Our precompute transform explicitly calculates and stores $d^{\ell}_{mn}(\theta)$ for a given spin number $n$, which produces dramatically accelerated transforms at runtime at the cost of $\mathcal{O}(L^3)$ memory overhead. Our on-the-fly transform evaluates $d^{\ell}_{mn}(\theta)$ for a given spin number $n$ and single harmonic mode $m$ (see Sec. \ref{sec:recursions}) at each recursive step (curved arrow) projecting onto harmonic space by $\sum_{\theta}$ or the intermediate representation by $\sum_{\ell}$.}
  \label{fig:otf_recursion}
\end{figure}
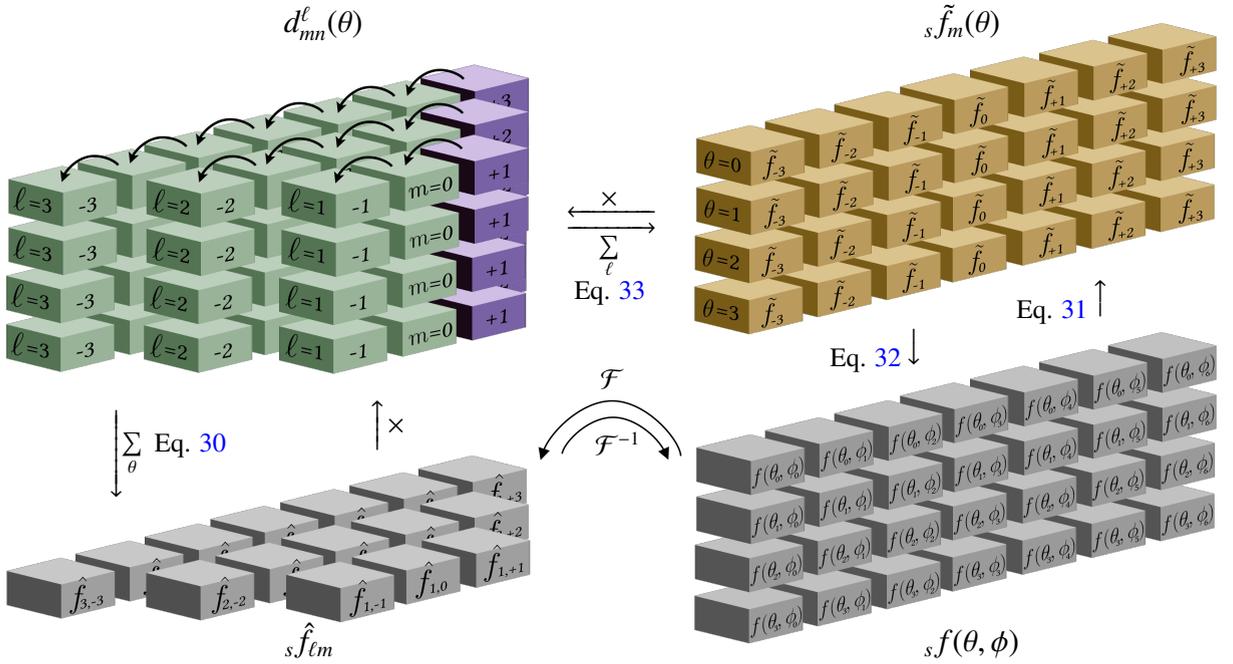

To scale to high resolutions (from $L \gtrsim 1024$, to $L \sim 10,000$ and beyond), the Wigner $d$-function components $d^{\ell}_{m, -s}(\theta)$ must instead be computed, used, and discarded recursively on-the-fly. Suppose one adopts the recursive algorithm defined in \eqn{\ref{eq:wigner-d-recursion}}, then evaluation of \eqn{\ref{eq:disc_sh_expanded}} is best solved by interleaving the summation over $\theta$ with the recursive step over order $m$. In the following we adopt the shorthand $\boldmath{D}^{\ell \theta}_{j}$ for the $j^{\text{th}}$ iterant during our recursive computation of $d^{\ell}_{m, -s}(\theta)$. Within a recursive step $j$ we: (i) evaluate $\boldmath{D}^{\ell \theta}_{j+1}$ from $\boldmath{D}^{\ell \theta}_{j}$ and $\boldmath{D}^{\ell \theta}_{j-1}$ after which these arrays are incremented; (ii) renormalise $\boldmath{D}^{\ell \theta}_{j+1}$ as discussed in \sectn{\ref{sec:recursions}}; (iii) project onto $\boldmath{D}^{\ell \theta}_{j+1}$; and (iv) sum over all co-latitudes $\theta$. In this way our peak memory overhead is capped with complexity $\mathcal{O}(L^2)$ which is the memory required to store a single spherical image. Furthermore, we optionally precompute some additional internal terms to save computational time, while keeping memory overhead to $\mathcal{O}(L^2)$. The inverse transform is completely analogous.  This resulting program is outlined in \fig{\ref{fig:otf_recursion}}.

As noted above, the Wigner summation terms of the forward and inverse transforms, given by \eqn{\ref{eq:disc_sh_expanded}} and \eqn{\ref{eq:disc_ish_expanded}}, can be computed in a highly parallelised manner on hardware accelerators, \textit{i.e.} GPUs and TPUs, before reducing over $\theta$ or $\ell$.  We recover $\mathcal{O}(L^2)$ independent threads, which for high bandlimits $L \sim 10^4$ can reach $L \sim 10^8$.  We are therefore able to make good use of the very large number of threads available on modern hardware accelerators.  Furthermore, we can trivially distribute computations over multiple accelerators, \textit{e.g.} multiple GPU devices, through single program multiple data (SPMD) computation, which is handled seamlessly in \texttt{JAX}. Asymptotically, such distributed computation recovers a computational saving given by the number of GPU devices available, and so for large clusters the acceleration can be extreme (indeed, in practice we find very close to a linear computational saving; see \sectn{\ref{sec:software}}).  Due to the {$\mathcal{O}(L^2)$} distribution and parallelisation of our algorithms, given sufficient computational resources our transforms exhibit {$\mathcal{O}(L)$} time complexity.

\subsection{Wigner transforms}

To conclude this section we consider the computation of the forward Wigner transform of bandlimited signals $f \in \mathcal{B}_L(\sothree)$ denoted by $\mathcal{W} : \mathcal{B}_L(\text{SO}(3)) \rightarrow \mathbb{C}^{(4L^3-L)/3}$ and the inverse denoted by $\mathcal{W}^{-1} : \mathbb{C}^{(4L^3-L)/3} \rightarrow \mathcal{B}_L(\text{SO}(3))$.  As discussed in \sectn{\ref{sec:background:wigner_transform}} \citep{mcewen:so3} by relating the Wigner $D$-functions to the spin spherical harmonic basis functions, one can compute the Wigner transform through a series of spin spherical harmonic transforms and Fourier transforms (see \eqn{\ref{eq:wigner_sht_forward}}--\eqn{\ref{eq:wigner_sht_inverse}}).  We can therefore make use of the approach previously discussed to compute forward and inverse spherical harmonic transforms for arbitrary spin, making the association $\alpha \leftrightarrow \phi$ and $\beta \leftrightarrow \theta$, and simply include an additional 1D FFT with respect to $\gamma$.

By selecting an equiangular sampling in $\gamma$ we can appeal to the usual Nyquist sampling theorem for periodic functions on the circle $\mathbb{S}^1$.  The full quadrature weights can then be written as $q(\alpha, \beta, \gamma) = q_\Phi(\alpha) \: q_\Theta(\beta) \: q_\Gamma(\gamma)$, where $q_\Phi(\alpha) \: q_\Theta(\beta)$ vary depending on the spherical sampling theorem or scheme adopted (as discussed in \sectn{\ref{sec:algorithms:samplings}}).  Since for all spherical samplings we consider equiangular sampling in $\gamma$, the corresponding quadrature weights read $q_\Gamma(\gamma) = 2 \pi / (2 N - 1)$ for azimuthal bandlimit $N$.  Whether or not we recover a sampling theorem on $\sothree$ with exact Wigner transforms simply depends on whether we adopt a sampling scheme (\textit{e.g.} HEALPix) or theorem (\textit{e.g.} DW, MW, GL) on the sphere.

This formulation of the Wigner transform straightforwardly inherits the advantages of our \texttt{JAX} spin spherical harmonic implementation and provides further avenues for distribution. The internal forward or inverse spherical harmonic transform considered must be evaluated for each spin number $n$, however each of these operations is independent and can be parallelised and/or distributed. Therefore for our Wigner transforms one may choose to distribute over not only $\ell$ and $\beta$ but also $n$.

\section{Gradients of spherical transforms} \label{sec:gradients}

One of the primary goals of this work is not only to provide computational approaches for spherical transforms that can be deployed on hardware accelerators but that also facilitate the efficient computation of gradients.  Gradients are essential for differentiable programming applications. For example, machine learning models on the sphere \cite{cohen:2018,kondor:2018:clebsch, cobb:efficient_generalized_s2cnn, mcewen:scattering} often require spherical transforms that are differentiable so that the models may be trained by gradient-based optimisation algorithms.  Differentiable physical models are also required for hybrid data-driven and model-based approaches; in many cases an underlying component of such physical models includes spherical transforms, for example in cosmology \citep{dodelson:2003} and seismology \citep{dahlen:1998}.

In this section we present techniques to efficiently compute gradients associated with forward and inverse spherical transforms.  We discuss both automatic and manual differentiation and their respective merits, and derive explicit expressions for Jacobian-vector and vector-Jacobian products.  Finally we present a hybrid approach, where we adopt manual differentiation for some components and automatic differentiation for others.  Our hybrid approach avoids the large memory overhead of (reverse-mode) automatic differentiation, while also avoiding the cumbersome bespoke implementation of full manual differentiation.

\subsection{Automatic differentiation (AD) modes}

Modern automatic differentiation (AD) predominately relies on one of two gradient propagation schemes: forward- and reverse-mode. Forward-mode AD applies the chain rule sequentially to each primitive operation in a forward (primal) trace of a function, accumulating the primal and derivative (tangent) variables during a single forward pass \cite{wengert:1964}. Conversely, in reverse-mode AD forward and reverse passes are performed. During the forward pass primal variables are computed and propagated, with intermediate values stored in memory for use in the following reverse pass.  In the reverse pass the derivatives (cotangents) with respect to intermediate values are propagated backwards \cite{baydin:2018}. Reverse mode AD applied to training neural networks with  a scalar cost function is commonly referred to as back-propagation \cite{rumelhart:1986}.  Forward- and reverse-mode AD provide efficient ways to compute, respectively, Jacobian-vector products and vector-Jacobian (transposed Jacobian-vector) products for the traced function.

The choice of which mode is most efficient is determined by the problem at hand. Consider the case in which we wish to compute the derivative of a function $\mathcal{F} : \mathbb{R}^n \rightarrow \mathbb{R}^m$.  A full column of the Jacobian can be computed in a single pass of forward-mode AD, whereas a full row of the Jacobian can be computed in a single pass of reverse-mode AD \cite{baydin:2018}.  Consequently, when $n \ll m$ forward-mode AD is most efficient, whereas when $m \ll n$ reverse-mode is most efficient.  However, note that intermediate values must be stored for reverse-mode AD, increasing memory requirements, which is not the case for forward-mode AD.

In machine learning scenarios one typically wishes to optimise a scalar loss function that depends on a very large number of input parameters (e.g., millions or billions of weights of a neural network).  To train a model using a gradient-based optimiser it is therefore necessary to compute gradients of a function $\mathcal{F} : \mathbb{R}^n \rightarrow \mathbb{R}$, where the Jacobian is identified with the gradient vector $\nabla \mathcal{F} \in \mathbb{R}^n$.  Reverse-mode AD is thus typically the preferred approach for training machine learning models.   Nevertheless, we support both forward- and reverse-mode AD.

\subsubsection{Jacobian-vector product (JVP)}

Consider the general setting of a function $\mathcal{F} : \mathbb{R}^n \rightarrow \mathbb{R}^m$ and its Jacobian $\partial F(x)$ evaluated at $x\in\mathbb{R}^n$.  The Jacobian provides a linear map between the tangent space of the domain of $\mathcal{F}$ at $x$ (isomorphic to $\mathbb{R}^n$) to the tangent space of the codomain of $\mathcal{F}$ at $\mathcal{F}(x)$ (isomorphic to $\mathbb{R}^m$): $\partial \mathcal{F}(x) : \mathbb{R}^n \rightarrow \mathbb{R}^m$.  Computationally, the Jacobian-vector product (JVP) of the Jacobian at $x$ and a tangent vector $v \in \mathbb{R}^n$ is given by the functional mapping $\big (x,v\big) \mapsto \big (\mathcal{F}(x), \partial \mathcal{F}(x) v \big)$.
When composing multiple functions both the primal and tangent values are computed and propagated.  Since values are used as they are computed in a single forward pass, intermediate values do not need to be stored.

To compute the Jacobian explicitly, forward-mode AD is initialised by setting a single element of $v$ to unity, corresponding to the input variable of interest, and the remainder to zero (i.e.\ one hot encoding).  It is therefore clear that a full column of the Jacobian can be computed in a single pass of forward-mode AD and hence it is most efficient for $m \gg n$, i.e.\ for tall Jacobians, as commented above.

\subsubsection{Vector-Jacobian product (VJP)}

Consider again the function $\mathcal{F} : \mathbb{R}^n \rightarrow \mathbb{R}^m$. Now, however, consider an element $v \in \mathbb{R}^m$ of the cotangent space (i.e.\ the tangent space of the codomain of $\mathcal{F}$).  Computationally, the vector-Jacobian product (VJP) is given by the functional mapping $\big (x,v\big) \mapsto \big (\mathcal{F}(x), v^\text{T} \partial \mathcal{F}(x) \big)$.  One may alternatively consider the linear mapping of a VJP as the transpose of the linear mapping of a JVP: $\big (x,v\big) \mapsto \big (\mathcal{F}(x), \partial \mathcal{F}(x)^\text{T} v \big)$. The map $\partial \mathcal{F}(x)^\text{T} : \mathbb{R}^m \rightarrow \mathbb{R}^n$ provides a linear mapping on cotangent spaces.  Since the mapping is in the reverse direction to the application of the function $\mathcal{F}$, cotangents need to be propagated in reverse.

When composing multiple functions, for reverse-mode AD an initial forward pass is performed to compute and propagate the primals $\mathcal{F}(x)$, storing intermediate values of $x$.  On the reverse pass the cotangents are computed by $\partial \mathcal{F}(x)^\text{T} v$, making use of the intermediate values of $x$ to compute $\partial \mathcal{F}(x)^\text{T}$, and then accumulated in reverse.

To compute the Jacobian explicitly, reverse-mode AD is initialised by setting a single element of $v$ to unity, corresponding to the output variable of interest, and the remainder to zero.  It is therefore clear that a full row of the Jacobian can be computed in a single pass of reverse-mode AD and hence it is most efficient for $m \ll n$, i.e.\ for wide Jacobians, as commented above.

\subsection{Automatic versus manual differentiation}

By leveraging an AD framework, such as \texttt{JAX}, one can compute JVPs and VJPs simply through the framework.  This approach avoids the need for any gradient-related implementation, thus eliminating complexity and possibilities for errors.  However, as discussed above, reverse-mode AD induces a memory overhead that in the worst case can grow in proportion to the number of operations of the traced function \cite{baydin:2018}.  Furthermore, for machine learning applications, reverse-mode AD is the mode of interest since it is most efficient for gradient-based optimisation of scalar loss functions that depend on a very large number of input parameters.  For the spherical transforms considered in this article, we indeed observed that the computation of VJPs purely by reverse-mode AD exhibited a very high memory overhead (as discussed further below).

An alternative to automatic differentiation is manual differentiation, where explicit expressions for JVPs and VJPs are determined analytically and then implemented directly.  For linear transforms, such as the spherical transforms considered herein, JVPs and VPJs take relatively straightforward forms, where the JVP is simply the transform itself and the VJP is closely related to the adjoint transform.  Algorithms to compute adjoints of spherical transforms efficiently have been constructed previously \cite{mcewen:css2,wallis:s2_recon,price:s2inv}.  Such a direct implementation of manual derivatives requires bespoke implementation, which can be cumbersome (especially to support the subtleties of different spherical samplings) and thus can be prone to errors.  Nevertheless, a forward pass to compute VJPs is not required, eliminating the memory overhead that is otherwise necessary.

\subsection{Manual JVPs and VJPs}
\label{sec:gradients:manual}

We derive analytic expressions for the JVP and VJP (and adjoint) of spherical  transforms.  For simplicity, but without loss of generality, we focus on the forward and inverse spherical harmonic transforms, although the expressions presented can be trivially extended to Wigner transforms.  For notational brevity we drop the explicit spin subscript in what follows, although the results presented directly hold for spin functions on the sphere.

\subsubsection{Forward spin spherical harmonic transform}

Consider the forward spin spherical harmonic transform $\mathcal{F} : \mathcal{B}_L(\mathbb{S}^2) \rightarrow \mathbb{C}^{L^2}$.
For computational purposes we must of course consider discretised representations of signals, which for sampling theorems corresponds to the space of bandlimited signals $f \in {B}_L(\mathbb{S}^2)$.  When adopting a sampling scheme that does not exhibit a formal sampling theorem (e.g.\ \texttt{HEALPix}), the space of signals is not formally ${B}_L(\mathbb{S}^2)$.  Nevertheless, we avoid making an explicit distinction to avoid complicating notation.  Furthermore, while for \eqn{\ref{eq:sht_cts_forward_discrete}} we distinguished between the exactness of sampling theorems and the approximation of sampling schemes, we consider the computed harmonic coefficients here $\hat{f} \in \mathbb{C}^{L^2}$ (rather than underlying true coefficients of a bandlimited signal) and so expressions that follow are computationally equal, rather than approximate.

The elements of the Jacobian of the forward transform $\mathcal{F}$ follow from \eqn{\ref{eq:sht_cts_forward_discrete}} and read
\begin{equation} \label{eq:jacobian_entries_forward}
  \Bigl (\partial \mathcal{F}(f) \Bigr)_{\ell m}^{\theta \phi}
  = \frac{\partial \hat{f}_{\ell m}}{\partial f(\theta, \phi)}
  = q(\theta, \phi) \: {}_sY^{*}_{\ell m}(\theta, \phi)
  .
\end{equation}
Critically, notice that the Jacobian is independent of the signal $f$, i.e.\ $\partial\mathcal{F}(f) = \partial\mathcal{F}$.  Consequently, intermediate primal values do not need to be stored in reverse-mode AD (nevertheless, other memory overheads do arise, as discussed below).
The JVP of a tangent $v \in T_{f}\mathcal{B}_L(\mathbb{S}^2) \cong \mathcal{B}_L(\mathbb{S}^2)$, where $T_{f}$ denotes the tangent space evaluated at $f$, is then given by
\begin{equation}
  \Bigl(\partial \mathcal{F} v \Bigr)_{\ell m}
  = \sum_{\theta \phi} \:
  q(\theta, \phi) \:
  v(\theta, \phi)
  \:{}_sY^{*}_{\ell m}(\theta, \phi)
  ,
\end{equation}
which as expected is nothing more than the transform itself, i.e.
\begin{equation}
  \partial \mathcal{F} v
  = \mathcal{F} v
  .
  % = \hat{v},
\end{equation}
The VJP of a cotangent $\hat{v} \in T_{\mathcal{F}(f)}\mathbb{C}^{L^2} \cong \mathbb{C}^{L^2}$, where $T_{\mathcal{F}(f)}$ denotes the cotangent space evaluated at $\mathcal{F}(f)$, is given by
\begin{equation}
  \Bigl(\partial \mathcal{F}^\text{T} \hat{v} \Bigr)_{\theta \phi}
  = \sum_{\ell m} \:
  q(\theta, \phi) \:
  \hat{v}_{\ell m}
  \:{}_sY^{*}_{\ell m}(\theta, \phi) ,
\end{equation}
which may be expressed in terms of an inverse transform $\mathcal{F}^{-1}$ by
\begin{equation} \label{eq:vjp_forward_sht}
  \partial \mathcal{F}^\text{T} \hat{v}
  = ( \mathcal{Q} \: \mathcal{F}^{-1} \hat{v}^\ast )^\ast
  ,
\end{equation}
where $\mathcal{Q}$ is the diagonal operator acting on spatial functions that corresponds to multiplication by the quadrature weights $q(\theta, \phi)$. It is apparent that a VJP can be computed by an inverse spherical harmonic transform of a complex conjugated cotangent vector, followed by the application of quadrature weights and subsequent conjugation.

Adjoint operators of transforms are commonly required (e.g.\ when solving inverse problems posed as variational regularisation problems by convex optimisation techniques; \cite{mcewen:css2,wallis:s2_recon,price:s2inv}). The VJP is closely related to the adjoint operator of the transform up to complex conjugation.  For completeness we note the adjoint of the forward spherical harmonic transforms $\mathcal{F}$ can be represented as
\begin{equation} \label{eq:adjoint_forward_sht}
  \mathcal{F}^\text{H} = \mathcal{Q} \: \mathcal{F}^{-1}.
\end{equation}

While we have focused on the forward spin spherical harmonic transform $\mathcal{F}$ the above results extend trivially to the forward Wigner transform.  It is simply necessary to replace the spherical harmonic transform $\mathcal{F}$ with the Wigner transform for discretised signals $\mathcal{W}: \mathcal{B}_L(\text{SO}(3)) \rightarrow \mathbb{C}^{(4L^3-L)/3}$ and to let $\mathcal{Q}$ represent application of the quadrature weights $q(\alphaup,\betaup,\gammaup)$ on $\text{SO}(3)$.

\subsubsection{Inverse spin spherical harmonic transform}

Let us now consider the inverse spin spherical harmonic transform $\mathcal{F}^{-1} : \mathbb{C}^{L^2} \rightarrow \mathcal{B}_L(\mathbb{S}^2) $. Again, for computational purposes we must of course consider discretised representations of signals.

The elements of the Jacobian of the inverse transform $\mathcal{F}^{-1}$ follow from \eqn{\ref{eq:sht_cts_inverse}} and simply read
\begin{equation}
  \Bigl (\partial \mathcal{F}^{-1}(\hat{f}) \Bigr)^{\ell m}_{\theta \phi}
  = \frac{\partial f(\theta, \phi)}{\partial \hat{f}_{\ell m}}
  = {}_sY_{\ell m}(\theta, \phi) .
\end{equation}
The Jacobian of the inverse transform differs to the Jacobian of the forward transforms only due to complex conjugation and the absence of quadrature weights (\textit{cf.}~\eqn{\ref{eq:jacobian_entries_forward}}).  Notice again that the Jacobian is independent of the signal $\hat{f}$, i.e.\ $\partial\mathcal{F}^{-1}(\hat{f}) = \partial\mathcal{F}^{-1}$.  Consequently, intermediate primal values again do not need to be stored in reverse-mode AD (but again other memory overheads do arise).
The JVP of a tangent $\hat{v} \in T_{\hat{f}}\mathbb{C}^{L^2} \cong \mathbb{C}^{L^2}$ is then given by
\begin{equation}
  \Bigl(\partial \mathcal{F}^{-1} \hat{v} \Bigr)_{\theta \phi}
  = \sum_{\ell m} \:
  \hat{v}_{\ell m}
  \:{}_sY_{\ell m}(\theta, \phi)
  ,
\end{equation}
which again as expected is nothing more than the transform itself, i.e.
\begin{equation}
  \partial \mathcal{F}^{-1} \hat{v}
  = \mathcal{F}^{-1} \hat{v}
  .
  % = \hat{v},
\end{equation}
The VJP of a cotangent $v \in T_{\mathcal{F}^{-1}(\hat{f})}\mathcal{B}_L(\mathbb{S}^2) \cong \mathcal{B}_L(\mathbb{S}^2)$ is given by
\begin{equation}
  \Bigl(\partial \mathcal{F}^{-\text{T}} {v} \Bigr)_{\ell m}
  = \sum_{\theta \phi} \:
  v(\theta, \phi)
  \:{}_sY_{\ell m}(\theta, \phi)
  ,
\end{equation}
where we have adopted the notation $\mathcal{F}^{-\text{T}} = (\mathcal{F}^{-1})^\text{T}$.  The VJP may be expressed in terms of a forward transform $\mathcal{F}$ by
\begin{equation}
  \partial \mathcal{F}^{-\text{T}} {v}
  = ( \mathcal{F} \mathcal{Q}^{-1} {v}^\ast )^\ast
  .
\end{equation}
It is apparent that a VJP can be computed by the application of quadrature weights to a complex conjugated cotangent vector, followed by a forward spherical harmonic transform and subsequent conjugation.

For completeness, the adjoint of the inverse transform is closely related to the VJP and can be represented as
\begin{equation}
  \mathcal{F}^{-\text{H}} = \mathcal{F} \mathcal{Q}^{-1}
  .
\end{equation}

\begin{figure}%[!b]
  \centering
  \includegraphics[width=\linewidth]{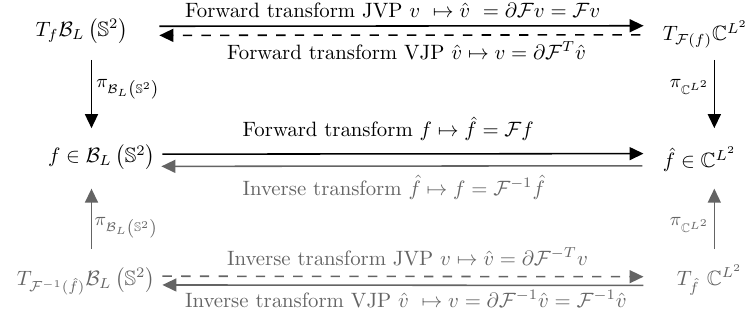}
  \caption{Commutative diagram describing JVPs and VJPs for both the forward (top loop in black) and inverse (bottom loop in grey) spherical transforms. These relations hold for both scalar and spin spherical harmonic transforms and also for Wigner transforms.  Note that in this diagram $\pi_{\mathcal{X}}$ denotes the generalised projection of a function onto $\mathcal{X}$. In each case the tangent (cotangent) space is isomorphic with the target space.}
  \label{fig:custom_gradient_commutative_diagram}
\end{figure}

Again, the above results extend trivially to the inverse Wigner transforms simply by replacing $\mathcal{F}^{-1}$ by $\mathcal{W}^{-1}: \mathbb{C}^{(4L^3-L)/3} \rightarrow \mathcal{B}_L(\text{SO}(3))$ and let $\mathcal{Q}$ represent application of the quadrature weights $q(\alphaup,\betaup,\gammaup)$ on $\text{SO}(3)$. A commutative diagram is provided in Figure \ref{fig:custom_gradient_commutative_diagram} to more clearly summarise our formulation of the JVPs and VJPs associated with the various spherical harmonic transforms.

\subsection{Hybrid automatic and manual differentiable spherical transforms}

In \sectn{\ref{sec:gradients:manual}} we derive analytic expressions for JVP and VJPs of spherical harmonic and Wigner transforms, which can be expressed simply in terms of forward or inverse transforms and application of quadrature weights (with appropriate complex conjugation).  Due to the linear nature of the generalised Fourier transforms considered, we also show that intermediate primal values do not need to be stored for reverse-mode AD, substantially alleviating memory overheads.

In principle AD would seem like the preferred approach since it avoids the complexity and implementation overhead of manual differentiation.  However, in practice the recursive nature of the computation of the Wigner $d$-functions requires a scan operation.  The scan carry requires $\mathcal{O}(L^2)$ memory and the total number of iterations over which one must scan is $\mathcal{O}(L)$.  Hence, in JAX for reverse-mode AD $\mathcal{O}(L^3)$ memory is required, which is prohibitive at high bandlimits.

It would therefore seem that manual differentiation is the preferred approach to avoid significant memory overhead.  However, the simple nature of the explicit expressions for the JVPs and VPJs given in \sectn{\ref{sec:gradients:manual}} hides the subtlety surrounding the details of different spherical samplings (e.g.\ the specific details discussed in \sectn{\ref{sec:algorithms:samplings}}, particularly for \texttt{HEALPix} and MW samplings).

We therefore consider a hybrid approach.  We adopt manual differentiation for the core internal components of spherical transforms related to recursions but adopt AD for surrounding components of the transforms that deal with the subtleties of spherical samplings.  In this manner we avoid the prohibitive memory overhead of a pure AD approach, while also avoiding the subtleties of cumbersome direct implementations of a manual differentiation approach. A similar hybrid gradient approach was recently considered in \citet{li2024differentiable} to mitigate a similar JAX memory issue whilst differentiating through cosmological simulations.

\subsubsection{Forward spin spherical harmonic transforms}

Consider the forward transform $\mathcal{F} = \mathcal{F}_\Theta \mathcal{F}_\Phi : f \mapsto \hat{f}$ in terms of the separate transforms with respect to $\phi$ and $\theta$, respectively $\mathcal{F}_\Phi : f \mapsto \tilde{f}$ (\eqn{\ref{eq:disc_sh_fft}}) and $\mathcal{F}_\Theta : \tilde{f} \mapsto \hat{f}$ (\eqn{\ref{eq:disc_sh_expanded}}).
Since the computation of $\mathcal{F}_\Theta$ requires evaluation of the Wigner $d$-functions by recursion, we compute gradients of this transform by explicit manual gradients, avoiding the memory overheads of AD.  Remaining computations, which effectively capture the computation of $\mathcal{F}_\Phi$ through multiple weighted 1D FFTs, are computed by AD, avoiding the subtleties of specific sampling schemes.  In the following we therefore focus on the computation of manual gradients for $\mathcal{F}_\Theta$.

The elements of the Jacobian of the forward transform $\mathcal{F}_\Theta$ read
\begin{equation}
  \Bigl (\partial \mathcal{F}_\Theta(\tilde{f}) \Bigr)_{\ell}^{\theta}
  = \frac{\partial \hat{f}_{\ell m}}{\partial \tilde{f}_m(\theta)}
  = (-1)^s \sqrt{\frac{2\ell+1}{4\pi}}
  q_\Theta(\theta) \:
  d^{\ell}_{m, -s}(\theta)
  .
\end{equation}
Again, notice that the Jacobian is independent of the signal $\tilde{f}$, i.e.\ $\partial\mathcal{F}_\theta(\tilde{f}) = \partial \mathcal{F}_\Theta$.
The JVP is then given by
\begin{equation}
  \Bigl(\partial \mathcal{F}_\Theta \tilde{v} \Bigr)_{\ell}
  =
  \sum_{\theta} \:
  (-1)^s \sqrt{\frac{2\ell+1}{4\pi}} \:
  q_\Theta(\theta) \:
  \tilde{v}_m(\theta) \:
  d^{\ell}_{m, -s}(\theta)
  ,
\end{equation}
which, as expected due to linearity, is simply given by the transform itself, i.e.\
\begin{equation}
  \partial \mathcal{F}_\Theta = \mathcal{F}_\Theta
  .
\end{equation}
The VJP is given by
\begin{equation}
  \Bigl(\partial \mathcal{F}_\Theta^\text{T} \hat{v} \Bigr)_{\theta}
  = \sum_{\ell} \:
  (-1)^s \sqrt{\frac{2\ell+1}{4\pi}}
  q_\Theta(\theta) \:
  \hat{v}_{\ell m} \:
  d^{\ell}_{m, -s}(\theta)
  ,
\end{equation}
which may be expressed by
\begin{equation}
  \partial \mathcal{F}_\Theta^\text{T} = \mathcal{Q}_\Theta \mathcal{F}_\Theta^{-1} = \mathcal{F}_\Theta^\text{H}
  ,
\end{equation}
where $\mathcal{F}_\Theta^{-1}$ is given by \eqn{\ref{eq:disc_ish_expanded}}, $\mathcal{Q}_\Theta$ is the diagonal operator corresponding to multiplication by the quadrature weights $q_\Theta(\theta)$, and $\mathcal{F}_\Theta^\text{H}$ denotes the corresponding adjoint operator.  Note the similarity to \eqn{\ref{eq:vjp_forward_sht}} and \eqn{\ref{eq:adjoint_forward_sht}}, although here complex conjugations are not required due to the reality of the transform (which follows from the reality of the Wigner $d$-functions).

\subsubsection{Inverse spin spherical harmonic transforms}

Consider the inverse transform $\mathcal{F}^{-1} = \mathcal{F}_\Phi^{-1} \mathcal{F}_\Theta^{-1}  : \hat{f} \mapsto f$ in terms of the separate transforms with respect to $\theta$ and $\phi$, respectively $\mathcal{F}^{-1}_\Theta : \hat{f} \mapsto \tilde{f}$ (\eqn{\ref{eq:disc_ish_expanded}}) and $\mathcal{F}^{-1}_\Phi : \tilde{f} \mapsto f$ (\eqn{\ref{eq:disc_ish_fft}}).  For identical reasons to the forward transform, we compute gradients of $\mathcal{F}^{-1}_\Theta$ manually and gradients of $\mathcal{F}^{-1}_\Phi$ by AD. In the following we therefore focus on the computation of the manual gradients of $\mathcal{F}^{-1}_\Theta$.

The elements of the Jacobian of the inverse transform $\mathcal{F}^{-1}_\Theta$ read
\begin{equation}
  \Bigl (\partial \mathcal{F}^{-1}_\Theta(\tilde{f}) \Bigr)^{\ell}_{\theta}
  = \frac{\partial \tilde{f}_m(\theta)}{\partial \hat{f}_{\ell m}}
  = (-1)^s \sqrt{\frac{2\ell+1}{4\pi}}
  d^{\ell}_{m, -s}(\theta)
  ,
\end{equation}
which is independent of the signal itself, i.e.\ $\partial\mathcal{F}^{-1}_\theta(\tilde{f}) = \partial \mathcal{F}^{-1}_\Theta$.
The JVP is then given by
\begin{equation}
  \Bigl(\partial \mathcal{F}^{-1}_\Theta \hat{v} \Bigr)_{\theta}
  =
  \sum_{\ell} \:
  (-1)^s \sqrt{\frac{2\ell+1}{4\pi}} \:
  \hat{v}_{\ell m} \:
  d^{\ell}_{m, -s}(\theta)
  ,
\end{equation}
which, as expected, is simply given by the transform itself, i.e.
\begin{equation}
  \partial \mathcal{F}^{-1}_\Theta = \mathcal{F}^{-1}_\Theta .
\end{equation}
The VJP is given by
\begin{equation}
  \Bigl(\partial \mathcal{F}^{-\text{T}}_\Theta \tilde{v} \Bigr)_{\ell}
  =
  \sum_{\theta} \:
  (-1)^s \sqrt{\frac{2\ell+1}{4\pi}} \:
  \tilde{v}_{m}(\theta) \:
  d^{\ell}_{m, -s}(\theta)
  ,
\end{equation}
which may be expressed by
\begin{equation}
  \partial \mathcal{F}^{-\text{T}}_\Theta = \mathcal{F}_\Theta \mathcal{Q}_\Theta^{-1} = \mathcal{F}_\Theta^{-\text{H}} ,
\end{equation}
where recall $\mathcal{F}_\Theta$ is given by \eqn{\ref{eq:disc_sh_expanded}} and we have adopted the notation $\mathcal{F}^{-\text{H}} = (\mathcal{F}^{-1})^\text{H}$.

\subsubsection{Wigner transforms}

The extension to compute gradients for Wigner transforms follows straightforwardly.  As discussed in \sectn{\ref{sec:background:wigner_transform}}, Wigner transforms can be computed via a series of spherical harmonic transforms of varying spin, combined with a Fourier transform for $\gamma$.  Following the hybrid differentiable approach discussed directly above, we compute explicit manual gradients for transforms with respect to $\beta$ and adopt AD for transforms with respect to both $\alpha$ and $\gamma$.  This yields the best of both approaches, avoiding memory overheads associated with AD of the recursions of Wigner $d$-functions, while also avoiding the computational complexities associated with manual differentiation for different sampling schemes.

\section{Software and evaluation} \label{sec:software}

A key aim of this work is the development of a modern open-source Python library for the accelerated computation of generalised Fourier transforms on the sphere and rotation group and their derivatives.  We target deployment on one or more modern hardware accelerators, such as GPUs and TPUs.
By leveraging the parallelised computation of Wigner functions presented in \sectn{\ref{sec:recursions}} and the general algorithmic framework presented in \sectn{\ref{sec:algorithms}}, we ensure our algorithms are able to well-utilise hardware accelerators.  Moreover, our algorithmic framework ensures we are largely agnostic to the choice of approach used to sample the sphere.
Critically, to support the increasing interest surrounding differentiable programming, gradients must be able to be computed efficiently.  By paying careful attention to the computation of gradients through a hybrid automatic and manual differentiation approach as described in \sectn{\ref{sec:gradients}}, we avoid the memory overhead of full automatic differentiation while also avoiding the complexities of full manual differentiation.
Finally, as with all good software packages, our library must be simple to use and easy to pick up.

A variety of differentiable programming and machine learning ecosystems exist in Python that could be considered for the implementation of our software, \textit{e.g.}\ \texttt{TensorFlow} and \texttt{PyTorch}, both of which exhibit a mature collection of Python APIs.  However, over time these libraries have become somewhat rigid due to the nature of heavily templated object oriented software focused on machine learning and hence the development of other bespoke functionality is hindered by, for example, significant boilerplate code. Consequently, we choose to develop our software in \texttt{JAX}, which is a functional differentiable programming Python framework recently developed by Google \cite{jax:2018:github}. \texttt{JAX} is particularly well-suited to scientific software development, where flexibility and rapid development are highly desirable.

With the above in mind we introduce \texttt{S2FFT}\footnote{\url{https://github.com/astro-informatics/s2fft}} {\href{https://github.com/astro-informatics/s2fft}{\faGithub}}, which is an open-source \texttt{JAX} package implementing the parallelised Wigner recursions presented in \sectn{\ref{sec:recursions}} and generalised Fourier transforms on the sphere and rotation group presented in \sectn{\ref{sec:algorithms}} and their derivatives as presented in \sectn{\ref{sec:gradients}}.  The software code has been developed following software engineering best practices, exhibits an extensive test suite and is well documented.
In the remainder of this section we briefly discuss the spherical sampling schemes supported and JAX hardware acceleration, before presenting benchmarking results in terms of precision and computational speed.

\subsection{Sampling agnostic transforms}

As discussed, our software is largely agnostic to the choice of spherical sampling considered.  We simply require an isolatitudinal sampling, which most popular spherical samplings follow due to the considerable computational savings that it affords through a separation of variables, and corresponding quadrature weights.
As a starting point Driscoll \& Healy \cite[][]{driscoll:1994}, McEwen \& Wiaux (both MW and MWSS) \cite{mcewen:fssht,daducci:ssdmri,ocampo:disco}, Gauss-Lengendre \cite[\textit{e.g.}][]{shukowsky:1986,mcewen:fssht}, and \texttt{HEALPix} \cite{gorski:2005} samplings are supported.
To simplify further open-sourced contributions, our code is designed with integration of future sampling schemes in mind, should the need arise. It is worth noting that once sample locations and associated quadrature weights are defined, all downstream functionality should work seamlessly.

\subsection{Hardware acceleration}

By design \texttt{JAX} functions can straightforwardly be deployed on hardware accelerators, \textit{e.g.}\ GPU and TPUs. In fact, the ease with which existing Python codebases, particularly those which make frequent use of \texttt{numpy}, may be abstracted for GPU deployment is one of the primary advantages of \texttt{JAX}.
In addition, we integrate the single program multiple data (SPMD) functionality accessible in \texttt{JAX} by the \texttt{\lstinline{pmap}} API. Specifically, for core transforms we provide an \texttt{spmd} flag that when set distributes expensive operations evenly across available devices.  As we show when benchmarking (\sectn{\ref{sec:software:benchmarking}}) we recover approximately linear acceleration in the number of devices considered.  This distribution is straightforwardly engineered to balance the compute load across devices (see \sectn{\ref{sec:algorithms}}).  Of course, for very low bandlimits the communication overhead incurred during distribution is significant. However, for even moderate to low bandlimits this communication is sub-dominant and significant computational savings are realised. Currently, the codebase provides tested and documented support for intra-node distribution (\textit{i.e.} multiple GPUs on a single node), with inter-node distribution to be added in future (\textit{i.e.} multiple nodes each with multiple GPUs).

\subsection{Benchmarking} \label{sec:software:benchmarking}

We present comprehensive benchmarking of the core transforms provided by our \texttt{S2FFT} software.
First, we quantify the precision of our transforms by computing the round-trip error and find errors to be at the level of numerical precision for the sampling theorems considered.
Second, we compute the average wall-clock computational speed of transforms, in both single and multiple GPU scenarios, for both our on-the-fly algorithms (computing Wigner $d$-functions on-the-fly) and precompute algorithm (computing and storing Wigner $d$-functions \textit{a priori}).  We compare computational time to the \texttt{SSHT}\footnote{\url{https://github.com/astro-informatics/ssht}} \cite{mcewen:fssht} and \texttt{SO3}\footnote{\url{https://github.com/astro-informatics/so3}} \cite{mcewen:so3} codes that compute spherical harmonic and Wigner transforms, respectively, for which underlying algorithms are implemented in C.  An extensive comparison against a wide variety of existing spherical transform software codes is beyond the scope of this study (furthermore, while there are numerous spherical harmonic transforms codes there are relatively few codes capable of also computing Wigner transforms).  In any case, comparison against other codes can be inferred from contemporary papers \citep[see \emph{e.g.}][]{mcewen:fssht, reinecke:2013:libsharp}.
In addition, all algorithms implemented in \texttt{S2FFT} are validated through extensive unit and regression tests.

\subsubsection{Precision}
Natively, our software supports both 32-bit and 64-bit floating point precision, however due to the unstable nature of three term recursions \cite{gautschi:1967} as outlined in \sectn{\ref{sec:recursions}} it is strongly recommended to adopt 64-bit precision. To mimic the most likely use-case, in our benchmarking we default to 64-bit precision, which is provided by \texttt{JAX} as an extended data type.
Both Driscoll \& Healy sampling \cite{driscoll:1994} and both McEwen \& Wiaux sampling schemes \cite{mcewen:fssht} afford sampling theorems on the sphere and therefore transforms based on such samplings are theoretically exact to machine precision.
%For 64-bit floating point precision this corresponds to errors of $\mathcal{O}(10^{-16})$. 

The protocol for assessing accuracy in practice is to first generate a random set of bandlimited generalised Fourier coefficients $\hat{f}$ from which a round-trip error metric $\mathbb{E}[\| \hat{f} \,-\, \mathcal{F}\mathcal{F}^{-1}\hat{f}\|_2]$ is evaluated, where the expectation is computed across 10 randomised experiments. Note that for a Wigner transform one clearly should replace $\mathcal{F}$ with $\mathcal{W}$. When considering Wigner transforms we enforced an azimuthal bandlimit of $N=5$, corresponding to $9$ equally spaced directions within each tangent plane,  which is a typical use case.
The results of these experiments are presented in \tbl{\ref{tab:sh_error_results}} and \tbl{\ref{tab:wig_error_results}} and also summarised in the bottom panels of \fig{\ref{fig:sh_benchmarking}} and \fig{\ref{fig:wt_benchmarking}}. In all cases our transforms are exact to 64-bit machine precision.
Note also that the gradient algorithms derived herein, and implemented in \texttt{S2FFT}, are validated against finite differences using the \texttt{check\_grads} functionality provided within \texttt{JAX}.

In the era of high precision science, access to theoretically exact transforms is of great importance \cite{bicep2:I:2014, planck2016-l01, pires:2020:euclid, amaro:2017:laser}. The best way to perform high precision science is to avoid introducing numerical errors in the first place.  For other applications one may wish to adopt 32-bit floating point precision, which can further accelerate compute and will half memory overhead. Such applications may include geometric machine learning on the sphere \cite{bronstein:2021:geometric, cohen:2018:spherical, kondor:2018:clebsch,  cobb:efficient_generalized_s2cnn}, where 32-bit precision (or lower) is often adopted.
% When adopting 32-bit precision one may expect to recover commensurate errors, \textit{i.e.}\ errors beginning at approximately $\mathcal{O}(10^{-8})$ rather than $\mathcal{O}(10^{-16})$.

\begin{table}
   \newcolumntype{C}[1]{>{\centering\let\newline\\\arraybackslash\hspace{0pt}}m{#1}}
   \centering % to have the caption near the table
   \begin{tabular}{C{0.05\textwidth}  C{0.12\textwidth} C{0.12\textwidth} C{0.12\textwidth} C{0.12\textwidth} C{0.12\textwidth} C{0.12\textwidth}} \toprule
      \multicolumn{7}{c}{Spherical harmonic transform round-trip error}                                              \\ \midrule
         & \multicolumn{3}{c}{On-the-fly transform} & \multicolumn{3}{c}{Precompute transform}                       \\
      \cmidrule(lr){5-7}
      \cmidrule(lr){2-4}

      L  & MW                                       & MWSS                                     & DH & MW & MWSS & DH \\ \midrule

      $8$
         & $3.6 \times10^{-16}$
         & $1.7 \times10^{-16}$
         & $5.1 \times10^{-16}$
         & $4.6 \times10^{-16}$
         & $4.3 \times10^{-16}$
         & $4.3 \times10^{-16}$                                                                                      \\

      $16$
         & $3.7 \times10^{-16}$
         & $2.7 \times10^{-16}$
         & $6.3 \times10^{-16}$
         & $5.4 \times10^{-16}$
         & $4.5 \times10^{-16}$
         & $4.5 \times10^{-16}$                                                                                      \\

      $32$
         & $7.5 \times10^{-16}$
         & $6.3 \times10^{-16}$
         & $3.5 \times10^{-16}$
         & $7.3 \times10^{-16}$
         & $7.2 \times10^{-16}$
         & $7.2 \times10^{-16}$                                                                                      \\

      $64$
         & $1.2 \times10^{-15}$
         & $1.1 \times10^{-15}$
         & $6.7 \times10^{-16}$
         & $1.2 \times10^{-15}$
         & $1.2 \times10^{-15}$
         & $1.2 \times10^{-15}$                                                                                      \\

      $128$
         & $2.3 \times10^{-15}$
         & $2.3 \times10^{-15}$
         & $1.3 \times10^{-15}$
         & $2.5 \times10^{-15}$
         & $2.4 \times10^{-15}$
         & $2.4 \times10^{-15}$                                                                                      \\

      $256$
         & $4.7 \times10^{-15}$
         & $5.0 \times10^{-15}$
         & $2.6 \times10^{-15}$
         & $4.7 \times10^{-15}$
         & $4.7 \times10^{-15}$
         & $4.7 \times10^{-15}$                                                                                      \\

      $512$
         & $1.0 \times10^{-14}$
         & $9.8 \times10^{-15}$
         & $4.6 \times10^{-15}$
         & $9.8 \times10^{-15}$
         & $9.7 \times10^{-15}$
         & $8.9 \times10^{-15}$                                                                                      \\

      $1024$
         & $1.9 \times10^{-14}$
         & $1.9 \times10^{-14}$
         & $9.3 \times10^{-15}$
         & $1.7 \times10^{-14}$
         & $1.5 \times10^{-14}$
         & $1.2 \times10^{-14}$                                                                                      \\

      $2048$
         & $3.7 \times10^{-14}$
         & $3.8 \times10^{-14}$
         & $1.9 \times10^{-14}$
         &
      -- & --                                       & --                                                             \\

      $4096$
         & $7.5 \times10^{-14}$
         & $7.7 \times10^{-14}$
         & $3.8 \times10^{-14}$
         &
      -- & --                                       & --                                                             \\

      $8192$
         & $1.5 \times10^{-13}$
         & $1.5 \times10^{-13}$
         & $8.3 \times10^{-14}$
         &
      -- & --                                       & --
      \\ \midrule
   \end{tabular}
   \caption{Round-trip error for scalar spherical harmonic transforms, \textit{i.e.}\ $\mathbb{E}[\| \hat{f} - \mathcal{F}\mathcal{F}^{-1}\hat{f} \|_2]$. Random bandlimited functions are generated, which are passed sequentially through an inverse and then forward spherical harmonic transform.  Errors are averaged across 10 realisations.  We consider samplings of the sphere that exhibit a sampling theorem with exact spherical harmonic transforms.  As expected, accuracy is therefore at the level of machine precision.} \label{tab:sh_error_results}
\end{table}

\begin{table}
  \newcolumntype{C}[1]{>{\centering\let\newline\\\arraybackslash\hspace{0pt}}m{#1}}
  \centering % to have the caption near the table
  \begin{tabular}{C{0.05\textwidth}  C{0.12\textwidth} C{0.12\textwidth} C{0.12\textwidth} C{0.12\textwidth}} \toprule
    \multicolumn{5}{c}{Wigner transform round-trip error}                                               \\ \midrule
      & \multicolumn{2}{c}{On-the-fly transform} & \multicolumn{2}{c}{Precompute transform}             \\
    \cmidrule(lr){4-5}
    \cmidrule(lr){2-3}

    L & MW                                       & MWSS                                     & MW & MWSS \\ \midrule

    $8$
      & $1.6 \times10^{-15}$
      & $1.3 \times10^{-15}$
      & $1.3 \times10^{-15}$
      & $1.2 \times10^{-15}$                                                                            \\

    $16$
      & $1.2 \times10^{-15}$
      & $1.0 \times10^{-15}$
      & $1.1 \times10^{-15}$
      & $1.1 \times10^{-15}$                                                                            \\

    $32$
      & $1.3 \times10^{-15}$
      & $1.2 \times10^{-15}$
      & $1.5 \times10^{-15}$
      & $1.2 \times10^{-15}$                                                                            \\

    $64$
      & $1.5 \times10^{-15}$
      & $1.4 \times10^{-15}$
      & $1.8 \times10^{-15}$
      & $1.4 \times10^{-15}$                                                                            \\

    $128$
      & $2.2 \times10^{-15}$
      & $2.0 \times10^{-15}$
      & $3.6 \times10^{-15}$
      & $2.0 \times10^{-15}$                                                                            \\

    $256$
      & $2.9 \times10^{-15}$
      & $2.8 \times10^{-15}$
      & $4.2 \times10^{-15}$
      & $2.9 \times10^{-15}$                                                                            \\

    $512$
      & $4.2 \times10^{-15}$
      & $4.3 \times10^{-15}$
      & $6.4 \times10^{-15}$
      & $4.2 \times10^{-15}$                                                                            \\

    $1024$
      & $5.7 \times10^{-15}$
      & $5.9 \times10^{-15}$
      & --                                       & --                                                   \\

    $2048$
      & $8.1 \times10^{-15}$
      & $8.2 \times10^{-15}$
      & --                                       & --                                                   \\

    $4096$
      & $1.2 \times10^{-14}$
      & $1.3 \times10^{-14}$
      & --                                       & --                                                   \\ \bottomrule
  \end{tabular}
  \caption{Round-trip error for Wigner transforms with azimuthal bandlimit $N=5$, \textit{i.e.}\ $\mathbb{E}[\| \hat{f} - \mathcal{W}\mathcal{W}^{-1}\hat{f} \|_2]$.  Random bandlimited functions are generated, which are passed sequentially through an inverse and then forward Wigner transform.  Errors are averaged across 10 realisations.  We consider samplings of the rotation group that exhibit a sampling theorem with exact Wigner transforms.  As expected, accuracy is therefore at the level of machine precision.} \label{tab:wig_error_results}
\end{table}

\begin{figure}
   \begin{subfigure}{.33\textwidth}
      \centering
      \includegraphics[width=\linewidth, height=11cm,keepaspectratio, trim={0cm 0cm 0.2cm 0cm}, clip]{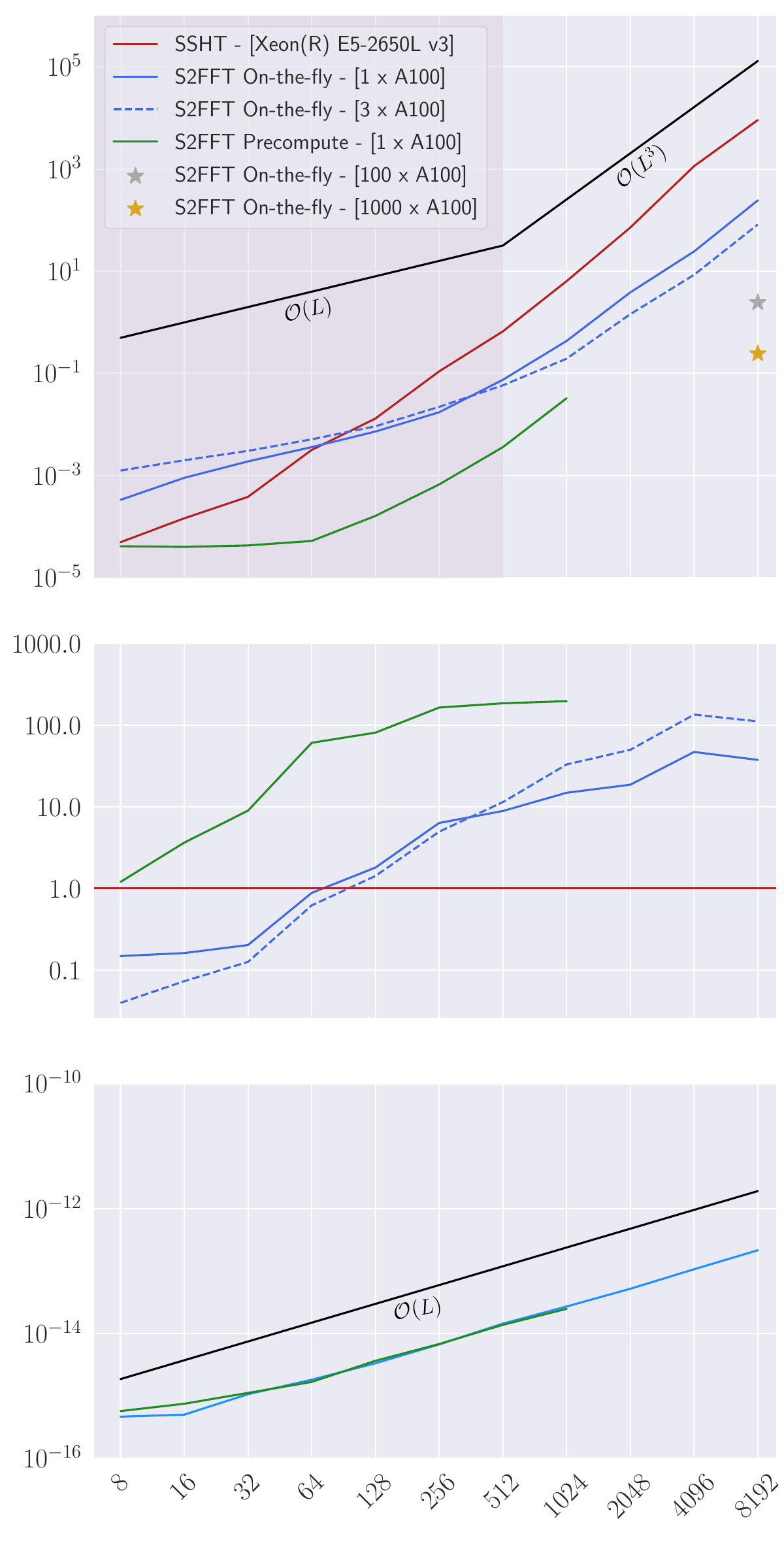}
      \put(-88,0){\rotatebox{0}{\footnotesize Bandlimit $L$}}
      \caption{MW sampling}
      % \label{fig:sub1}
   \end{subfigure}%
   \hspace{-0.2cm}
   \begin{subfigure}{.33\textwidth}
      \centering
      \includegraphics[width=\linewidth, height=11cm,keepaspectratio, trim={2cm 0cm 0.2cm 0cm}, clip]{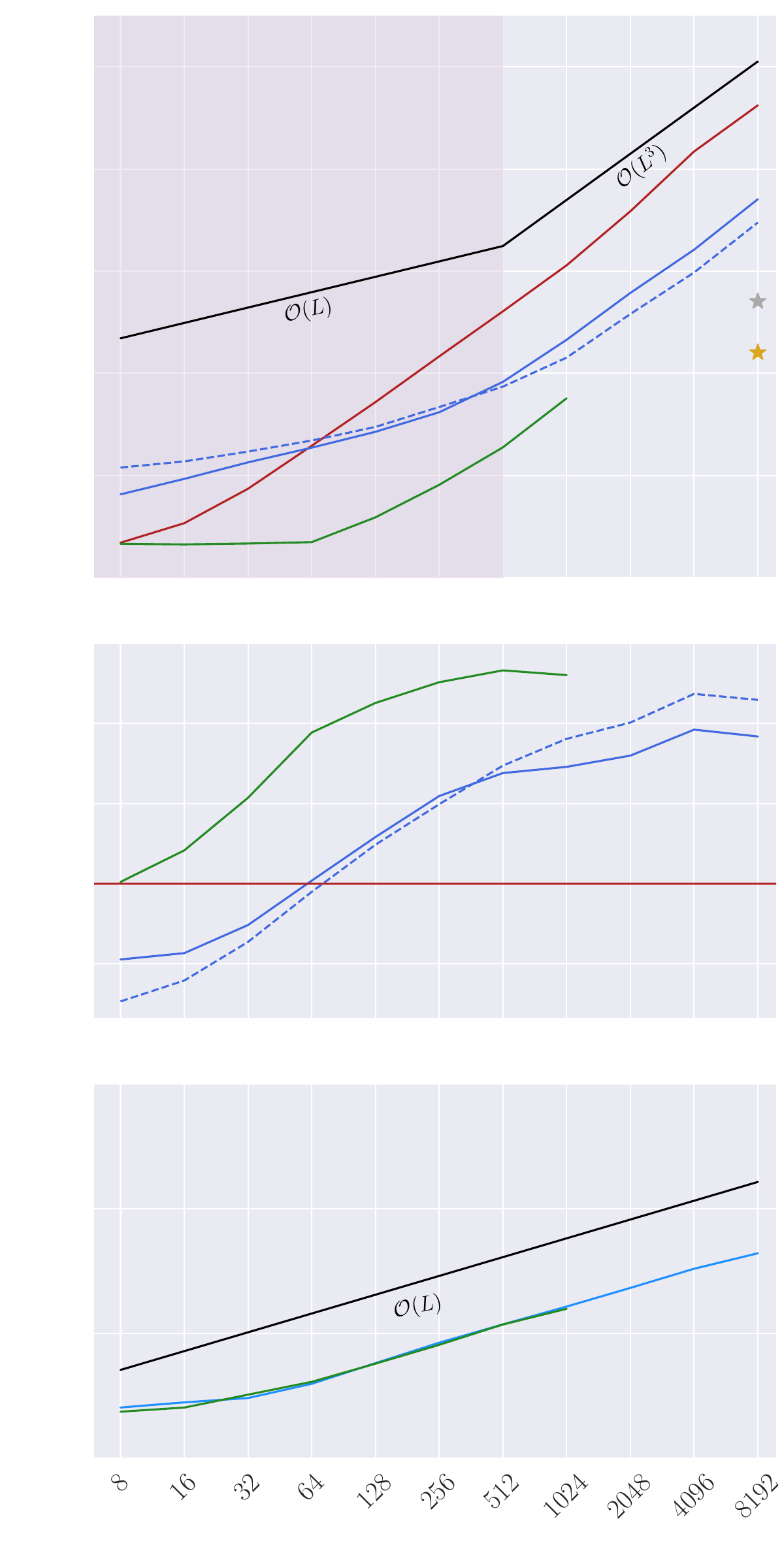}
      \put(-88,0){\rotatebox{0}{\footnotesize Bandlimit $L$}}
      \caption{MWSS sampling}
      % \label{fig:sub2}
   \end{subfigure}
   \hspace{-0.55cm}
   \begin{subfigure}{.33\textwidth}
      \centering
      \includegraphics[width=\linewidth, height=11cm,keepaspectratio, trim={2cm 0cm 0.2cm 0cm}, clip]{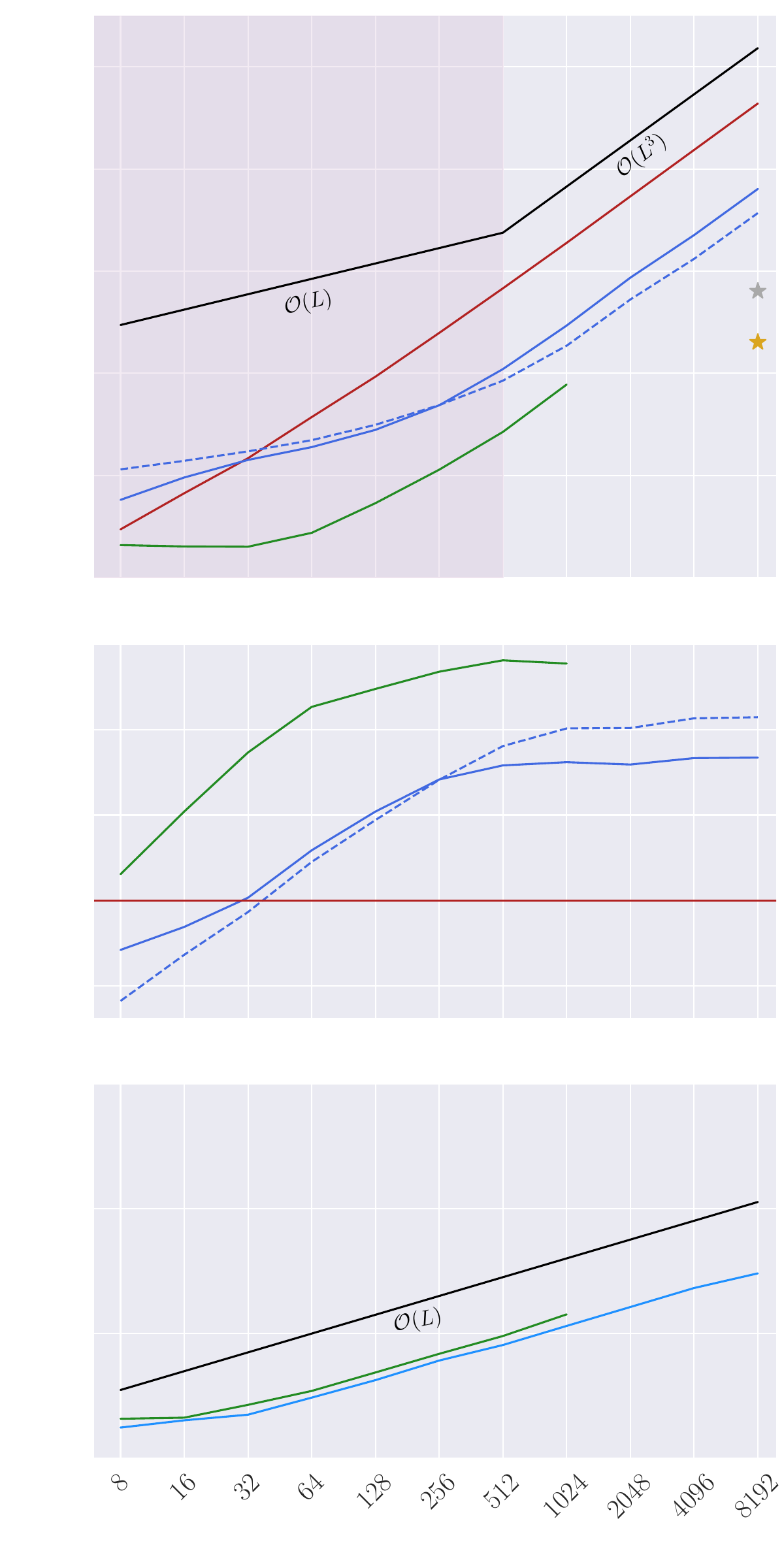}
      \put(-88,0){\rotatebox{0}{\footnotesize Bandlimit $L$}}
      \caption{DH sampling}
      % \label{fig:sub3}
   \end{subfigure}
   % \put(-90,15){\rotatebox{0}{\footnotesize Bandlimit $L$}}
   % \put(-240,15){\rotatebox{0}{\footnotesize Bandlimit $L$}}
   % \put(-390,15){\rotatebox{0}{\footnotesize Bandlimit $L$}}
   \put(-456,66){\rotatebox{90}{\footnotesize Error}}
   \put(-456,140){\rotatebox{90}{\footnotesize Acceleration}}
   \put(-456,230){\rotatebox{90}{\footnotesize Compute Time (s)}}
   \caption{Numerical benchmarking results for the spherical harmonic transforms implemented in \texttt{S2FFT} compared to the C implementation of \texttt{SSHT}. We consider samplings of the sphere that afford a sampling theorem and hence exact spherical harmonic transforms.  \texttt{S2FFT} provides multiple operational modes: a precompute approach with faster compute at $\mathcal{O}(L^3)$ memory overhead and an on-the-fly transform with negligible memory overhead (see \sectn{\ref{sec:algorithms}}).
      \textbf{Top:} Average compute time required for a spherical harmonic transform for each operational modality.
      \textbf{Middle:} Relative acceleration of a given modality with respect to their \texttt{SSHT} counterpart.
      \textbf{Bottom:} Averaged round-trip numerical error. In each case we recover machine precision. We also empirically observe $\mathcal{O}(L)$ error scaling (in line with previous work \cite{mcewen:fssht}). The silver and gold stars represent the performance one may expect to realise should one distribute our on-the-fly algorithms across 100 and 1000 GPUs respectively.}
   \label{fig:sh_benchmarking}
\end{figure}
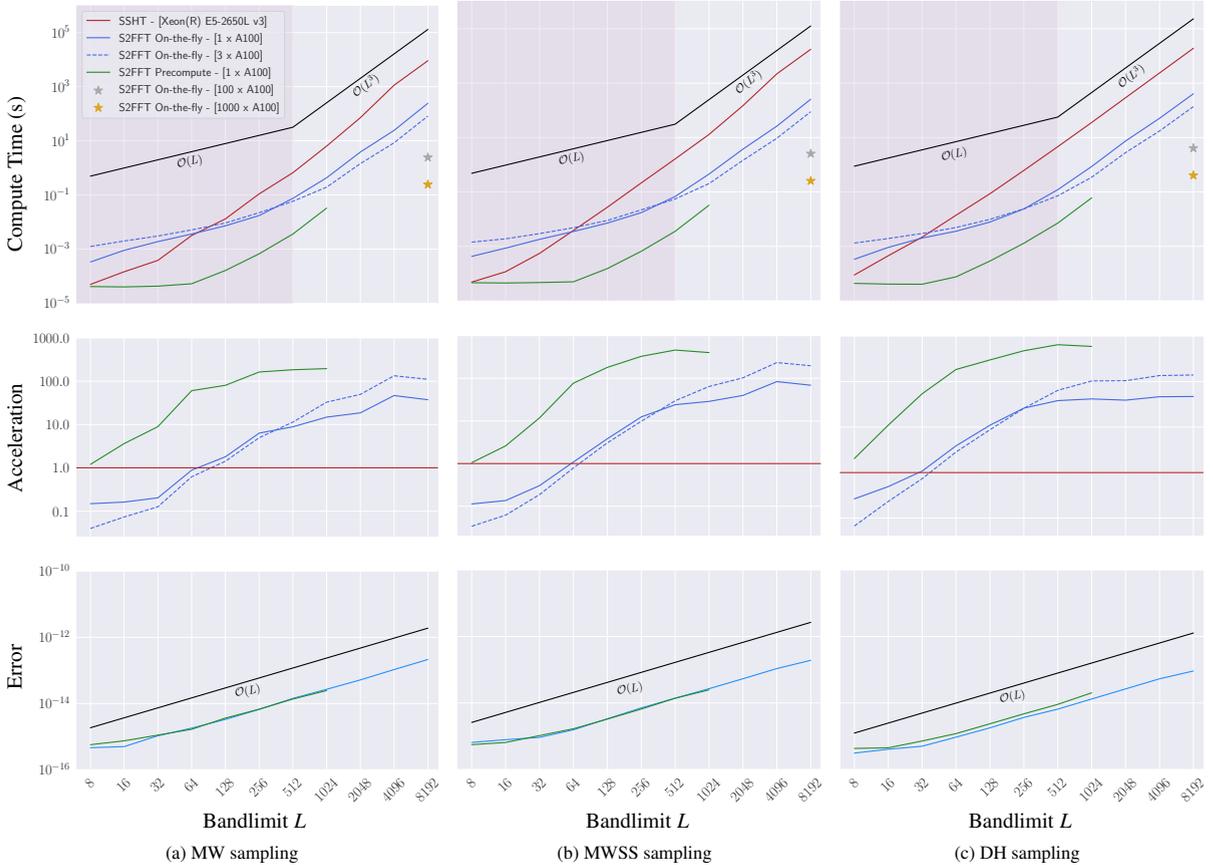

\subsubsection{Computational speed}

To benchmark computational speed our transforms are run over both a single and three NVIDIA A100 GPUs, each with 40GB on board memory.  For comparison the existing \texttt{SSHT} and \texttt{SO3} C codes are run on a Xeon(R) E5-2650L v3 multithreaded CPU with a dedicated machine.

A summary of computational times are presented in \fig{\ref{fig:sh_benchmarking}} for a complex scalar spherical harmonic transform, with raw numerical results tabulated in \tbl{\ref{tab:sh_raw_numerical_results}}.  We provide results for both our on-the-fly and precompute transforms; for the latter we also quote memory requirements for the precompute.  Our software handles functions of arbitrary spin, and therefore performance does not degrade with spin number, in contrast to other approaches.  Results for our implementation of the Wigner transform are presented in \fig{\ref{fig:wt_benchmarking}}, with raw numerical results provided in \tbl{\ref{tab:wig_results}}.

Unsurprisingly, for very low $L$ we find that when running on GPUs our transforms are slightly slower than those provided by \texttt{SSHT}, which is primarily due to scheduling and memory communication overheads that can throttle execution on GPUs. However, for $L\geq 64$ our transforms begin to achieve superior speed, reaching as much as a 234-fold and 459-fold acceleration for the on-the-fly and precompute transforms respectively.
When running across multiple GPUs we see degraded performance at very low $L$, due to the additional inter-device communication overhead, however we asymptotically recover a further acceleration by a factor of the number of available devices.  In practice for high bandlimits we achieve very close to optimal linear scaling with increasing number of GPUs due to the highly parallelised and balanced nature of our algorithms.  For example, for spherical harmonic transforms at $L=8192$ when moving from one to three GPUs we achieve an additional acceleration of 2.96, 2.97, 2.87 for DH, MW and MWSS sampling respectively.  For the Wigner transform at $L=4096$ and $N=5$ we achieve an additional acceleration of 2.99.  With this in mind, provided access to sufficiently many GPUs our transforms exhibit an effective linear time complexity.

\begin{table}%[!h]
  \newcolumntype{C}[1]{>{\centering\let\newline\\\arraybackslash\hspace{0pt}}m{#1}}
  \centering % to have the caption near the table
  \begin{tabular}{C{0.05\textwidth} C{0.1\textwidth} C{0.11\textwidth} C{0.1\textwidth} C{0.1\textwidth} C{0.1\textwidth} C{0.11\textwidth} C{0.1\textwidth}} \toprule
    \multicolumn{8}{c}{Spherical harmonic transform computational time (ms)}                                                       \\ \midrule
           &                                 & \multicolumn{4}{c}{On-the-fly transform} & \multicolumn{2}{c}{Precompute transform} \\
    \cmidrule(lr){7-8}
    \cmidrule(lr){3-6}
    $L$
           & Time \texttt{SSHT}
           & Time \texttt{S2FFT}
           & Speed-Up 1 $\times$ GPU
    %     & 3 $\times$ GPU
           & Speed-Up 3 $\times$ GPU
           & 3 $\times$ GPU Ratio
           & Time \texttt{S2FFT}
           & Speed-Up 1 $\times$ GPU \\ \midrule
       %     & Memory (MB) \\ \midrule

    $8$    & \hspace{5pt}$8.8 \times10^{-2}$
           & \hspace{5pt}$3.3 \times10^{-1}$
           & 0.26
    %     & $1.3 \times10^{0}$              
           & 0.07
           & 0.27
           & \hspace{5pt}$4.4 \times10^{-2}$ 
           & 2.03 \\
       %     & \hspace{5pt}$1.5 \times10^{-2}$ \\

    $16$   & \hspace{5pt}$4.5 \times10^{-1}$
           & \hspace{5pt}$9.2 \times10^{-1}$
           & 0.49
    %     & $1.9 \times10^{0}$              
           & 0.23
           & 0.47
           & \hspace{5pt}$4.1 \times10^{-2}$ 
           & 11.0 \\
       %     & \hspace{5pt}$1.3 \times10^{-1}$ \\

    $32$   & $2.2 \times10^{0}$
           & $2.0 \times10^{0}$
           & 1.08
    %     & $3.0 \times10^{0}$              
           & 0.73
           & 0.68
           & \hspace{5pt}$4.0 \times10^{-2}$ 
           & 53.9\\
       %     & $1.0 \times10^{0}$ \\

    $64$   & $1.4 \times10^{1}$
           & $3.6 \times10^{0}$
           & 3.87
    %     & $4.9 \times10^{0}$              
           & 2.83
           & 0.73
           & \hspace{5pt}$7.6 \times10^{-2}$ 
           & 184\\
       %     & $8.3 \times10^{0}$ \\

    $128$  & $8.6 \times10^{1}$
           & $7.8 \times10^{0}$
           & 11.0
    %     & $9.8 \times10^{0}$              
           & 8.74
           & 0.80
           & \hspace{5pt}$2.9 \times10^{-1}$ 
           & 299\\
       %     & $6.7 \times10^{1}$ \\

    $256$  & $6.2 \times10^{2}$
           & $2.4 \times10^{1}$
           & 26.1
    %     & $2.4 \times10^{1}$              
           & 25.8
           & 0.99
           & $1.3 \times10^{0}$              
           & 475\\
       %     & $5.4 \times10^{2}$ \\

    $512$  & $4.6 \times10^{3}$
           & $1.2 \times10^{2}$
           & 38.1
    %     & $7.2 \times10^{1}$              
           & 64.1
           & 1.68
           & $7.1 \times10^{0}$              
           & 646\\
       %     & $4.3 \times10^{3}$ \\

    $1024$ & $3.6 \times10^{4}$
           & $8.6 \times10^{2}$
           & 41.5
    %     & $3.5 \times10^{2}$              
           & 103
           & 2.48
           & $6.0 \times10^{1}$              
           & 592\\
       %     & $3.4 \times10^{4}$ \\

    $2048$ & $2.9 \times10^{5}$
           & $7.5 \times10^{3}$
           & 39.0
    %     & $2.8 \times10^{3}$              
           & 104
           & 2.67
           & --
           & -- \\

    $4096$ & $2.4 \times10^{6}$
           & $5.1 \times10^{4}$
           & 46.4
    %     & $1.7 \times10^{4}$              
           & 135
           & 2.91
           & --
           & -- \\

    $8192$ & $1.9 \times10^{7}$
           & $4.1 \times10^{5}$
           & 47.0
    %     & $1.4 \times10^{5}$              
           & 139
           & 2.96
           & --
           & -- \\ \midrule

    $8$    & \hspace{5pt}$5.0 \times10^{-2}$
           & \hspace{5pt}$3.3 \times10^{-1}$
           & 0.15
    %     & $1.3 \times10^{0}$              
           & 0.04
           & 0.27
           & \hspace{5pt}$4.1 \times10^{-2}$ 
           & 1.20\\
       %     & \hspace{5pt}$7.7 \times10^{-3}$ \\

    $16$   & \hspace{5pt}$1.5 \times10^{-1}$
           & \hspace{5pt}$9.0 \times10^{-1}$
           & 0.16
    %     & $2.0 \times10^{0}$              
           & 0.07
           & 0.44
           & \hspace{5pt}$4.0 \times10^{-2}$ 
           & 3.63\\
       %     & \hspace{5pt}$6.3 \times10^{-2}$ \\

    $32$   & $3.8 \times10^{-1}$
           & $1.9 \times10^{0}$
           & 0.20
    %     & $3.0 \times10^{0}$              
           & 0.13
           & 0.65
           & \hspace{5pt}$4.3 \times10^{-2}$ 
           & 8.94\\
       %     & \hspace{5pt}$5.2 \times10^{-1}$ \\

    $64$   & $3.2 \times10^{0}$
           & $3.6 \times10^{0}$
           & 0.88
    %     & $5.1 \times10^{0}$              
           & 0.62
           & 0.70
           & \hspace{5pt}$5.2 \times10^{-2}$ 
           & 60.5\\
       %     & $4.2 \times10^{0}$ \\

    $128$  & $1.3 \times10^{1}$
           & $7.3 \times10^{0}$
           & 1.80
    %     & $9.2 \times10^{0}$              
           & 1.42
           & 0.79
           & \hspace{5pt}$1.6 \times10^{-1}$ 
           & 80.6\\
       %     & $3.3 \times10^{1}$ \\

    $256$  & $1.1 \times10^{2}$
           & $1.7 \times10^{1}$
           & 6.32
    %     & $2.2 \times10^{1}$              
           & 4.95
           & 0.78
           & \hspace{5pt}$6.7 \times10^{-1}$ 
           & 163\\
       %     & $2.7 \times10^{2}$ \\

    $512$  & $6.6 \times10^{2}$
           & $7.5 \times10^{1}$
           & 8.85
    %     & $5.8 \times10^{1}$              
           & 11.4
           & 1.29
           & $3.6 \times10^{0}$              
           & 184\\
       %     & $2.1 \times10^{3}$ \\

    $1024$ & $6.4 \times10^{3}$
           & $4.3 \times10^{2}$
           & 14.8
    %     & $1.9 \times10^{2}$              
           & 32.9
           & 2.22
           & $3.3 \times10^{1}$              
           & 196\\
       %     & $1.7 \times10^{4}$ \\

    $2048$ & $7.2 \times10^{4}$
           & $3.9 \times10^{3}$
           & 18.6
    %     & $1.4 \times10^{3}$              
           & 49.7
           & 2.67
           & --
           & -- \\

    $4096$ & $1.1 \times10^{6}$
           & $2.4 \times10^{4}$
           & 46.7
    %     & $8.5 \times10^{3}$              
           & 134
           & 2.87
           & --
           & -- \\

    $8192$ & $9.1 \times10^{6}$
           & $2.4 \times10^{5}$
           & 37.4
    %     & $8.2 \times10^{4}$              
           & 111
           & 2.97
           & --
           & -- \\ \midrule

    $8$    & \hspace{5pt}$4.8 \times10^{-2}$
           & \hspace{5pt}$4.3 \times10^{-1}$
           & 0.11
    %     & $1.4 \times10^{0}$              
           & 0.03
           & 0.27
           & \hspace{5pt}$4.6 \times10^{-2}$ 
           & 1.05\\
       %     & \hspace{5pt}$8.6 \times10^{-3}$ \\

    $16$   & \hspace{5pt}$1.2 \times10^{-1}$
           & \hspace{5pt}$8.6 \times10^{-1}$
           & 0.14
    %     & $1.9 \times10^{0}$              
           & 0.06
           & 0.43
           & \hspace{5pt}$4.5 \times10^{-2}$ 
           & 2.60\\
       %     & \hspace{5pt}$6.7 \times10^{-2}$ \\

    $32$   & $5.5 \times10^{-1}$
           & $1.8 \times10^{0}$
           & 0.30
    %     & $2.9 \times10^{0}$              
           & 0.19
           & 0.63
           & \hspace{5pt}$4.7 \times10^{-2}$ 
           & 11.8\\
       %     & \hspace{5pt}$5.3 \times10^{-1}$ \\

    $64$   & $3.8 \times10^{0}$
           & $3.5 \times10^{0}$
           & 1.09
    %     & $4.8 \times10^{0}$              
           & 0.79
           & 0.72
           & \hspace{5pt}$5.0 \times10^{-2}$ 
           & 76.9\\
       %     & $4.2 \times10^{0}$ \\

    $128$  & $2.7 \times10^{1}$
           & $7.2 \times10^{0}$
           & 3.82
    %     & $8.9 \times10^{0}$              
           & 3.06
           & 0.80
           & \hspace{5pt}$1.5 \times10^{-1}$ 
           & 180\\
       %     & $3.4 \times10^{1}$ \\

    $256$  & $2.1 \times10^{2}$
           & $1.7 \times10^{1}$
           & 12.4
    %     & $2.2 \times10^{1}$              
           & 9.80
           & 0.79
           & \hspace{5pt}$6.6 \times10^{-1}$ 
           & 327\\
       %     & $2.7 \times10^{2}$ \\

    $512$  & $1.6 \times10^{3}$
           & $6.8 \times10^{1}$
           & 24.0
    %     & $5.5 \times10^{1}$              
           & 29.7
           & 1.24
           & $3.6 \times10^{0}$              
           & 459\\
       %     & $2.1 \times10^{3}$ \\

    $1024$ & $1.3 \times10^{4}$
           & $4.5 \times10^{2}$
           & 28.7
    %     & $2.0 \times10^{2}$              
           & 64.2
           & 2.24
           & $3.3 \times10^{1}$              
           & 400\\
       %     & $1.7 \times10^{4}$ \\

    $2048$ & $1.5 \times10^{5}$
           & $3.8 \times10^{3}$
           & 39.6
    %     & $1.5 \times10^{3}$              
           & 102
           & 2.58
           & --
           & -- \\

    $4096$ & $2.2 \times10^{6}$
           & $2.6 \times10^{4}$
           & 83.6
    %     & $9.4 \times10^{3}$              
           & 234
           & 2.80
           & --
           & -- \\

    $8192$ & $1.8 \times10^{7}$
           & $2.6 \times10^{5}$
           & 68.7
    %     & $8.9 \times10^{4}$              
           & 197
           & 2.87
           & --
           & -- \\ \bottomrule
  \end{tabular}
  \put(-460,90){\rotatebox{90}{DH Sampling}}
  \put(-460,-60){\rotatebox{90}{MW Sampling}}
  \put(-460,-200){\rotatebox{90}{MWSS Sampling}}
  \caption{Computational time of the \texttt{S2FFT} scalar spherical harmonic transform when distributed over a single and three NVIDIA A100 GPUs compared to \texttt{SSHT} running on a multithreaded Xeon(R) E5-2650L v3 CPU. We provide raw round-trip wall-clock time benchmarks and quote the relative acceleration provided by our algorithms. Note that the 3 $\times$ GPU ratio corresponds to the ratio of the acceleration realised when running \texttt{S2FFT} over three compared to one GPU.  This ratio  approaches the ideal ratio of three that corresponds to optimal linear scaling with increasing number of GPUs due to the highly parallelised and balanced nature of our algorithms.} \label{tab:sh_raw_numerical_results}
\end{table}
\begin{table}%[h!]
       \newcolumntype{C}[1]{>{\centering\let\newline\\\arraybackslash\hspace{0pt}}m{#1}}
       \centering % to have the caption near the table
       \begin{tabular}{C{0.05\textwidth} C{0.1\textwidth} C{0.11\textwidth} C{0.1\textwidth} C{0.1\textwidth} C{0.1\textwidth} C{0.11\textwidth} C{0.1\textwidth}} \toprule
              \multicolumn{8}{c}{Wigner transform computational time (ms)}                                                                   \\ \midrule
                     &                                 & \multicolumn{4}{c}{On-the-fly transform} & \multicolumn{2}{c}{Precompute transform} \\
              \cmidrule(lr){7-8}
              \cmidrule(lr){3-6}
              $L$
                     & Time \texttt{SO3}
                     & Time \texttt{S2FFT}
                     & Speed-Up 1 $\times$ GPU
              %     & 3 $\times$ GPU
                     & Speed-Up 3 $\times$ GPU
                     & 3 $\times$ GPU Ratio
                     & Time \texttt{S2FFT}                  
                     & Speed-Up 1 $\times$ GPU \\ \midrule
                     % & Memory (MB)                                                                                                           \\ \midrule

              $8$    & \hspace{5pt}$3.0 \times10^{-1}$
                     & $3.9 \times10^{0}$
                     & 0.08
              %     & $2.4 \times10^{0}$              
                     & 0.12
                     & 1.5
                     & \hspace{5pt}$4.3 \times10^{-2}$ 
                     & 6.98\\
                     % & \hspace{5pt}$8.4 \times10^{-2}$                                                                                       \\

              $16$   & $1.1 \times10^{0}$
                     & $7.7 \times10^{0}$
                     & 0.14
              %     & $4.0 \times10^{0}$              
                     & 0.28
                     & 2.0
                     & \hspace{5pt}$3.9 \times10^{-2}$ 
                     & 28.7\\
                     % & \hspace{5pt}$7.0 \times10^{-1}$                                                                                       \\

              $32$   & $3.4 \times10^{0}$
                     & $1.6 \times10^{1}$
                     & 0.21
              %     & $7.2 \times10^{0}$              
                     & 0.47
                     & 2.24
                     & \hspace{5pt}$9.2 \times10^{-2}$ 
                     & 36.3\\
                     % & $5.7 \times10^{0}$                                                                                                    \\

              $64$   & $2.9 \times10^{1}$
                     & $3.2 \times10^{1}$
                     & 0.93
              %     & $1.4 \times10^{1}$              
                     & 2.11
                     & 2.27
                     & \hspace{5pt}$3.1 \times10^{-1}$ 
                     & 93.9\\
                     % & $4.6 \times10^{1}$                                                                                                    \\

              $128$  & $1.3 \times10^{2}$
                     & $6.8 \times10^{1}$
                     & 1.96
              %     & $2.7 \times10^{1}$              
                     & 4.93
                     & 2.52
                     & $1.4 \times10^{0}$              
                     & 94.4\\
                     % & $3.7 \times10^{2}$                                                                                                    \\

              $256$  & $1.0 \times10^{3}$
                     & $1.6 \times10^{2}$
                     & 6.26
              %     & $6.8 \times10^{1}$              
                     & 14.9
                     & 2.38
                     & $5.9 \times10^{0}$              
                     & 171\\
                     % & $2.9 \times10^{3}$                                                                                                    \\

              $512$  & $6.0 \times10^{3}$
                     & $6.2 \times10^{2}$
                     & 9.73
              %     & $2.4 \times10^{2}$              
                     & 24.9
                     & 2.56
                     & $3.0 \times10^{1}$              
                     & 200\\
                     % & $1.9 \times10^{4}$                                                                                                    \\

              $1024$ & $5.8 \times10^{4}$
                     & $4.5 \times10^{3}$
                     & 12.9
              %     & $1.6 \times10^{3}$              
                     & 36.1
                     & 2.80
                     & --
                     & --                                                                                                               \\

              $2048$ & $6.4 \times10^{5}$
                     & $3.5 \times10^{4}$
                     & 18.4
              %     & $1.2 \times10^{4}$              
                     & 53.6
                     & 2.91
                     & --
                     & --                                                                                                               \\

              $4096$ & $1.0 \times10^{7}$
                     & $2.8 \times10^{5}$
                     & 37.1
              %     & $9.3 \times10^{4}$              
                     & 111
                     & 2.99
                     & --
                     & --                                                                                                              \\ \bottomrule
       \end{tabular}
       \vspace{-5pt}
       \caption{Computational time of the \texttt{S2FFT} Wigner transform with azimuthal bandlimit $N=5$ when distributed over a single and three NVIDIA A100 GPUs compared to \texttt{SO3} running on a multithreaded Xeon(R) E5-2650L v3 CPU. We provide raw round-trip wall-clock time benchmarks and quote the relative acceleration provided by our algorithms. Note that the 3 $\times$ GPU ratio corresponds to the ratio of the acceleration realised when running \texttt{S2FFT} over three compared to one GPU.  This ratio  approaches the ideal ratio of three that corresponds to optimal linear scaling with increasing number of GPUs due to the highly parallelised and balanced nature of our algorithms.} \label{tab:wig_results}
       \vspace{-15pt}
\end{table}

\begin{figure}
   \centering
   \begin{subfigure}{.5\textwidth}
      \centering
      \includegraphics[width=\linewidth, height=11cm,keepaspectratio, trim={0cm 0cm 0.23cm 0cm}, clip]{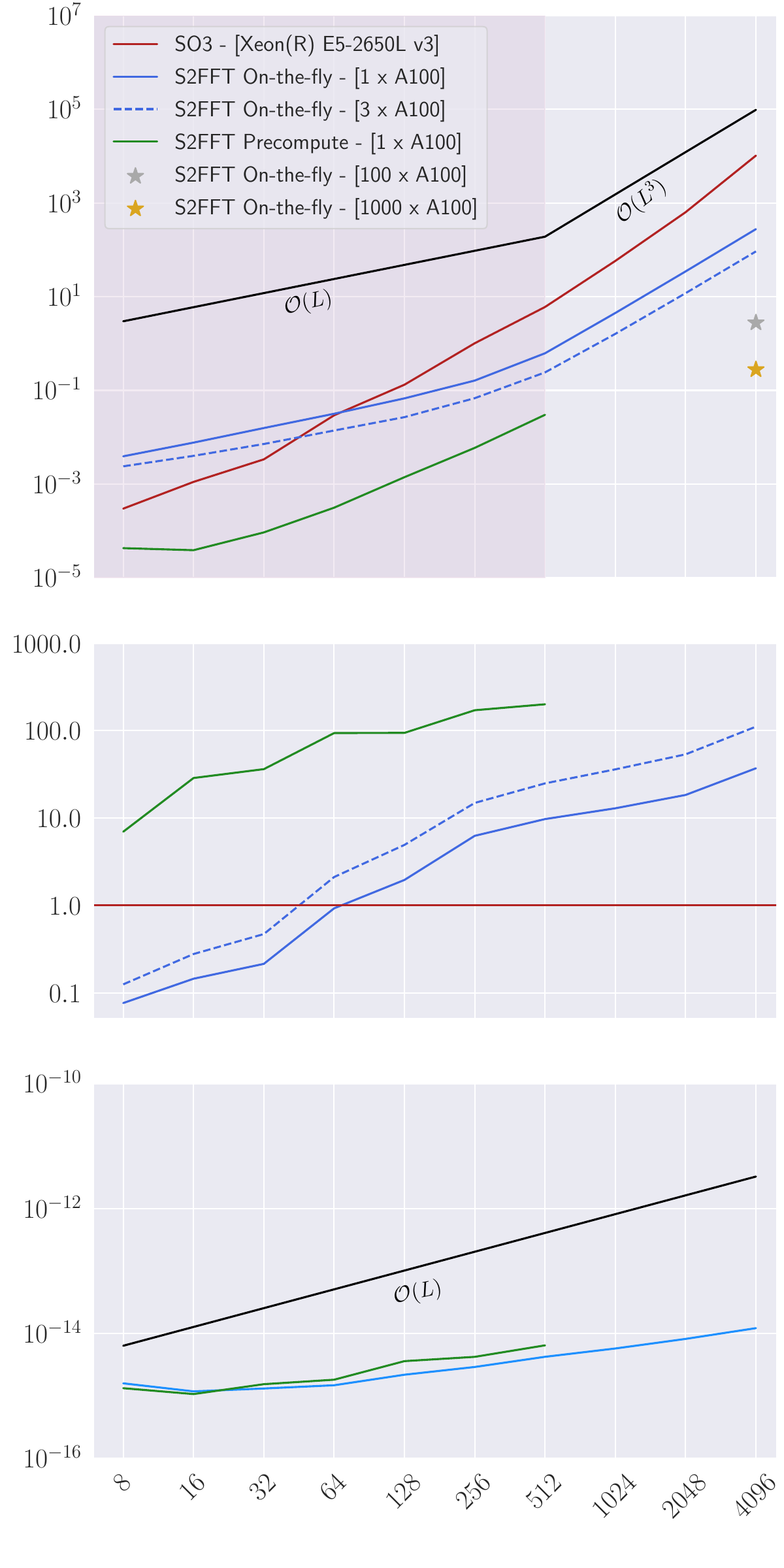}
      \put(-88,0){\rotatebox{0}{\footnotesize Bandlimit $L$}}
      \caption{MW sampling}
      % \label{fig:sub1}
   \end{subfigure}%
   \hspace{-2.6cm}
   \begin{subfigure}{.5\textwidth}
      \centering
      \includegraphics[width=\linewidth, height=11cm,keepaspectratio, trim={2cm 0cm 0.23cm 0cm}, clip]{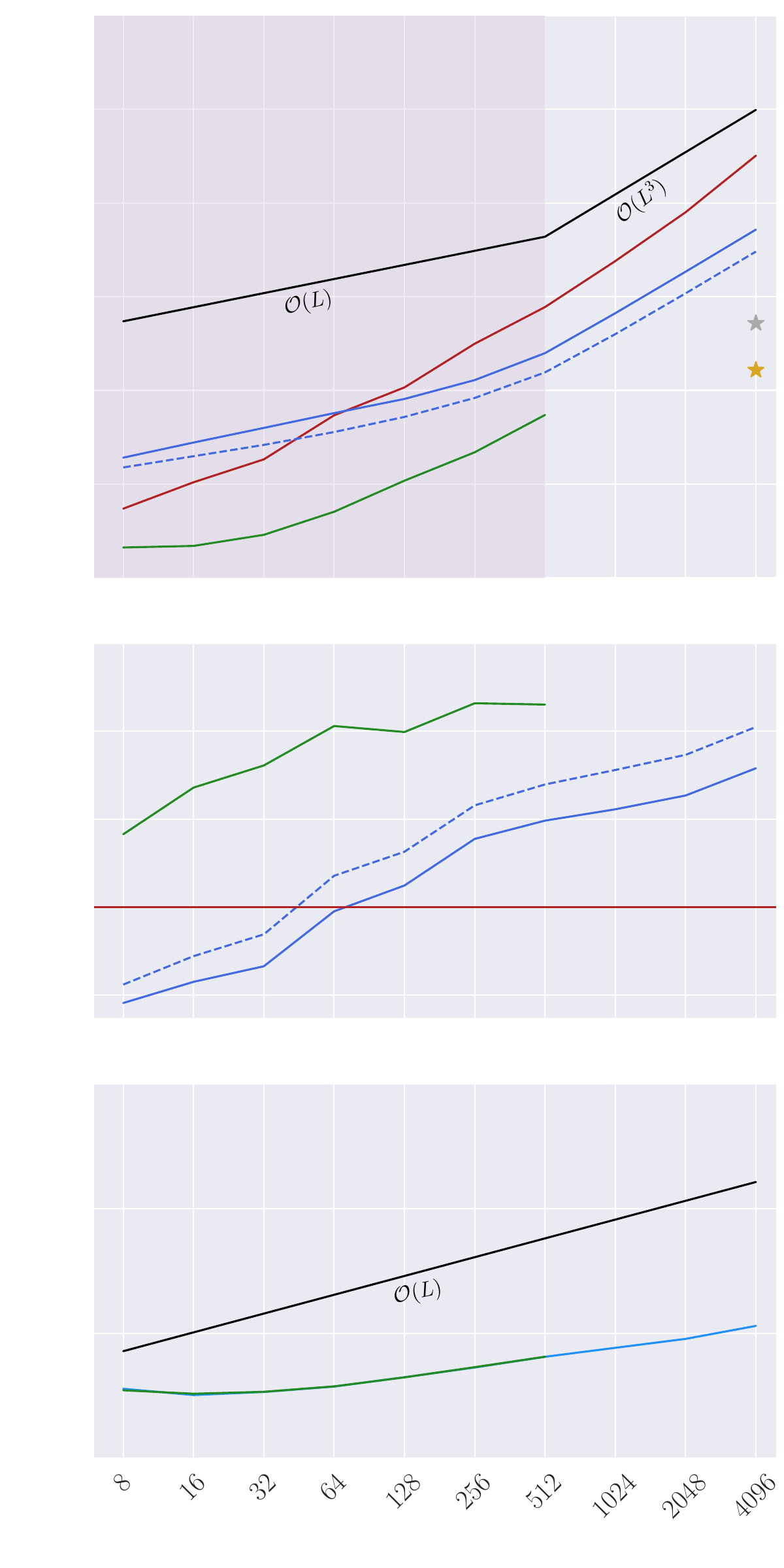}
      \put(-88,0){\rotatebox{0}{\footnotesize Bandlimit $L$}}
      \caption{MWSS sampling}
      % \label{fig:sub2}
   \end{subfigure}
   % \put(-132,15){\rotatebox{0}{\footnotesize Bandlimit $L$}}
   % \put(-292,15){\rotatebox{0}{\footnotesize Bandlimit $L$}}
   \put(-365,66){\rotatebox{90}{\footnotesize Error}}
   \put(-365,140){\rotatebox{90}{\footnotesize Acceleration}}
   \put(-365,230){\rotatebox{90}{\footnotesize Compute Time (s)}}
   \caption{Numerical benchmarking results for the Wigner transforms implemented \texttt{S2FFT} compared to the C implementation of \texttt{SO3}. For this analysis we consider MW and MWSS \cite{mcewen:fssht} samplings of the rotation group. As these samplings afford sampling theorems on the rotation group, their corresponding Wigner transforms are exact. In these comparisons we fix the azimuthal bandlimit to $N=5$, a typical use case corresponding to 9 $\gamma$ rotation angles. \texttt{S2FFT} provides multiple operational modes: a precompute approach with faster compute at $\mathcal{O}(NL^3)$ memory overhead and an on-the-fly transform with negligible memory overhead (see \sectn{\ref{sec:algorithms}}).
      \textbf{Top:} Average compute time required for a Wigner transform for each operational modality.
      \textbf{Middle:} Relative acceleration of a given modality with respect to their \texttt{SO3} counterpart.
      \textbf{Bottom:} Averaged round-trip numerical error. In each case we recover machine precision. We also empirically observe at most $\mathcal{O}(L)$ error scaling (in line with previous work \cite{mcewen:so3}). The silver and gold stars represent the performance one may expect to realise should one distribute our on-the-fly algorithms across 100 and 1000 GPUs respectively.}
   \label{fig:wt_benchmarking}
\end{figure}

\section{Conclusions} \label{sec:conclusion}

We have developed differentiable and accelerated algorithms to compute generalised Fourier transforms on the sphere and rotation group and their gradients.  We present a recursive algorithm for the calculation of Wigner $d$-functions that is both stable to high harmonic degrees and extremely parallelisable. This level of parallelisation is achieved as we recurse along harmonic order $m$ independently for all other indices, which is stable due to a custom on-the-fly normalisation scheme. Leveraging this recursion we develop fast and exact spin spherical harmonic and Wigner transforms that are well-suited for the high throughput computing provided by modern hardware accelerators.  We develop a hybrid
% automatic and manual 
differentiation approach so that gradients can be computed efficiently, avoiding the memory overhead of full automatic differentiation while also avoiding the complexities of full manual differentiation.  Both forward- and reverse-mode differentiation is supported.

To support the above advances, we develop and release \texttt{S2FFT} {\href{https://github.com/astro-informatics/s2fft}{\faGithub}}, an open-source software library implemented in the \texttt{JAX} differentiable programming framework.  Our transforms provide accelerated computation over one or multiple hardware accelerators (\textit{i.e.}\ GPUs and TPUs), accelerated and memory-efficient gradient computation, and are largely agnostic to the spherical sampling considered, with support already for DH \cite[Driscoll \& Healy;][]{driscoll:1994}, MW \cite[McEwen \& Wiaux;][]{mcewen:fssht,mcewen:so3}, MWSS \cite[McEwen \& Wiaux symmetric sampling;][]{mcewen:fssht, daducci:ssdmri, ocampo:disco}, and \texttt{HEALPix} \cite{gorski:2005} sampling, with plans to add others in the near future.
\texttt{S2FFT} supports both real and complex functions of arbitrary spin within a single easy to use API. As such, and in line with our design ethos, \texttt{S2FFT} is lightweight, straightforward to install, well tested and documented, and can be integrated into existing differentiable pipelines easily. For moderate resolution application we also provide a precompute mode, with enhanced computational speed at the cost of added memory requirements.

\texttt{S2FFT} is comprehensively benchmarked up to bandlimit $L=8192$ (higher bandlimits are certainly feasible but have not been considered to date due to the computational burden of further benchmarking).
The transforms and gradient computations implemented are validated, where for sampling theorems that admit exact spherical transforms the error of a round-trip transform is observed to be of order of machine precision.
In addition we benchmark the accelerated computation time of \texttt{S2FFT} against alternative C codes implementing spherical harmonic \cite{mcewen:fssht} and Wigner \cite{mcewen:so3} transforms.
Our transforms provide up to a 234-fold and 400-fold acceleration, respectively, for the on-the-fly and precompute transforms.  Moreover, our spin spherical harmonic transforms are engineered to distribute compute across many accelerators.  For high bandlimits we achieve very close to optimal linear scaling with increasing number of GPUs due to the highly parallelised and balanced nature of our algorithms, \textit{i.e.}\ when considering $3$ GPUs we realise a $3\times$ acceleration compared to running on a single GPU.  Provided access to sufficiently many GPUs our transforms thus exhibit an unprecedented effective linear time complexity.

With researchers becoming increasingly interested in differentiable programming for scientific applications, we hope these foundational tools will be of great use in coming years. The differentiability afforded by our transforms means that, for example, scattering transforms on the sphere \citep{mcewen:scattering}, which rely on spherical harmonic, Wigner, and spherical wavelet transforms, can be extended to differentiable transforms, which then opens up their use as statistical generative models, as has been considered previously in the planar setting \citep[\textit{e.g.}][]{cheng:2021:weak}.  Indeed, we have already leveraged the differentiable spherical harmonic and Wigner transforms developed in the current article to realise differentiable and accelerated wavelet \citep{price:s2wav} and scattering \citep{mouset:2024} transforms on the sphere, and corresponding generative models.
Furthermore, the acceleration afforded by our transforms means that, for example, sub-kilometre resolution and below is now feasible for global spherical modelling or analysis of the Earth's climate or geophysical properties through data-driven or hybrid methods.  These are just a small number of potential uses of the differentiable and accelerated spherical transforms developed in the current article; we hope they will find many more uses in the sciences and beyond.

\section*{CRediT authorship contribution statement}
Author contributions are specified below, following the Contributor Roles Taxonomy (CRediT\footnote{\url{https://www.elsevier.com/authors/policies-and-guidelines/credit-author-statment}}).\\
\textbf{Matthew~A.~Price:}
\noindent Conceptualisation, Methodology, Software, Investigation, Validation, Writing (Original Draft, Review \& Editing)\\
\textbf{Jason~D.~McEwen:}
\noindent Conceptualisation, Methodology, Software, Investigation, Validation, Writing (Original Draft, Review \& Editing).
\section*{Declaration of competing interest}
The authors declare they have no known competing financial interests or personal relationships that could have appeard to influence the work reported in this paper.

\section*{Data availability}
All algorithms and transforms presented in this work are collectively released to the public in a professionally developed open-source software package \texttt{S2FFT}\footnote{\url{https://github.com/astro-informatics/s2fft}} {\href{https://github.com/astro-informatics/s2fft}{\faGithub}}. All results presented in this paper are entirely and transparently reproducible, provided access to appropriate computing hardware. This software is an ongoing development project and we strongly encourage community involvement going forward.

\section*{Acknowledgments}
The authors would like to thank Martin Reinecke for discussions relating to \texttt{HEALPix} spherical harmonic transforms and to Fran\c{c}ois Lanusse for discussions relating to \texttt{JAX} development.
We developed the software presented in this article in collaboration with Advanced Research Computing (ARC) at UCL through an open source software sustainability initiative. This work is supported by EPSRC (grant number EP/W007673/1).

%%Vancouver style references.
\bibliographystyle{model1-num-names}
\bibliography{bib, mybibs_new, bib_myname}

% \appendix
% \include{appendixA}

%
% \section{Price-McEwen Wigner recursion stability}
% Stability of our recursive algorithm
%
% \section{Extended benchmarking results}
% Additional test details and runs to make the body more readable.

\end{document}